\documentclass[showpacs,preprintnumbers,amsmath,amssymb]{revtex4-1}

\usepackage{bm}
\usepackage{mathrsfs}
\usepackage{subfigure}
\usepackage{graphicx}
\usepackage{epstopdf}
\usepackage{color}
\newcommand{\md}{\mathrm{d}}
\newcommand{\Ltd}{\Lambda_{\rm 3D}}
\newcommand{\Lfd}{\Lambda_{\rm 4D}}
\newcommand{\Lpv}{\Lambda_{\rm PV}}
\newcommand{\Lpt}{\Lambda_{\rm PT}}

\begin{document}
\title{Regularization dependence on phase diagram
in Nambu--Jona-Lasinio model}

\author{T. Inagaki}
\affiliation{
Information Media Center, Hiroshima University,
Higashi-Hiroshima, Hiroshima 739-8521, Japan
}
\author{D. Kimura}
\affiliation{
General Education, Ube National College of Technology,
Ube, Yamaguchi 755-8555, Japan
}
\author{H. Kohyama}
\affiliation{
Department of Physics,
National Taiwan University,
Taipei, Taiwan 10617
}

\date{\today}

%%%%%%%%%%%%%%%%%%%%%%%%%%%%
\begin{abstract}
We study the regularization dependence on meson properties and
the phase diagram of quark matter by using the two flavor
Nambu--Jona-Lasinio model. We find that the meson properties and
the phase structure do not show drastically difference depending the
regularization procedures. We also find that the location or the
existence of the critical end point highly depends on the regularization
methods and the model parameters. Then we think that regularization
and parameters are carefully considered when one investigates the
QCD critical end point in the effective model studies. 
\end{abstract}
%%%%%%%%%%%%%%%%%%%%%%%%%%%%

%\pacs{11.30.Qc, 12.39.-x}

\maketitle

%%%%%%%%%%%%%%%%%%%%%%%%%%%%
% 1. INTRODUCTION          %
%%%%%%%%%%%%%%%%%%%%%%%%%%%%
\section{INTRODUCTION}
%%%%%%
The phase structure of quark matter on finite temperature and density
has actively been studied for decades~\cite{Fukushima:2010bq}. Under
usual condition, meaning low temperature and density, quarks are
confined inside hadrons and they never be able to observed as a
single particle. On the other hand, due to the nature of the asymptotic
freedom~\cite{Gross:1973id},
quarks and gluons can be free from the confinement at high temperature
and density, because the coupling
strength becomes weak at high energy. It is, therefore, expected that
quark matter undergoes the confined/deconfined phase transition at some
temperature and density.
%[Refs. if any].
This is important subject both in theoretical
and experimental studies since it crucially relates to the quark matter
properties at relativistically high energy collisions and extremely dense
stellar objects such as neutron stars.
%{\bf [Refs]}.

The first principle for quarks and gluons is Quantum Chromodynamics
(QCD) which is a non-Abelian gauge field theory for fermions.
Our goals is to evaluate the phase structure based on this first
principle QCD,
however, it is difficult to extract theoretical predictions due to the nature
of complicated strongly interacting system. One of the most reliable 
approaches is to use the discretised version of QCD called the Lattice 
QCD (LQCD) in which theoretical calculation is performed on the discrete
spacetime~\cite{Wilson:1974sk}. Although the LQCD works well at finite
temperature $T$ for
small chemical potential $\mu \simeq 0$, there is the technical difficulty
called the ``sign problem" when one tries to investigate the system at
intermediate chemical potential. There effective models maybe nicely
adopted because some models can consistently treat the system at finite
temperature and chemical potential.

For the sake of evaluating the phase structure of quark
matter at finite temperature and chemical potential, we will employ the
Nambu--Jona-Lasinio (NJL) model~\cite{NJL} which is the most
frequently used one in this context (there are a lot of nice review papers
on the model, see,  e.g.~\cite{Vogl:1991qt,Klevansky:1992qe,
Hatsuda:1994pi,Buballa:2005rept,Huang:2004ik}.)
The model is constructed by incorporating the four point
quark interaction into the model Lagrangian, so it is not renormalizable
due to this higher dimensional operator.
Therefore, the physical predictions
of the model inevitably depend on the regularization procedure and
the model parameters chosen. The resulting phase diagram on the
$T-\mu$ plane is as well affected by the parameters and regularization
prescriptions. So it is an important issue to study whether the phase structure
obtained in one regularization method is consistent with the ones from
different regularization methods.

In this paper, we are going to study the phase structure of quark
matter in the NJL model with various regularization
ways, which are three
dimensional ($3$D) momentum cutoff, four dimensional ($4$D)
momentum cutoff, Pauli-Villars (PV) regularization, proper-time (PT) 
regularization, and the dimensional regularization (DR). The $3$D cutoff
scheme is the most popular method in this model and a lot of works have been
done in this way. The $4$D cutoff method preserves the
Lorentz symmetry in which space and time are treated on equal footing.
The Pauli-Villars regularization is based on the subtraction
of the amplitude considering the virtually heavy particle to suppress
the unphysical high energy
contribution coming from loop integrals
\cite{Pauli:1949zm, Itzykson1980,Cheng1984}.
The proper-time regularization makes integrals finite through the 
exponentially dumping factor
\cite{Schwinger:1951nm,Itzykson1980}.
The dimensional regularization analytically continues
the spacetime dimension in the loop integrals to a non-integer value,
then try to obtain finite contribution from the 
integrals \cite{'tHooft:1972fi}.
Beside from the frequently
used $3$D cutoff way, there have been a lot of works by using the
%%%%%%
$4$D \cite{Klevansky:1992qe,Hatsuda:1994pi},
%%%%%%
PV \cite{Klevansky:1992qe,Kahana:1992jm,Osipov:2004bj,Moreira:2010bx}, 
%%%%%%
PT \cite{Klevansky:1992qe,Suganuma:1990nn, Klimenko:1990rh,
Klimenko:1991he,Gusynin:1994re,Inagaki:1997nv,
Inagaki:2003yi,Inagaki:2004ih,Inagaki:2003ac,Cui:2014hya}, and
%%%%%%
DR \cite{Krewald:1991tz,jafarov:2004,jafarov:2006,
inagaki:2008,Inagaki:2011uj,Inagaki:2012re}.
The physical consequences depend on the regularization
\cite{fujihara:2009}.
%%%%%%

%
%
%I think it is nice that we align more related works based on various
%{\bf regularizations
%to find a regularization independent properties of NJL model
%and to construct a suitable effective model of QCD.}
%{\bf SUGGESTION: ``Please add more descriptions and
%references relating to regularizations."}

This paper is organized as follows; Section \ref{sec_model} introduces
the model Lagrangian, and show the model treatment on the meson
properties and the explicit formalism at finite temperature and chemical
potential.  In Sec. \ref{sec_reg}, we present various regularization
procedures, $3$D, $4$D, PV, PT and DR prescriptions with explicit
equations. We then perform the parameter fitting in Sec. \ref{sec_para}.
In Sec. \ref{sec_meson}, the numerical results of the meson properties
are shown. We then draw the phase diagrams with several parameter sets
using various regularization methods in Sec. \ref{sec_pd}.
We also study
the phase diagram with the parameters fixed under the condition with
the same constituent quark mass in Sec. \ref{sec_m_fixed}.
In Sec. \ref{sec_discussion}, we give the discussions on the obtained
results. Finally, we write the concluding remarks in Sec. \ref{sec_conclusion}.
Several detailed calculations are shown in Appendix.

%%%%%%%%%%%%%%%%%%%%%%%%%%%%
\section{Two flavor NJL model}
\label{sec_model}
%%%%%%%%%%%%%%%%%%%%%%%%%%%%
In this paper we consider two light quarks with equal mass.
The model has $SU_{L}(2)\otimes SU_R(2)$ flavor symmetry at
the massless limit, $m\rightarrow 0$.

%%%%%%%%%%%%%%%%%%%%%%%%%%%%
\subsection{The Lagrangian and gap equation}
%%%%%%%%%%%%%%%%%%%%%%%%%%%%
The Lagrangian of the two flavor NJL model is given by
%%%%%%
\begin{eqnarray}
\mathcal{L}=
   \bar{\psi} (i\! \not\! \partial - \hat{m}) \psi
   +G \left[ (\bar{\psi} \psi )^2 
    +( \bar{\psi} i \gamma_{5} \tau^{a} \psi )^2 \right],
\label{lag}
\end{eqnarray}
%%%%%%
where $\hat{m}$ is the diagonal mass matrix $\hat{m}={\rm diag}(m_u,m_d)$
and $G$ is the
effective coupling strength of the four point interaction.
We set $m_d=m_u$ in this paper.
The application of  the mean-field approximation
%%%%%%
\begin{eqnarray}
   \langle \overline{\psi} \psi \rangle  
   \simeq -\frac{\sigma}{2G}
\label{MFA}
\end{eqnarray}
%%%%%%
leads the following mean-field Lagrangian
%%%%%%
\begin{eqnarray}
   \tilde{\mathcal{L}}=
   \bar{\psi} (i\! \not\! \partial - m^*) \psi
   -\frac{\sigma^2}{4G},
\label{MFA-lag}
\end{eqnarray}
%%%%%%
with the constituent mass $m^* = m_u + \sigma$. 
The flavor symmetry is broken down,
$SU_{L}(2)\otimes SU_{R}(2) \rightarrow SU_{L+R}(2)$,
by non-vanishing current quark mass, $m_u$, and dynamically
generated $\sigma$.
Thanks to the simple form of the Lagrangian, one can easily
evaluate the effective potential,
${\mathcal V}_{\rm eff}=-\ln Z/V$ where $Z$ is the partition
function
%%%%%%
\begin{align}
  &Z = \int \!{\mathcal D}[\psi] \exp 
       \left[ i \int \md^4 x \,
          \tilde {\mathcal L} 
       \right]
\end{align}
%%%%%%
and $V$ is the volume of the system. After some algebra, we see
%%%%%%
\begin{eqnarray}
{\mathcal V}_{\rm eff}(\sigma) &=& \frac{\sigma^2}{4G}
 -\int \frac{\md^4k}{i(2\pi)^4} \ln \det (\not\! k - m^*).
\label{potential}
%&=& \frac{\sigma^2}{4G}
% -2 N_c N_f \int \frac{d^4 k}{i(2\pi)^4} 
% \ln(m^{* 2} - k^2 ).
\end{eqnarray}
%%%%%%
The detailed derivation of the effective potential is presented 
in \cite{Huang:2004ik}.

The gap equation is obtained through the extreme condition of the
potential with respect to $\sigma$, namely,
%%%%%%
\begin{eqnarray}
  \frac{\partial {\mathcal V}_{\rm eff}}{\partial \sigma} = 0.
\label{diff_V}
\end{eqnarray}
%%%%%%
This condition leads the following gap equation
%%%%%%
\begin{eqnarray}
  \sigma &=& 2N_f G \cdot i {\rm tr}S(m^*) ,
\end{eqnarray}
%%%%%%
with the number of flavors $N_f$ and
%%%%%%
\begin{eqnarray}
  i {\rm tr} S(m^*) 
  &=& -  {\rm tr} \int \frac{\md^4 k}{i(2\pi)^4}
  \frac{1}{ \not\! k - m^* + i\varepsilon} ,
\end{eqnarray}
%%%%%%
where trace takes the spinor and color indices.  This is the key
equation in the model because it determines the values of the
chiral condensate $\langle \overline{\psi} \psi \rangle$
and the constituent quark mass $m^*$.

%%%%%%%%%%%%%%%%%%%%%%%%%%%%
\subsection{Meson properties}   %
%%%%%%%%%%%%%%%%%%%%%%%%%%%%
The properties of the pion and sigma meson can be studied based on
the model with the determined chiral condensate. The interacting 
Lagrangian of the pion and quarks is written by
%%%%%%
\begin{eqnarray}
\mathcal{L}_{\pi qq}=
   i g_{\pi qq} \bar{\psi} \gamma_5  \tau \cdot \pi \psi,
\label{lag}
\end{eqnarray}
%%%%%%
where $\tau_i$ are $2\times 2$ matrices in the flavor space and $\pi^i$
represent the pion fields. The explicit expression is
$\tau \cdot \pi = \tau_- \pi^- + \tau_+ \pi^+ + \tau_0 \pi^0$,
with $\tau_{\pm} = (\tau_1 \pm \tau_2)/\sqrt{2}$ and $\tau_0 = \tau_3$
where $\tau_i$ are the Pauli matrices.

By applying the random phase approximation, we can write the pion
propagator as the summation of the geometrical series of the
one-loop diagram, which gives
%%%%%%
\begin{eqnarray}
 \Delta_\pi(p^2)
 = \frac{g_{\pi qq}^2}{p^2 - m_\pi^2}
 \simeq \frac{2G}{1-2G\Pi^\pi(p^2)},
\label{pion_prop}
\end{eqnarray}
%%%%%%
where $\Pi^\pi$ is the following quark loop contribution 
%%%%%%
\begin{align}
\Pi^{\pi} (p^2)
 =-2 \int \!\! \frac{\md^4 k}{i(2\pi)^4} {\rm tr}
  \left[ \gamma_5 S(k) \gamma_5 S(k-p)
  \right],
\end{align}
%%%%%%
with the quark propagator
\begin{equation}
  S(k) = \frac{1}{ \not\! k - m^* + i \epsilon}.
\end{equation}
%%%%%%
The explicit derivation of Eq.~(\ref{pion_prop}) is discussed in the review
paper~\cite{Klevansky:1992qe}. The pion mass is calculated at the pole
position of the propagator, so the condition reads
%%%%%%
\begin{eqnarray}
 \left[ {1-2G\Pi^\pi(p^2)} \right]\bigl|_{p^2 = m_\pi^2} =0.
\label{pion_mass}
\end{eqnarray}
%%%%%%
It should be noted that the residue at the pole $p^2=m_\pi^2$ coincides 
with  the square of the coupling strength $g_{\pi qq}^2$ so we have the 
relation
%%%%%%
\begin{eqnarray}
 g_{\pi qq}^2
 =\left( \frac{\partial \Pi^\pi}{\partial p^2} \right) ^{-1}
 \biggl|_{p^2 = m_\pi^2}. 
\label{g_piqq}
\end{eqnarray}
%%%%%%

In the similar manner, the sigma meson mass is evaluated at the pole
of the propagator,
%%%%%%
\begin{eqnarray}
 \Delta_\sigma(p^2)
 = \frac{g_{\sigma qq}^2}{p^2 - m_\sigma^2}
 \simeq \frac{2G}{1-2G\Pi^\sigma(p^2)},
\label{pion_sigma}
\end{eqnarray}
%%%%%%
with
%%%%%%
\begin{align}
\Pi^{\sigma} (p^2)
 =-2 \int \!\! \frac{\md^4 k}{i(2\pi)^4} {\rm tr}
  \left[ S(k) S(k-p)
  \right].
\end{align}
%%%%%%
Therefore the condition which determines the sigma meson mass
becomes
%%%%%%
\begin{eqnarray}
 \left[ {1-2G\Pi^\sigma(p^2)} \right]\bigl|_{p^2 = m_\sigma^2} =0.
\label{sigma_mass}
\end{eqnarray}
%%%%%%

The pion decay constant is calculated from the following equation
%%%%%%
\begin{align}
  i \delta_{ij} p^{\mu} f_{\pi}
  =
  \langle 0| \bar{\psi} \frac{\tau_j}{2} \gamma^{\mu}
             \gamma_5 \psi  |\pi_i \rangle.
\end{align}
%%%%%%
The explicit form becomes
%%%%%%
\begin{align}
  p^{\mu}f_{\pi} = \frac{1}{2} 
  \int \frac{\md^4 k}{i (2\pi)^4}
  \, {\rm tr}
   \left[ \gamma^{\mu} \gamma_5 g_{\pi qq} S(k) \gamma_5 S(k-p)
   \right].
\label{decay}
\end{align}
%%%%%%

Thus the equations Eqs.~(\ref{pion_mass}), (\ref{sigma_mass})
and (\ref{decay}) are the ones which determine the pion mass,
sigma mass, and the pion decay constant.

%Then the pion decay constant is evaluated as
% \begin{align}
%  f_{\pi}
%  = 4 N_{\mathrm c} g_{\pi qq}M 
%          \int \frac{\md^3 k}{(2\pi)^3}
%          \frac{1}{E_k(4E_k^2-m_{\pi}^2)} \,.
% \end{align}

%%%%%%%%%%%%%%%%%%%%%%%%%%%%
\subsection{Explicit formalism at finite temperature}   %
%%%%%%%%%%%%%%%%%%%%%%%%%%%%
Since our purpose is to study the phase structure on temperature $T$
and chemical potential $\mu$, we need to extend the equations to
finite temperature. According to the imaginary time formalism,
the integral region of the
temporal component becomes finite due to the periodic or anti-periodic
condition of fields as
%%%%%%
\begin{align}
  Z = \int \!{\mathcal D}[\psi] \exp 
       \left[ \int_0^{\beta} \!\md  \tau \int \! d^3 {\bf x} 
          \Bigl(\tilde {\mathcal L} + \mu \bar{\psi}\gamma_0 \psi \Bigr)
       \right].
\label{partition}
%   = Z_{\rm const.} Z_q \,, \\
%  &Z_{\rm const} =  \exp
%       \left[ -\int_0^{\beta} \!\! \int \!\! d^3 \! {\bf x} \,
%             \bigl( G \phi^2 \bigr)
%                               \right] \,,
%  \quad
%  Z_q =  \int \!{\mathcal D}q \exp
%         \left[ \int_0^{\beta} \!\! \int \!\! d^3 \! {\bf x} \,
%           \Bigl( \bar{q} (i \not\! p -M + \mu \gamma_0) q \Bigr)
%         \right].
\end{align}
%%%%%%
where $\tau$ is imaginary time and $\beta$ is the inverse temperature
$1/T$. Consequently, continuous integral in the temporal direction is 
replaced by the following discrete summation,
%%%%%%
\begin{align}
  \int \frac{\md^4 k}{i(2\pi)^4} F(k_0, {\bf k})
  \to   T\sum_{n=-\infty}^{\infty} 
   \int \frac{\md^3 k}{(2\pi)^3} F(i\omega_n+\mu, {\bf k}),
\label{finite_t}
\end{align}
%%%%%%
where $\omega_n = 2n\pi T$ or $(2n+1)\pi T$ depending on the
statistical property of field, i.e., for bosons or fermions, and the chemical
potential seen in Eq. (\ref{partition}) appears in the way $i\omega_n + \mu$.
In this paper, we only treat fermionic quark loop contributions
then $\omega_n = (2n+1)\pi T$ is always the case.

With the help of the formalism Eq.(\ref{finite_t}), we see that the
gap equation at finite temperature becomes
%%%%%%
\begin{align}
 &\sigma = 2N_f G \cdot [{\rm tr} S^{0} + {\rm tr} S^{T}], \\ 
 %%%%%%
 & {\rm tr} S^{0} = - N_c m^* \!\! \int \frac{\md^3 k}{(2\pi)^3}
       \frac{1}{2E}, 
       \label{trS0} \\
 & {\rm tr} S^{T} =  N_c m^* \!\! \int \frac{\md^3 k}{(2\pi)^3}
       \frac{1}{2E}
     \left[ 
       \sum_{\pm} f(E^{\pm})
     \right],
     \label{trST}
\end{align}
%%%%%%
%where $\beta$ is the inverse temperature $1/T$,
where $N_c$ is the number of colors,
$E = \sqrt{k^2
+ m^{*2}}$, $E^{\pm}=E \pm \mu$ and $f(E)=1/(1+e^{\beta E})$.
It is important to note
that the contributions can be expressed by the summation of the $T$
independent part (${\rm tr}S^0$) and $T$ dependent part (${\rm tr}S^T$).
This characteristic is general
if one takes the infinite number of the frequency summation
in finite temperature field theory and
crucial when we apply the regularization procedures to the
appearing integrals.

Since the gap equation is derived by differentiating the effective potential
with respect to $\sigma$, then the effective potential can be obtained
by integrating the gap equation (see, for example, \cite{Inagaki:1995}),
%%%%%%
\begin{eqnarray}
{\mathcal V}(\sigma) = \frac{\sigma^2}{4G}
  -N_f \int^{\sigma}_0 \md \sigma^{\prime}\ 
  i{\rm tr} S(m_u+\sigma^{\prime}) .
\label{integral_gap}
\end{eqnarray}
%%%%%%
where we have dropped the suffix in ${\mathcal V}_{\rm eff}$ and just
written ${\mathcal V}$ for notational simplicity. Thereafter the effective
potential at finite temperature ${\mathcal V} = {\mathcal V}^\sigma
+ {\mathcal V}^0 +{\mathcal V}^T$ is evaluated as
%%%%%%
\begin{align}
 &{\mathcal V}^\sigma
 = \frac{\sigma^2}{4G},\\
 &{\mathcal V}^0
 =    -2N_f N_c \int \frac{\md^3 k}{(2\pi)^3} E, \\
 &{\mathcal V}^T
 =    -2N_f N_c T\int \frac{{\rm d}^3 k}{(2\pi)^3}
           \sum_{\pm}
             \ln \Bigl[1 + e^{-\beta E^{\pm}} \Bigr].
\label{thermo}
\end{align}
%%%%%%
It is important to note that, if we apply some regularizations,  the results
between the direct calculation from Eq.~(\ref{potential}) and the one after integrating the gap equation may become different, because
regularization essentially means the subtraction and there are several
ways of subtractions. Therefore, in this paper, we persistently use the
latter way shown in Eq.~(\ref{integral_gap}) so that the model treatment
becomes consistent.
It should be noticed that the finite temperature correction, ${\mathcal V}^T$,
contains no divergent integral. A finite result can be obtained for the
finite temperature correction without applying any regularizations.

Next, we carry on the integral in the meson properties; the one loop
contribution can be written as
%%%%%%
\begin{equation}
 \Pi^{\pi} (p^2)
  =  -\frac{2 {\rm tr} S }{m^*} + p^2 I(p^2) ,
\end{equation}
%%%%%%
with
%%%%%%
\begin{equation}
I(p^2)=4N_c \int \frac{\md^4 k}{i(2\pi)^4}
  \frac{1}{ ( k^2-m^{*2} ) [ (k-p)^2-m^{*2} ]}.
\end{equation}
%%%%%%
Since ${\rm tr}S$ is already evaluated above, the remaining task is
to calculate $I (=I^0 + I^T)$, and it becomes
%%%%%%
\begin{align}
 &I^0(p^2)
  = 4N_c \int \frac{\md^3 k}{(2\pi)^3}
     \frac{1}{E(4E^2-p^2)} ,
  \label{I0} \\
 &I^T(p^2)
  = -4N_c \int \frac{\md^3 k}{(2\pi)^3}
   \frac{ \sum_{\pm} f(E^{\pm}) }{E(4E^2-p^2)}.
  \label{IT}
\end{align}
%%%%%%
Similarly, the one-loop diagram of the scalar channel can be
written as
%%%%%%
\begin{align}
 \Pi^{\sigma}(p^2)
 = -\frac{2 {\rm tr} S }{m^*}   
    +(p^2-4m^{*2})\, I(p^2).
\end{align}
%%%%%%
We now have already evaluated all the ingredients of $\Pi^\sigma$
above in Eqs. (\ref{trS0}), (\ref{trST}), (\ref{I0}) and (\ref{IT}), so we
do not need further calculations.

Finally, let us derive the equation for the pion decay constant. After a
bit of algebra we obtain the relation,
%%%%%%
\begin{align}
 f_\pi = g_{\pi qq} m^* I(0).
\end{align}
%%%%%%
Here we evaluate $f_\pi$ at $p^2=0$ following~\cite{Klevansky:1992qe}.

%%%%%%%%%%%%%%%%%%%%%%%%%%%%
\section{Regularization procedures}      %
\label{sec_reg}
%%%%%%%%%%%%%%%%%%%%%%%%%%%%
Since the integrals obtained in the previous section include infinities,
we need to apply some regularization so that the model leads finite
quantities. As mentioned in the introduction, the model is
not renormalizable, then the model predictions inevitably depend on
regularization procedures chosen. Here, we shall present possible
regularization methods in this section.

%%%%%%%%%%%%%%%%%%%%%%%%%%%%
\subsection{Three dimensional cutoff scheme}    %
%%%%%%%%%%%%%%%%%%%%%%%%%%%%
The idea of the three dimensional (3D) cutoff is simple; one drops
high frequency mode by introducing the cutoff scale $\Ltd$ into the 
integrals. We work in the 3-dimensional polar coordinate system and
cut the radial coordinate as
%%%%%%
\begin{align}
  \int \frac{\md^4 k}{(2\pi)^4}
  \to 
  \int \frac{\md k_0}{2\pi}
  \int^{\Ltd}_0 \!\! \frac{k^2 \md k}{(2\pi)^3}\int d\Omega_{3}.
\end{align}
%%%%%%

By performing the integrals, we have for the gap equation
$\sigma = 2N_f G \,{\rm tr}S$,
%%%%%%
\begin{eqnarray}
  {\rm tr}S^0_{\rm 3D} 
  &=&  \frac{N_c m^*}{2\pi^2}
    \left( \Ltd \sqrt{\Ltd^2 + m^{*2}}
    -m^{*2} \ln \frac{\Ltd + \sqrt{\Ltd^2 + m^{*2}}}{m^*} 
   \right), \\
 {\rm tr}S^T_{\rm 3D}  
  &=&
   -\frac{N_c m^*}{\pi^2} \int^{\Ltd}_0 \!\!\! \md k\ 
     \frac{k^2}{E} \left[ \sum_{\pm}f(E^{\pm}) \right] .
\end{eqnarray}
%%%%%%%
The effective potential can also be calculated as
\begin{eqnarray}
  {\mathcal V}^0_{\rm 3D} (\sigma)
  %&=& \frac{\sigma^2}{4G}
  % -\frac{N_c N_f}{\pi^2} \int^\Lambda_0 dk\ 
  % k^2 \sqrt{k^2 + \sigma'^2} \\
  &=& - \frac{N_c N_f}{8\pi^2}
   \left[  \Ltd   \sqrt{\Ltd^2 + m^{*2}}
        (2\Ltd^2 + m^{* 2}) - m^{*4} 
        \ln \frac{\Ltd \sqrt{\Ltd^2 + m^{*2}}}{m^*} 
   \right], \\
 {\mathcal V}_{\rm 3D}^T (\sigma)
  &=& -\frac{N_c N_f T}{\pi^2}
   \int^{\Ltd}_0 \md k\ k^2 
   \left[ \sum_{\pm} \ln(1+e^{-\beta E^{\pm}}) 
  \right].
\end{eqnarray}
%%%%%%%%
The quark loop integral in the meson properties $I(p^2)$ reads
%%%%%%
\begin{align}
 &I_{\rm 3D}^0
  = \frac{2N_c}{\pi^2} \int_0^{\Ltd} \md k\
     \frac{k^2}{E(4E^2-p^2)} ,
  \label{I0_3D} \\
 &I_{\rm 3D}^T
  = -\frac{2N_c}{\pi^2} \int_0^{\Ltd} \md k\
   \frac{k^2}{E(4E^2-p^2)}
     \left[   \sum_{\pm}f(E^{\pm})
     \right].
  \label{IT_3D}
\end{align}
%%%%%%
Note that the integral diverges around $4E^2 \simeq p^2$, and we
apply the principal integral to avoid this divergence~\cite{Asakawa:1989bq}.
It may be worth mentioning that the integral can be performed analytically
when $p^2=0$ for $T=0$, then one has for the pion decay constant,
%%%%%%
\begin{eqnarray}
  f_{\pi{\rm 3D}}^2 = \frac{N_c m^{*2}}{2\pi^2}
   \left( - \frac{\Ltd}{\sqrt{\Ltd^2 + m^{*2}}}
    +\ln \frac{\Ltd + \sqrt{\Ltd^2 + m^{*2}}}{m^*} 
   \right). 
\end{eqnarray}
%%%%%%
We thus obtain the required quantities in evaluating the phase diagram
and meson properties.

%%%%%%%%%%%%%%%%%%%%%%%%%%%%
\subsection{Four dimensional cutoff scheme}      %
%%%%%%%%%%%%%%%%%%%%%%%%%%%%
In the four dimensional (4D) cutoff regularization scheme, we introduce
the cutoff scale $\Lfd$ in the Euclidean space after performing the Wick
rotation,
%%%%%%
\begin{align}
  \int \frac{\md^4 k_E}{(2\pi)^4}
  \to 
  \int^{\Lfd}_0 \!\! \frac{k_E^3 \md k_E}{(2\pi)^4}\int d\Omega_4.
\end{align}
%%%%%%
This is well known four dimensional cutoff method for $T=0$ case.
As the natural extension to finite temperature, we introduce the cutoff
scale by
%%%%%%
\begin{eqnarray}
  \int \frac{\md^4 k_E}{(2\pi)^4}
  \to 
   T \sum_{n=-L_4-1}^{L_4}
    \int_0^{ \sqrt{\Lfd^2-\omega_n^2} } \frac{k^2 \md k}{(2\pi)^3}\int d\Omega_3,
\end{eqnarray}
%%%%%%
where $L_4$ is the maximum integer which does not exceed
$\Lfd / (2\pi T) - 1/2$. 

In the 4D cutoff way, it is difficult to divide the contribution into the
temperature independent and dependent parts, since there is also
cutoff in the frequency summation.

The explicit form of ${\rm tr}S$ and the effective potential become
%%%%%%
\begin{eqnarray}
 {\rm tr}S_{\rm 4D}
  &=& \frac{N_c m^* T}{2\pi^2} 
   \sum_{n=-L_4-1}^{L_4}
    \int_0^{ \sqrt{\Lfd^2-\omega_n^2} } \!\! \md k k^2    
    \frac{1}{(\omega_n^-)^2 +E^2},\\
 {\mathcal V}_{\rm 4D} (\sigma)
  &=& 
 -\frac{N_c N_f T}{4\pi^2} 
  \sum_{n=-L_4-1}^{L_4}
    \int_0^{ \sqrt{\Lfd^2-\omega_n^2} } \!\! \md k k^2    
 \ln((\omega_n^-)^2 + E^2  ),
\end{eqnarray}
%%%%%%
%%%%%%
% \begin{eqnarray}
%  I_{\rm 4D}(p^2)
%  &=& 
%   \,\, \color{red}{{\rm NOT \,\,\, YET.}}
%\end{eqnarray}
%%%%%%
where $\omega_n^- = \omega_n - i\mu$.

For $T=0$, the integral can be performed analytically by using the Feynman
parameter method,
%%%%%%
\begin{eqnarray}
 {\rm tr} S_{\rm 4D}^0
  &=& \frac{N_c m^*}{\pi^2}
  \left[ \Lfd^2  -m^{*2} \ln \left( \frac{\Lfd^2+m^{*2}}{m^{*2}} \right) 
  \right], \\
    {\mathcal V}^0_{\rm 4D} (\sigma)
  &=& - \frac{N_c N_f}{8\pi^2}
   \left[  \Lfd^2 m^{*2}
       - m^{*4} \ln \frac{\Lfd^2 + m^{*2} }{m^{*2}}
       + \Lfd^{*4}  \ln (\Lfd^2 + m^{*2})
   \right].
\end{eqnarray}
%%%%%%
One should give the special attention in calculating $I_{\rm 4D}(p^2)$,
because the integral includes divergence to be cured as seen in the 3D cutoff
case. The analytic expression of $I^0_{\rm 4D}(p^2)$ will be given in
appendix \ref{app_I0}

Again we show the explicit form for the pion decay constant at
$T=0$,
%%%%%%
\begin{eqnarray}
  f_{\pi{\rm 4D}}^2 = \frac{N_c m^{*2}}{4\pi^2}
   \left[ -\frac{\Lfd^2}{\Lfd^2 + m^{*2}}
   +\ln \left( \frac{\Lfd^2+m^{*2}}{m^{*2}} \right) \right] . 
\end{eqnarray}

%%%%%%%%%%%%%%%%%%%%%%%%%%%%
\subsection{Pauli-Villars regularization}
%%%%%%%%%%%%%%%%%%%%%%%%%%%%
In this regularization, the divergences from loop integrals are
subtracted by introducing virtually heavy particles as
%%%%%%
\begin{equation}
  \frac{1}{k^2-m^2} \, \longrightarrow \,
  \frac{1}{k^2-m^2}
 -\sum_i \frac{a_i}{k^2-\Lambda_i^2}.
\end{equation}
%%%%%%
This manipulation induces virtual frictional force so that the
contribution from unphysical high frequency mode is suppressed.

In evaluation the gap equation, we apply the following subtraction
%%%%%%
\begin{equation}
  \frac{1}{p^2-m^{*2}}
 - \frac{a_1}{p^2-\Lambda_1^2}
 - \frac{a_2}{p^2-\Lambda_2^2},
\end{equation}
%%%%%%
where
%%%%%%
\begin{equation}
  a_1 = \frac{m^{*2}-\Lambda_2^2}{\Lambda_1^2 - \Lambda_2^2}, \quad
  a_2 = \frac{\Lambda_1^2-m^{*2}}{\Lambda_1^2 - \Lambda_2^2}.
\end{equation}
%%%%%%
By setting the cutoff scales $\Lambda_1 = \Lambda_2 = \Lpv$ after the
subtraction, we have
%%%%%%
\begin{align}
{\rm tr}S_{\rm PV}^0
  &= \frac{N_c m^*}{4\pi^2} 
     \left( \Lpv^2  - m^{*2} + m^{*2} \ln \frac{m^{*2}}{\Lpv^{*2}}
     \right) , \\
%%%%%
  {\rm tr} S_{\rm PV}^T
  &= -2N_c m^* \int \frac{\md^3 p}{(2\pi)^3} 
    \left[
      \frac{f(E_m^{\pm})}{E_m} 
      - \left( 1 + \frac{\Lpv^2-m^{*2}}{2p^2} \right)
        \frac{f(E_{\Lambda}^{\pm})}{E_{\Lambda}}    
    \right].
\end{align}
%%%%%%
where $E_m = \sqrt{k^2 + m^{*2}}$ and
$E_{\Lambda} = \sqrt{k^2 + \Lpv^2}$.
By integrating the above equation, we obtain the effective potential
%%%%%%
\begin{align}
  {\mathcal V}_{\rm PV}^0
  &= - \frac{N_c N_f}{8\pi^2}
    \left[ \Lpv^2 m^{*2} - \frac{3}{4}m^{*4} 
          +\frac{1}{2}m^{*4} \ln \frac{m^{*2}}{\Lpv^2}
    \right], \\
  {\mathcal V}_{\rm PV}^T
  &= - \frac{N_c N_f T}{\pi^2}
   \sum_{\pm} \int \md k\ \Bigl[ 
     k^2  \ln(1+e^{-\beta E_m^{\pm}})
    -\frac{m^2}{8}(4k^2 + 2\Lambda^2 -m^{*2})
      \frac{f(E_{\Lambda}^{\pm})}{E_{\Lambda}}
  \Bigr].
\label{potential_PV}
\end{align}
%%%%%%
Since the divergence coming from the integral $I(p^2)$ is order of $\log$,
one subtraction is enough to make it finite, so we get
%%%%%%
\begin{align}
  &I_{\rm PV}^0
  = \frac{2N_c}{\pi^2} \int \md k k^2
  \left[
     \frac{1}{E_m(4E_m^2-p^2)} 
    -\frac{1}{E_{\Lambda}(4E_{\Lambda}^2-p^2)} 
  \right],
  \label{I0_PV} \\
  &I_{\rm PV}^T
  = -\frac{2N_c}{\pi^2} \int \md k k^2
     \left[  
        \frac{ \sum_{\pm}f(E_m^{\pm})}{E_m(4E_m^2-p^2)}
       -\frac{ \sum_{\pm}
          f(E_{\Lambda}^{\pm})}{E_{\Lambda}(4E_{\Lambda}^2-p^2)}
     \right].
  \label{IT_PV}
\end{align}
%%%%%%

The pion decay constant at $T=0$ becomes
\begin{eqnarray}
f_{\pi {\rm PV}}^2 = \frac{N_c m^{*2}}{4\pi^2}
 \left( - 1 + \frac{\Lpv^2}{\Lpv^2 - m^{*2}}
  \ln \frac{\Lpv^2}{m^{*2}} \right) . 
\end{eqnarray}

%%%%%%%%%%%%%%%%%%%%%%%%%%%%
\subsection{Proper-time regularization}
%%%%%%%%%%%%%%%%%%%%%%%%%%%%
The basic idea of the proper-time regularization is based on the
following manipulation of the Gamma function,
%%%%%%
\begin{equation}
  \frac{1}{A^n} \to \frac{1}{\Gamma[n]}\int_{1/\Lpt^2}^{\infty} \md \tau\,\, 
  \tau^{n-1}e^{-A \tau},
\label{pro1}
\end{equation}
%%%%%%
where the lower cut $1/\Lpt^2$ induces the dumping factor into the
original propagator since, for example with $n=1$,
%%%%%%
\begin{equation}
  \frac{1}{k_E^2+m^{*2}} 
  \to 
  \int_{1/\Lpt^2}^{\infty}  \md \tau\, e^{-A \tau}=
  \frac{1}{k_E^2+m^{*2}} e^{-(k_E^2+m^{*2})/\Lpt^2}.
\end{equation}
%%%%%%
Therefore in this regularization high frequency contribution is
dumped by the factor $e^{-k_E^2/\Lpt^2}$, so the original
divergent integral turns out to be finite.
For $A$ contains a imaginary part, Eq. (\ref{pro1})
is modified as,
\begin{equation}
  \frac{1}{A^n} \to \frac{i^n}{\Gamma[n]}\int_{+0}^{\infty} \md \tau\,\, 
  \tau^{n-1}e^{-iA \tau}, \quad ({\rm Im}(A)<0, {\rm Re}(n)>0) .
\label{pro2}
\end{equation}

Under this procedure, the integral of ${\rm tr}S$ in the gap equation
becomes
%%%%%%
\begin{eqnarray}
{\rm tr}S_{\rm PT}^0
 &=& \frac{N_c m^*}{4\pi^2}
   \left[ \Lpt^2 e^{-m^{*2}/\Lpt^2} 
   + m^{*2} Ei(-m^{*2}/\Lpt^2)
   \right] , 
\label{gap_pro_0} \\
{\rm tr}S_{\rm PT}
 &=& \frac{N_c m^* T}{2 \pi^{3/2}} 
   \sum_{n=0}^\infty
   \int^\infty_{+0} \frac{d\tau}{\tau^{3/2}}
   \left[ e^{-i\pi/4} e^{-i\left\{(\omega_n^-)^2 + m^{*2}
   \right\}\tau } + c.c. \right] .
\label{gap_pro_t}
\end{eqnarray}
%%%%%%
where $Ei(-x)$ is the exponential-integral function.
For $m^{*2} \ll \Lpt^2$, Eq. (\ref{gap_pro_0}) is expanded as
\begin{eqnarray}
{\rm tr}S_{\rm PT}^0
 \simeq \frac{N_c m^*}{4\pi^2}
   \left[ \Lpt^2 - m^{*2} + m^{*2} \left(
   \ln \frac{m^{*2}}{\Lpt^2} + \gamma_E - \frac{m^{*2}}{2\Lpt^2} 
   \right)\right] .
\label{gap_pro_02}
\end{eqnarray}

We rotate the contour of the integration in
Eq. (\ref{gap_pro_t}) to the imaginary axis of $\tau$
\cite{Inagaki:2003yi,Inagaki:2004ih,Inagaki:2003ac}.
%%%%%%
For $\omega_0^2-\mu^2+m^{*2}>0$, the trace becomes
%%%%%%
\begin{eqnarray}
 {\rm tr}S_{\rm PT}
 = \frac{N_c m^* T}{\pi^{3/2}} 
    \sum_{n=0}^\infty
   \int^\infty_{1/\Lambda_{\rm PT}^2} \frac{d\tau}{\tau^{3/2}}
   \cos(2\omega_n \mu \tau) 
     e^{-(\omega_n^2 - \mu^2 + m^{*2})\tau},
\end{eqnarray}
%%%%%%
and for $\omega_0^2-\mu^2+m^{*2}<0$, 
% $(\mu > \sqrt{(\pi T)^2 + m^2})$,
%%%%%%
\begin{eqnarray}
 {\rm tr}S_{\rm PT}
&=& \frac{N_c m^* T}{\pi^{3/2}}  \left[
  % \theta(\sigma-\mu') \sum_{n=0}^\infty
  % \int^\infty_{1/\Lambda^2} \frac{d\tau}{\tau^{3/2}}
  % \cos(2\omega_n \mu \tau) 
  % e^{-(\omega_n^2 - \mu^2 +   \sigma'^2)\tau}
  % \right. \nonumber \\
   % \theta(\mu'-m^*) 
   \sum_{n>[N]}^\infty
   \int^\infty_{1/\Lambda_{\rm PT}^2} \frac{d\tau}{\tau^{3/2}}
   \cos(2\omega_n \mu \tau) 
   e^{-(\omega_n^2 - \mu^2 + m^{*2})\tau}  \right.
  \nonumber\\
&&-
   %\theta(\mu'-m^*) 
   \sum_{n=0}^{[N]} \left\{
   \int^\infty_{1/\Lambda_{\rm PT}^2} \frac{d\tau}{\tau^{3/2}}
   \sin(2\omega_n \mu \tau) 
   e^{(\omega_n^2 - \mu^2 + m^{*2})\tau} 
  \right.\nonumber\\
&&-\left.\left.  \Lambda_{\rm PT}\, {\rm Re} \left( e^{i\pi/4}
 \int^{\pi/2}_{-\pi/2} d\theta\, e^{-i\theta/2}
  \exp\left[-i \{ (\omega_n^-)^2 
  + m^{*2} \} e^{i\theta}/\Lambda_{\rm PT}^2 \right]
  \right) \right\} \right] ,
\label{eq_trSpt}
\end{eqnarray}
%%%%%%
where % $\mu' = \sqrt{\mu^2-(\pi T)^2}-m^2$ and 
$N = \{\sqrt{\mu^2-m^{*2}}/(\pi T)-1\}/2$ .
%%%%%%
%For $\sigma'^2 \ll \Lambda^2$,
% \begin{eqnarray}
% \langle \sigma \rangle &=& \frac{N_c N_f}{2\pi^2} G \sigma'
%  \left[ \Lambda^2 \left\{1 - \frac{\sigma'^2}{\Lambda^2}
%  +\frac12 \left(\frac{\sigma'^2}{\Lambda^2}\right)^2 + \cdots 
%  \right\} \right. \nonumber \\ 
%  &&\left. + \sigma'^2 \left\{ \ln \frac{\sigma'^2}{\Lambda^2} 
%  + \gamma_E - \frac{\sigma'^2}{\Lambda^2} + \frac{1}{2\cdot 2!}
%  \left(\frac{\sigma'^2}{\Lambda^2}\right)^2 + \cdots
% \right\}  \right]  \\
% &\simeq&  \frac{N_c N_f}{2\pi^2} G \sigma'
%  \left[ \Lambda^2 - \sigma'^2
%  + \sigma'^2 \left\{ \ln \frac{\sigma'^2}{\Lambda^2} 
%  + \gamma_E - \frac{\sigma'^2}{2\Lambda^2} \right\}  \right] .
% \end{eqnarray}
%%%%%
Similarly, the effective potential can be calculated through
%%%%%%
\begin{eqnarray}
{\mathcal V}_{\rm PT}(\sigma) = %\frac{\sigma^2}{4G}
\frac{N_c N_f T}{4\pi^{3/2}} \sum_{n=0}^\infty
 \int^\infty_{+0} \frac{d\tau}{\tau^{5/2}}
 \left[ e^{-3i\pi/4} e^{-i\left\{(\omega_n^-)^2 + m^{*2}
 \right\}\tau } + c.c. \right] .
\label{veff_pro_t}
\end{eqnarray}
%%%%%% 
For $\omega_0^2-\mu^2+m^{*2}>0$, one has
% $(\mu < \sqrt{(\pi T)^2 + m^2})$,
%%%%%%
\begin{eqnarray}
{\mathcal V}_{\rm PT}(\sigma) = %\frac{\sigma^2}{4G}
 \frac{N_c N_f T}{2\pi^{3/2}} \sum_{n=0}^\infty
 \int^\infty_{1/\Lambda_{\rm PT}^2} \frac{d\tau}{\tau^{5/2}}
 \cos(2\omega_n \mu \tau) e^{-(\omega_n^2 - \mu^2 + m^{*2})\tau},
\end{eqnarray}
%%%%%%
and for $\omega_0^2-\mu^2+m^{*2}<0$,
% $( \mu > \sqrt{(\pi T)^2 + m^2})$,
%%%%%%
\begin{eqnarray}
{\mathcal V}_{\rm PT}(\sigma) &=& %\frac{\sigma^2}{4G}
 \frac{N_c N_f T}{2\pi^{3/2} }\left[ 
%\theta(\sigma-\mu')\sum_{n=0}^\infty
% \int^\infty_{1/\Lambda^2} \frac{d\tau}{\tau^{5/2}}
% \cos(2\omega_n \mu \tau) e^{-(\omega_n^2 - \mu^2 + \sigma'^2)\tau}
%\right. \nonumber \\
% \theta(\mu'-\sigma)
 \sum_{n>[N]}^\infty
 \int^\infty_{1/\Lambda_{\rm PT}^2} \frac{d\tau}{\tau^{5/2}}
 \cos(2\omega_n \mu \tau) e^{-(\omega_n^2 - \mu^2 + m^{*2})\tau}
\right. \nonumber \\
&& %\theta(\mu'-\sigma)
+ \sum_{n=0}^{[N]} \left\{
 \int^\infty_{1/\Lambda_{\rm PT}^2} \frac{d\tau}{\tau^{5/2}}
 \sin(2\omega_n \mu \tau) e^{(\omega_n^2 - \mu^2 + m^{*2})\tau}
\right. \nonumber \\
&&+ \left.\left.  \Lambda^3\, {\rm Re} \left( e^{-i\pi/4}
 \int^{\pi/2}_{-\pi/2} d\theta\, e^{-3i\theta/2}
  \exp\left[-i \{ (\omega_n^-)^2 + m^{*2} \} e^{i\theta}/\Lambda_{\rm PT}^2 
  \right]
\right) \right\} \right] .
\end{eqnarray}
%%%%%%
$I_{\rm PT}$ can also be calculated by
%%%%%%
\begin{align}
  I_{\rm PT}(p^2) 
  =
  -\frac{N_c T}{\pi^{3/2}} \sum_{n=0}^\infty
  \int_0^{1/2} \md \alpha
  \int^\infty_{+0} \frac{d\tau}{\tau^{1/2}}
  \left[ e^{-3i\pi/4} e^{-i\left\{(\omega_n^-)^2 + \Delta
  \right\}\tau } + c.c. \right] .
\end{align}
%%%%%%
where $\alpha$ is the Feynman integration parameter and
$\Delta = m^{*2} -p^2/4 + \alpha^2 p^2$. 
%
% For $\omega_0^2-\mu^2+\Delta >0$,
%%%%%%
% \begin{align}
%  I_{\rm PT}(p^2) 
%  =
%  \frac{N_c T}{\pi^{3/2}} \sum_{n=0}^\infty
%  \int_0^1 \md \alpha
%  \int^\infty_{1/\Lambda_{\rm PT}^2} \frac{d\tau}{\tau^{1/2}}
%  \left[ \cos(2\omega_n \mu \tau)
%            e^{-( \omega_n^2-\mu^2 + \Delta) \tau } \right] ,
% \end{align}
%%%%%%
Then the integral can be written
%%%%%%
\begin{eqnarray}
  I_{\rm PT}(p^2) &=& %\frac{\sigma^2}{4G}
  \frac{2 N_c T}{\pi^{3/2}}   \int_0^{1/2} \md \alpha
  \sum_{n=0}^\infty 
  \left[ 
   % \sum_{n=[N]}^\infty 
   \theta(W_n(\alpha))
   \int^\infty_{1/\Lambda_{\rm PT}^2} \frac{d\tau}{\tau^{1/2}}
   \cos(2\omega_n \mu \tau) 
   e^{-(\omega_n^2 - \mu^2 + \Delta )\tau}
   \right. \nonumber \\
  && %\theta(\mu'-\sigma)
  +    \theta(-W_n(\alpha)) \left\{
   \int^\infty_{1/\Lambda_{\rm PT}^2} \frac{d\tau}{\tau^{1/2}}
   \sin(2\omega_n \mu \tau) 
   e^{(\omega_n^2 - \mu^2 + \Delta )\tau}
   \right. \nonumber \\
  &&- \left.\left. 
   \frac{1}{\Lambda} \, {\rm Re}
   \left(   e^{-i\pi/4}
     \int^{\pi/2}_{-\pi/2} d\theta\, e^{i\theta/2}
     \exp\left[-i \{ (\omega_n^-)^2 + \Delta \}
     e^{i\theta}/\Lambda_{\rm PT}^2 
            \right]
   \right)  \right\}  \right] ,
\label{eq_Ipt}
  % && %\theta(\mu'-\sigma)
  %+ \sum_{n=0}^{[N]} \left\{
  % \int^\infty_{1/\Lambda_{\rm PT}^2} \frac{d\tau}{\tau^{1/2}}
  % \sin(2\omega_n \mu \tau) e^{(\omega_n^2 - \mu^2 + \Delta )\tau}
  % \right. \nonumber \\
  % &&+ \left.\left.  \frac{1}{\Lambda} \, {\rm Re} \left( e^{-i\pi/4}
  % \int^{\pi/2}_{-\pi/2} d\theta\, e^{i\theta/2}
  % \exp\left[-i \{ (\omega_n^-)^2 + \Delta \} e^{i\theta}/\Lambda_{\rm PT}^2 
  % \right]
  % \right) \right\} \right] .
\end{eqnarray}
%%%%%%
where $W_n(\alpha) = \omega_n^2 + m^{*2} +(\alpha^2 -1/4)p^2 -\mu^2$.

The pion decay constant at $T=0$ reads the following simple form,
%%%%%%
\begin{eqnarray}
f_{\pi{\rm PT}}^2 = -\frac{N_c m^{*2}}{4\pi^2} Ei(-m^{*2}/\Lpt^2).
\end{eqnarray}
For $m^{*2} \ll \Lpt^2$, we have
\begin{eqnarray}
f_{\pi{\rm PT}}^2 \simeq \frac{N_c m^{*2}}{4\pi^2} 
 \left\{ -\gamma_E + \frac{m^{*2}}{\Lpt^2}
 + \ln\frac{\Lpt^2}{m^{*2}}
 \right\}.
\end{eqnarray}

%%%%%%%%%%%%%%%%%%%%%%%%%%%%
\subsection{Dimensional regularization}
%%%%%%%%%%%%%%%%%%%%%%%%%%%%
In the dimensional regularization method, we obtain finite quantities
through analytically continuing
the dimension in the loop integral 
to a non-integer value, $D$, as
%%%%%%
\begin{align}
  \int \frac{\md^4 k}{(2\pi)^4}
  \to M_0^{4-D}
  \int \frac{\md^D k}{(2\pi)^D},
\end{align}
%%%%%%
where the scale parameter $M_0$ is inserted so as to adjust the
mass dimension of physical quantities.
The method is well known since it preserves 
most of symmetries. 
Note that this result is the same as the result obtained
from the proper-time integral ($0<\tau<\infty$) and expressed
by the poles of the Gamma function.

The trace ${\rm tr}S$ in the gap equation reads
%%%%%%
\begin{align}
&{\rm tr} S_{\rm DR}^0
  = \frac{ -N_c  M_0^{4-D} m^*} {(2\pi)^{D/2}} 
      \Gamma \left( 1-\frac{D}{2} \right)
      (m^{*2})^{D/2-1}, \\
&{\rm tr} S_{\rm DR}^T 
  = -A_{D} m^* 
   \int  \md k \frac{k^{D-2} }{2E}
 \left[ \sum_{\pm} f(E^{\pm}) \right],
\end{align}
%%%%%%
where 
%%%%%%
\begin{equation}
  A_{D}= \frac{N_c 2^{2-D/2} M_0^{4-D}}
                      {\pi^{(D-1)/2}\Gamma((D-1)/2)}.
\end{equation}
%%%%%%
The effective potential becomes
%%%
\begin{align}
 &{\mathcal V}_{\rm DR}^0
  =  \frac{N_c N_f M_0^{4-D} }{2(2\pi)^{D/2}}
     \Gamma \Bigl(-\frac{D}{2}\Bigr) (m^{*2})^{D/2}, \\
 &{\mathcal V}_{\rm DR}^T
 =  -A_{D} N_f T
        \int {\md} k\ k^{D-2}
        \sum_{\pm} \ln \Bigl[1 + e^{-\beta E^{\pm}} \Bigr]. 
\end{align}
%%%%%%
In the similar manner, the integral $I(p^2)$ appearing in the
meson propagator is calculated as
%%%%%%
\begin{align}
& I_{\rm DR}^0
= A_{D}
   \int \md k 
   \frac{k^{D-2}}{E(4E^2-p^2)}, \\
& I_{\rm DR}^{T}
=A_{D} \int \md k 
  \frac{-k^{D-2}}{E(4E^2-p^2)}
  \left[ \sum_{\pm} f(E^{\pm}) \right].
\end{align}
%%%%%%
Note we need to perform the principal integration for
$m^{*2} < p^2/4$.

The pion decay constant at $T=0$ reads the following simple form,
%%%%%%
\begin{eqnarray}
f_{\pi{\rm DR}}^2 = \frac{N_c M_0^{4-D}}{(2\pi)^{D/2}} 
\Gamma\left(2-\frac{D}{2}\right) (m^{*2})^{D/2-1} .
\end{eqnarray}
We show a concrete example of ${\rm tr} S^0_{\rm DR}$ and 
$f_{\pi {\rm DR}}^2$ for $D \simeq 2,3,4$ in appendix \ref{app_DR}.

%%%%%%%%%%%%%%%%%%%%%%%%%%%%
\section{Model parameters}
\label{sec_para}
%%%%%%%%%%%%%%%%%%%%%%%%%%%%
Having obtained the equation which determines the pion mass and
pion decay constant, we are now ready to perform the parameter
fitting. 
In the previous section we suppose that all the cutoff scales 
are equal in each regularization to reduce the parameters. Thus
the model has three parameters: the cutoff scale $\Lambda$,
the current quark mass $m_u$ and the coupling strength $G$. Whereas
in the DR there appears one more parameter, 
so the total number becomes four: the current mass $m_u$,
dimension $D$, scale parameter $M_0$ and the coupling $G$, as
discussed in Ref.~\cite{Inagaki:2010nb}. In this section, we shall set the
model parameters by fitting the pion mass and decay constant.
The actual values we use are shown below
%%%%%%
\begin{equation}
  m_\pi = 135\,{\rm MeV}, \,\, f_\pi = 94\,{\rm MeV}.
\end{equation}
%%%%%%
For the case with the DR, we perform fitting with one more quantity,
the neutral pion decay constant to two photons, which will be discussed 
later.

%%%%%%%%%%%%%%%%%%%%%%%%%%%%
\subsection{Parameters in various regularizations}
%%%%%%%%%%%%%%%%%%%%%%%%%%%%
Here we align the model parameters in various regularizations in
this subsection.

%%%%%%
\begin{table}[h!]
\caption{Parameters in the 3D cutoff}
\label{table_3d_para}
%\begin{ruledtabular}
\begin{center}
\begin{tabular}{lcccc}
\hline
$m_u$(MeV) & $\Ltd$(MeV)  & $G \cdot 10^{-6}$(MeV${}^{-2}$) & $m^*$(MeV)
& $\langle \bar{u}u \rangle^{1/3}$(MeV)\\
\hline
3.0 & 942 & 2.00 & 220 & $-300$ \\
%%3.5 & 853 & 2.51 & 237  & -285 \\
4.0 & 781 & 3.09 & 255  & $-272$ \\
%%4.5 & 720 & 3.80 & 279  & -262 \\
5.0 & 665 & 4.71 & 311  & $-253$ \\
5.5 & 609 & 6.26 & 375  & $-245$ \\
\hline
\end{tabular}
\end{center}
%\end{ruledtabular}
\end{table}
%%%%%%%%%%%%%%%%%%%%%%%%%%
%%%%%%%%%%%%%%%%%%%%%%%%%%
\begin{table}[h!]
\caption{Parameters in the 4D cutoff}
\label{table_4d_para}
%\begin{ruledtabular}
\begin{center}
\begin{tabular}{ccccc}
\hline
$m_u$(MeV)  & $\Lfd$(MeV)  & $G \cdot 10^{-6}$(MeV${}^{-2}$) & $m^*$(MeV)  & $\langle \bar{u}u \rangle^{1/3}$(MeV) \\
\hline
3.0 & 1397 & 1.80 & 198  & $-300$ \\
%3.5 & 1274 & 2.20 & 208  & -285 \\
%4.0 & 1176 & 2.64 & 219  & -272 \\
%4.5 & 1095 & 3.12 & 230  & -262 \\
5.0 & 1027 & 3.64 & 242  & $-253$ \\
%5.5 & 969   & 4.22 & 255  & -245 \\
%6.0 & 919   & 4.86 & 269  & -238 \\
%6.5 & 875   & 5.60 & 286  & -232 \\
%7.0 & 835   & 6.45 & 306  & -226 \\
%7.5 & 800   & 7.49 & 332  & -221 \\
7.1 & 827   & 6.66 & 311  & $-225$ \\
8.0 & 768   & 8.88 & 369  & $-216$ \\
\hline
\end{tabular}
\end{center}
%\end{ruledtabular}
\end{table}
%%%%%%%%%%%%%%%%%%%%%%%%%%
%%%%%%%%%%%%%%%%%%%%%%%%%%
\begin{table}[h!]
\caption{Parameters in the Pauli-Villars regularization}
\label{table_pv_para}
%\begin{ruledtabular}
\begin{center}
\begin{tabular}{ccccc}
\hline
$m_u$(MeV)  & $\Lpv$(MeV)  & $G \cdot 10^{-6}$(MeV${}^{-2}$) & $m^*$(MeV) 
& $\langle \bar{u}u \rangle^{1/3}$(MeV) \\
\hline
3.0 & 1420 & 1.77 & 195  & $-300$ \\
%3.5 & 1302 & 2.15 & 203  & -285 \\
%4.0 & 1209 & 2.56 & 212  & -272 \\
%4.5 & 1133 & 2.99 & 220  & -262 \\
5.0 & 1071 & 3.45 & 229  & $-253$ \\
%5.5 & 1018 & 3.93 & 238  & -245 \\
%6.0 & 974 & 4.44 & 247  & -238 \\
%6.5 & 937 & 4.98 & 255  & -232 \\
%7.0 & 905 & 5.55 & 265  & -226 \\
%7.5 & 877 & 6.15 & 274  & -221 \\
%8.0 & 853 & 6.78 & 283  & -216 \\
9.94 & 780 & 9.56 & 311  & $-199$ \\
10.0 & 778 & 9.64 & 312  & $-198$ \\
15.0 & 729 & 19.4 & 417  & $-173$ \\
\hline
\end{tabular}
\end{center}
%\end{ruledtabular}
\end{table}
%%%%%%%%%%%%%%%%%%%%%%%%%%
%%%%%%%%%%%%%%%%%%%%%%%%%%
\begin{table}[h!]
\caption{Parameters in the proper-time regularization}
\label{table_pt_para}
%\begin{ruledtabular}
\begin{center}
\begin{tabular}{ccccc}
\hline
$m_u$(MeV)  & $\Lpt$(MeV) & $G\cdot 10^{-6}$(MeV${}^{-2}$) & $m^*$(MeV)  
& $\langle \bar{u}u \rangle^{1/3}$(MeV)  \\
\hline
3.0 & 1464 & 1.61 & 178 & $-300$ \\
%3.5 & 1342.03 & 1.95579 & 185.222 & -285.326 \\
%4.0 & 1244 & 2.30 & 191 & -272 \\
%4.5 & 1164.18 & 2.68174 & 198.294 & -262.395 \\
5.0 & 1097 & 3.07 & 204 & $-253$ \\
%5.5 & 1040.44 & 3.48165 & 211.345 & -245.414 \\
%6.0 & 991 & 3.91 & 217 & -238 \\
%6.5 & 949.02 & 4.36209 & 224.729 & -232.124 \\
%7.0 & 911 & 4.83 & 231 & -226 \\
%7.5 & 878.79 & 5.33226 & 238.697 & -221.311 \\
%8.0 & 849 & 5.85 & 245 & -216 \\
10.0 & 755 & 8.13 & 265 & $-198$ \\
12.6 & 680 & 12.1 & 311 & $-183$ \\
15.0 & 645 & 17.2 & 372 & $-173$ \\
\hline
\end{tabular}
\end{center}
%\end{ruledtabular}
\end{table}
%%%%%%%%%%%%%%%%%%%%%%%%%%
%%%%%%%%%%%%%%%%%%%%%%%%%%

Table \ref{table_3d_para}, \ref{table_4d_para}, \ref{table_pv_para}
and \ref{table_pt_para} show how the parameters change according
to the current quark mass $m_u$, where we first set the value of $m_u$
then search the parameters $\Lambda$ and $G$ which lead
$m_\pi=135$MeV and $f_\pi=94$MeV. One sees the tendency
that cutoff scale $\Lambda$ becomes smaller with increasing $m_u$,
while $G$ becomes larger according to $m_u$. We confirm that the
values of the cutoff and four point coupling are 
%similar; actually the difference is within factor $1.5$
O$(1)$ and O$(10^{-6})$ in MeV scale in these regularizations.  

We also showed the values of the constituent quark mass $m^*$
and the chiral condensate $\langle \bar{u}u \rangle^{1/3}$ which
are the predicted quantities in the models. We note that the values of
$m^*$ are about $200-300$MeV which are comparable to one third of
the proton mass. We also note that $m^*$ increases with respect to
$m_u$, while the absolute value of $\langle \bar{u}u \rangle$ decreases 
when $m_u$ becomes large. Since the relation  $m^* \propto G \langle 
\bar{u}u \rangle$ holds, even $\langle \bar{u}u \rangle$ becomes
smaller $m^*$ can be larger due to the large value of $G$, which is
actually the case in these regularizations.

%%%
We plot the parameters of each regularization in %Fig. \ref{fig_para}.
%%%%%%
Fig.1.
%%%%%%
The black circles denote the value which satisfy $m^*=311$MeV
for each regularization.
The relation between the cutoff scale $\Lambda$ and $G$
for the $4$D cutoff, Pauli-Villars regularization and
proper-time regularization resembles each other.
The relation between $m_u$ and $\Lambda$ for these regulatizations also
resembles each other. In the case of same value for $m^*$,
the $m_u$ dependence of $\Lambda$ is large. However, the values of
$\Lambda$ are close, $660-830$MeV.
In the case of same value for
$m^*$, the $m_u$ dependence on $G$ is large and the values of $G$ 
are separate in each regularization.
The relation between $m_u$ and
$G$ is different in each regularization.

%%%%%%%%%%%%%%%%%%%%%%%%%%%%
\begin{figure}
 \begin{center}
 %\vspace{0.5cm}
  %\hspace{1.0cm}
  \subfigure{
    \includegraphics[height=5cm,keepaspectratio]{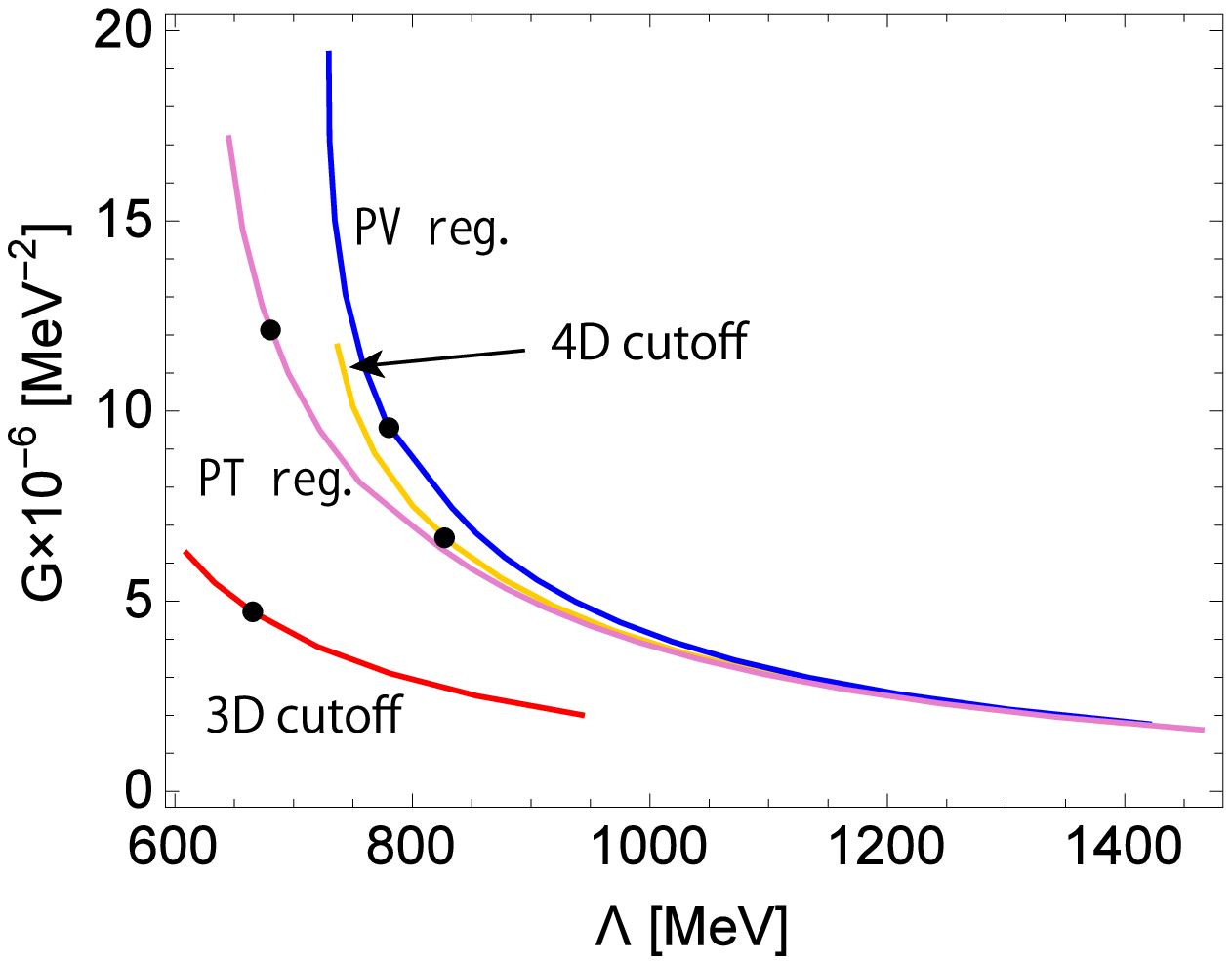}
  }
  %\hspace{1.0cm}
  \subfigure{
    \includegraphics[height=5cm,keepaspectratio]{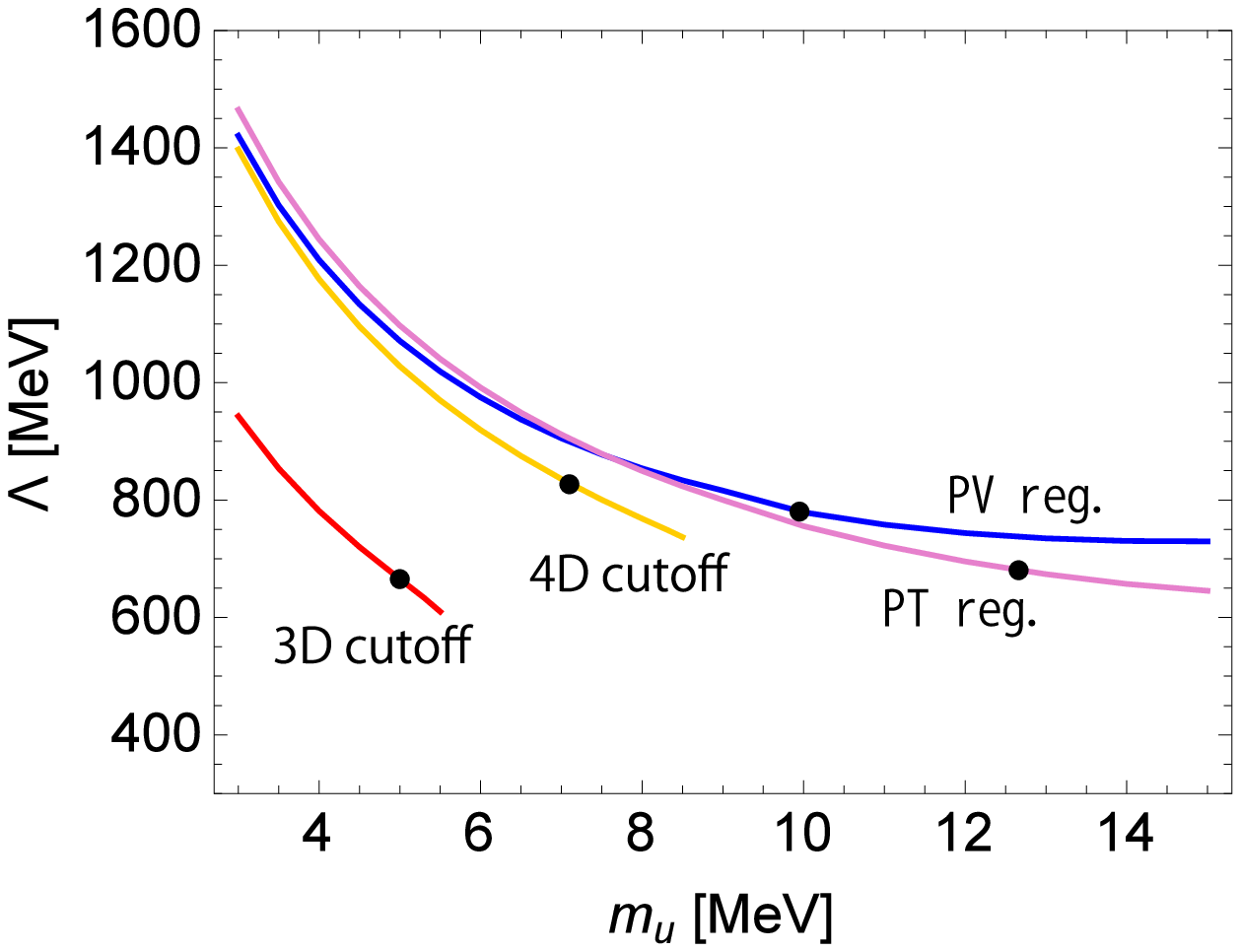}
  }

  \includegraphics[height=5cm,keepaspectratio]{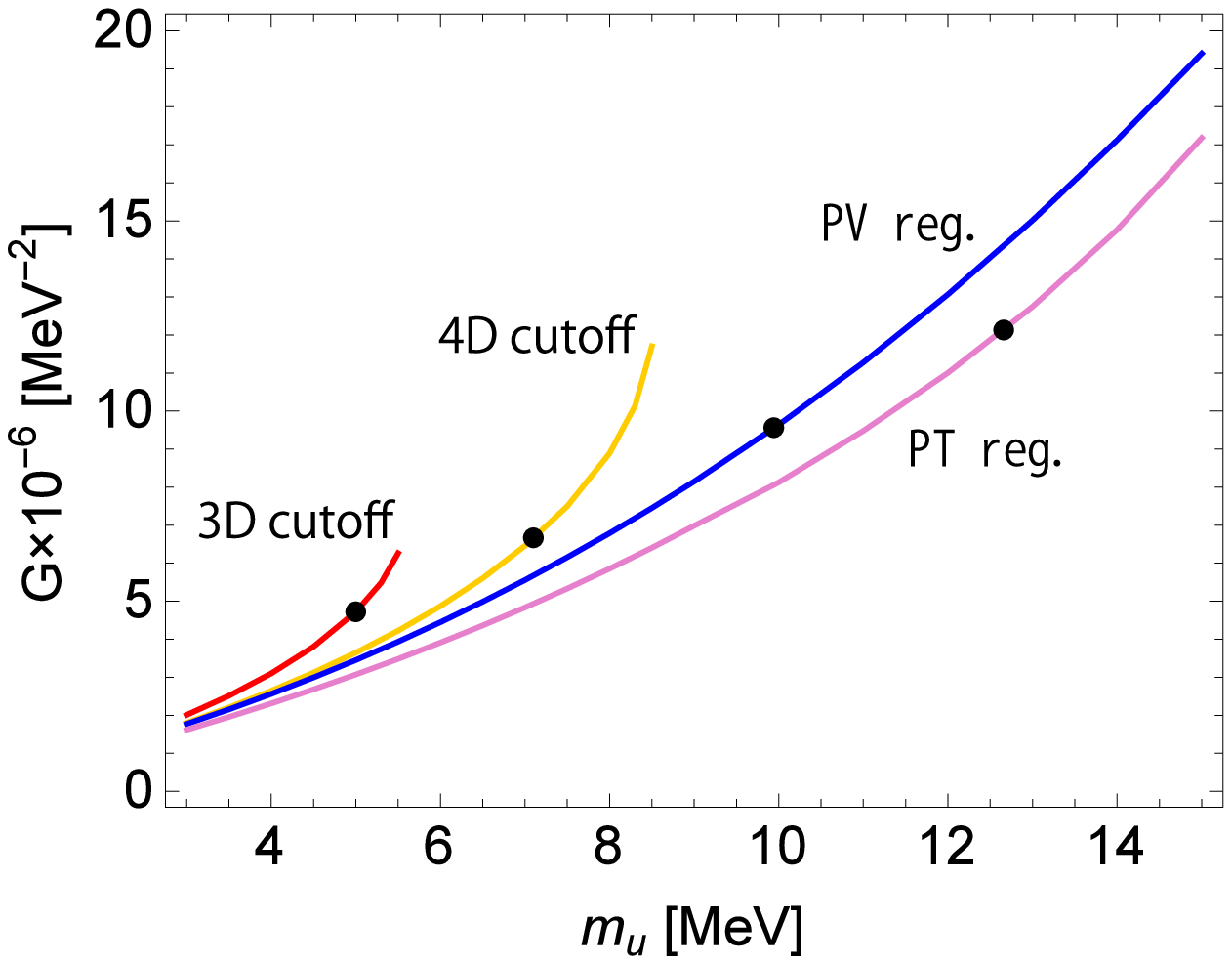}

  \caption{Parameters in each regularization}
\end{center}
\label{fig_para}
\end{figure}
%%%%%%%%%%%%%%%%%%%%%%%%%%%%

%%%%%%%%%%%%%%%%%%%%%%%%%%%%
\subsection{Parameter fitting in the dimensional regularization}
%%%%%%%%%%%%%%%%%%%%%%%%%%%%
For the sake of the parameter fitting in the DR, we present the
calculation of the pion to two photon decay rate,
$\Gamma_{\pi^0 \gamma \gamma}$, in this subsection. %$= 7.8\,{\rm eV}$
The decay rate can be evaluated through the following one-loop
amplitude, $T_{\mu\nu}(k_1,k_2)$,
%%%%%%
\begin{align}
  &T_{\mu\nu} (k_1,k_2) =
  4i \epsilon_{\mu\nu\rho\lambda} k_1^{\rho} k_2^{\lambda}
  \cdot T_{\gamma} \\
%%%%%%
  &T_{\gamma} = g_{\pi qq} e^2 \frac{N_c}{3} m^*
  M_0^{4-D} \int \!\! \frac{\md^D p}{i(2\pi)^D}
  \left[ 
    \frac{1}{(p-k_2)^2-m^{*2}}
    \frac{1}{p^2-m^{*2}}  
    \frac{1}{(p+k_1)^2-m^{*2}} 
  \right].
\end{align}
%%%%%%
where $e$ is the QED coupling constant and $k_1$ and $k_2$ are the
external momentum of emitted
photons so the square of the total momentum coincides with that
of the original pion, namely, $(k_1+k_2)^2 = m_\pi^2$. By using
$T_{\gamma}$, the decay rate, $\Gamma_{\pi^0 \gamma \gamma}$,
is expressed as
%%%%%%
\begin{align}
  \Gamma_{\pi \gamma \gamma}
  = \frac{m_\pi^3}{64 \pi} |T_{\gamma}|^2.
\end{align}
%%%%%%
The detailed derivation is presented in the paper \cite{Krewald:1991tz}.
After some algebra, one obtains
%%%%%%
\begin{align}
 &T_{\gamma} =
  -4g_{\pi qq} \frac{\alpha_e N_c m^*}{3\pi}    \,
  \frac{ \left( 4 \pi M_0^2 \right)^{2-D/2}}
          {(m^{*2})^{3-D/2}}
  \frac{\Gamma(2-D/2)}{\hat{m}_\pi^2}  
  \int_0^1 \md x \,\,
  \frac{1}{x} \left\{
    \left[ 1- x(1-x)\hat{m}_\pi^2 \right]^{-2+D/2} - 1
  \right\}.
\end{align}
%%%%%%
with $\alpha_e = e^2/(4\pi)$, and $\hat{m}_\pi^2 = m_\pi^2/m^{*2}$.

With the observables, $m_\pi=135{\rm MeV}$, $f_\pi=94{\rm MeV}$
and
\begin{equation}
 \Gamma_{\pi^0 \gamma \gamma} = 7.8\,{\rm eV},
\end{equation}
we perform the parameter fitting in the DR following \cite{Krewald:1991tz}. 
Table. \ref{table_dim_para} shows the fitted parameters in the DR case.
%%%%%%%%%%%%%%%%%%%%%%%%%%%%%%%%%%%%%%%%%%%%%%%%
\begin{table}[h!]
\caption{Parameters in the dimensional regularization}
\label{table_dim_para}
%\begin{ruledtabular}
\begin{center}
\begin{tabular}{cccccc}
\hline
$m_u$(MeV)  & $D$ & $GM_0^{4-D}$(MeV${}^{-2}$) & $M_0$(MeV)  
& $m^*$(MeV)  & $\langle \bar{u}u \rangle^{1/3}$(MeV) \\
\hline
%3.0 & 2.3695 & -0.011420 & 109.89985 & -568.434 & -298.539 \\
%4.0 & 2.4685 & -0.008262 & 103.08927 & -541.710 & -271.447 \\
%5.0 & 2.5582 & -0.005947 & 96.82379 & -517.577 & -252.194 \\
%6.0 & 2.6393 & -0.004335 & 91.06331 & -495.884 & -237.628 \\
%7.0 & 2.7144 & -0.003199 & 85.66064 & -475.634 & -225.827 \\
%8.0 & 2.7826 & -0.002410 & 80.63580 & -457.680 & -216.275 \\
3.0 & 2.37 & $-0.01134$ & 110 & $-570$ & $-299$ \\
5.0 & 2.56 & $-0.00588$ & 97 & $-519$ & $-253$ \\
8.0 & 2.78 & $-0.00241$ & 80 & $-459$ & $-217$ \\
20.9 & 3.32 & $-0.000229$ & 37 & $-311$ & $-160$ \\
\hline
\end{tabular}
\end{center}
%\end{ruledtabular}
\end{table}
%%%%%%%%%%%%%%%%%%%%%%%%%%%%%%%%%%%%%%%%%%%%%%%%
We note that both the constituent quark mass and chiral condensate
grow up with increasing $m_u$, which is the characteristic feature in
this regularization~\cite{Inagaki:2010nb}.

%%%%%%%%%%%%%%%%%%%%%%%%%%%%
\section{Meson properties}
\label{sec_meson}
%%%%%%%%%%%%%%%%%%%%%%%%%%%%
We have presented the required equations in the model, then set
the parameters for various regularizations. It is now ready for the
actual numerical analysis on the model predictions. Here we shall
show the thermal meson properties, which are the pion mass, the pion
decay constant, and the sigma meson mass.

At finite temperature, there are two ways of the application
of each regularization; one is to apply the regularization only for the
temperature independent contribution because the temperature
dependent contributions are always finite due to the characteristic
factor of the Fermi-Dirac distribution, i.e., $f(E)$. The other is to
apply the regularization both for the temperature independent and
dependent parts, since the regularization essentially relates to the
cutoff of the model so the introduction of the same cutoff clearly
determines the model scale. On the other hand, the former method
retains more symmetry of the model. Then the physical meaning of
these prescriptions are that the former one respects the model
symmetry, and the latter does the cutoff scale of the model. Since
the model is not renormalizable, the predictions depend on the
regularization ways and our purpose in this paper is to study the
regularization dependence on the model. Therefore we shall study
all the cases and compare the results among various 
regularizations.

%%%%%%%%%%%%%%%%%%%%%%%%%%%%
\subsection{Results with regularizing $T$-independent contribution}
%%%%%%%%%%%%%%%%%%%%%%%%%%%%
In this subsection, we show the results of meson properties based on
the procedure of applying the regularization only to the temperature
independent part. The required integrals are ${\rm tr}S$ and $I(p^2)$,
and we evaluate the following combination,% for ${\rm tr}S$
%%%%%%%
\begin{align}
  {\rm tr}S = {\rm tr}S_{\rm Reg}^0 + {\rm tr}S^T
\end{align}
%%%%%%%%
where ${\rm tr}S_{\rm Reg}^0$ is ${\rm tr}S$ for $T=0$. The lower index
indicates %the ${\rm tr}S^0$ in 
each regularization, namely,  ${\rm tr}S_{\rm 3D}^0$, ${\rm tr}S_{\rm 4D}^0$,
${\rm tr}S_{\rm PV}^0$, ${\rm tr}S_{\rm PT}^0$ and
${\rm tr}S_{\rm DR}^0$. For the temperature dependent part
${\rm tr}S^T$, here we use the form shown in  Eq. (\ref{trST}). Similarly
for $I(p^2)$, we use the equivalent expression, 
%%%%%%%
\begin{align}
  I(p^2) = I_{\rm Reg}^0(p^2) + I^T(p^2),
\end{align}
%%%%%%%%
with $I_{\rm Reg}^0$ for each regularization way.

% Results for the Pion Mass
%%%%%%%%%%%%%%%%%%%%%%%%%%%%
%\input fig_pion.tex
% --- figure --- %
\begin{figure}[h!]
 \begin{center}
 %\vspace{-0.5cm}
 %\hspace{1.0cm}
  \subfigure{
    \includegraphics[height=4.6cm,keepaspectratio]{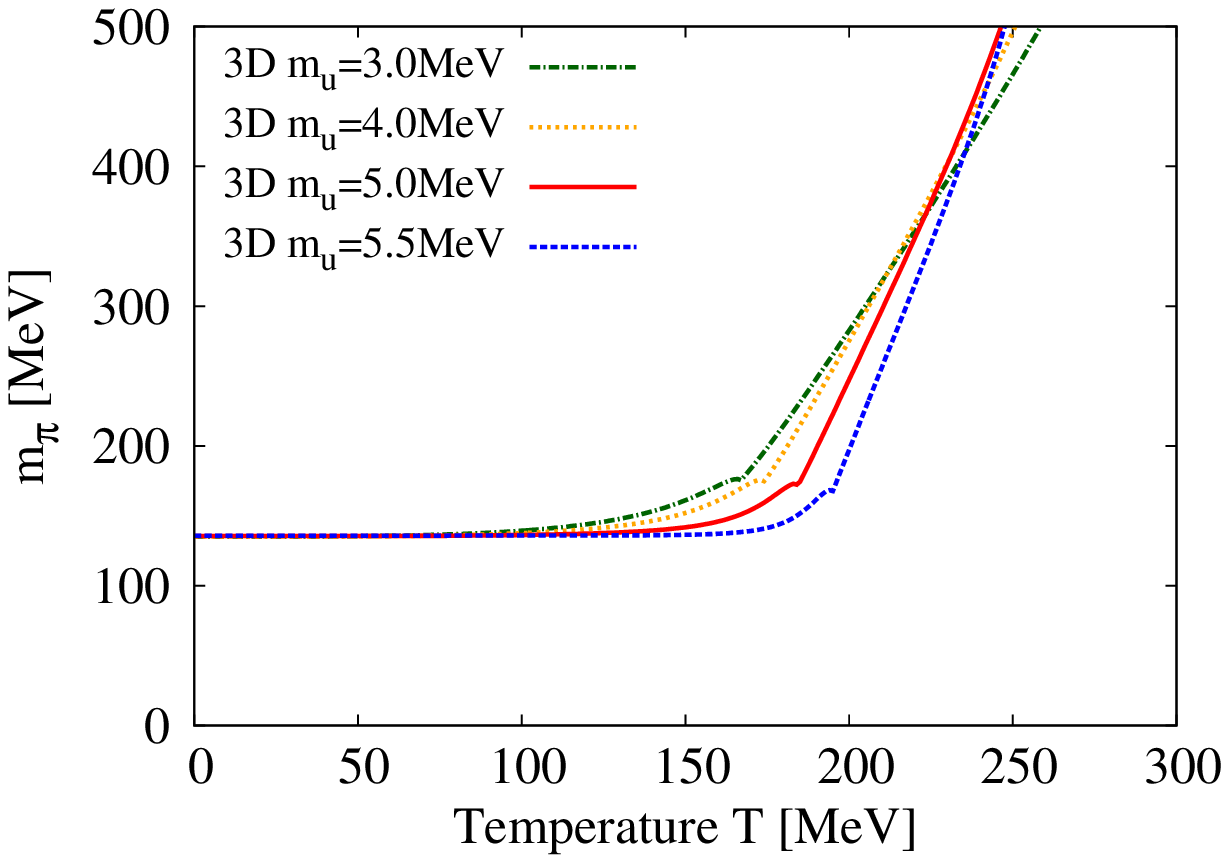} 
  }
  %%\hspace{1.0cm}
  \subfigure{
    \includegraphics[height=4.6cm,keepaspectratio]{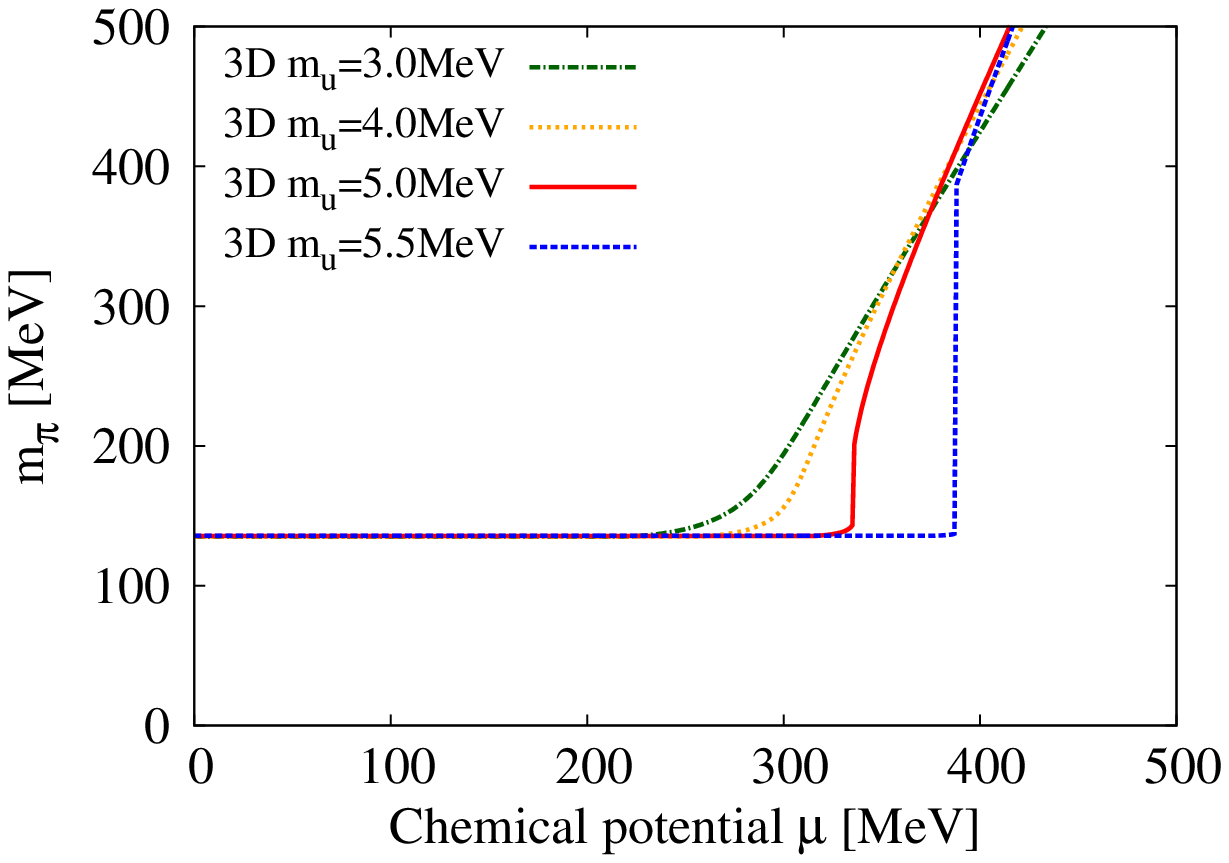} 
  }
  %\hspace{1.0cm}
  \subfigure{
    \includegraphics[height=4.6cm,keepaspectratio]{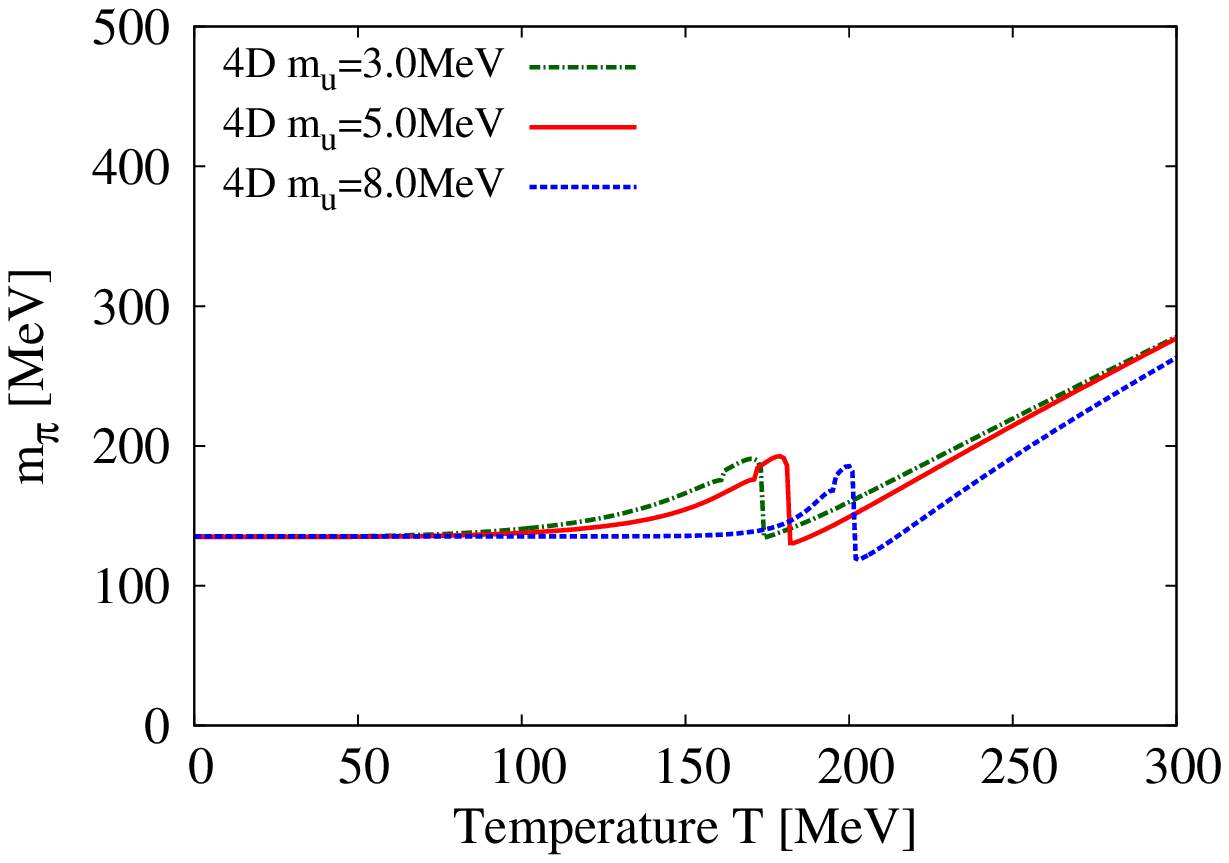} 
  }
  %%\hspace{1.0cm}
  \subfigure{
    \includegraphics[height=4.6cm,keepaspectratio]{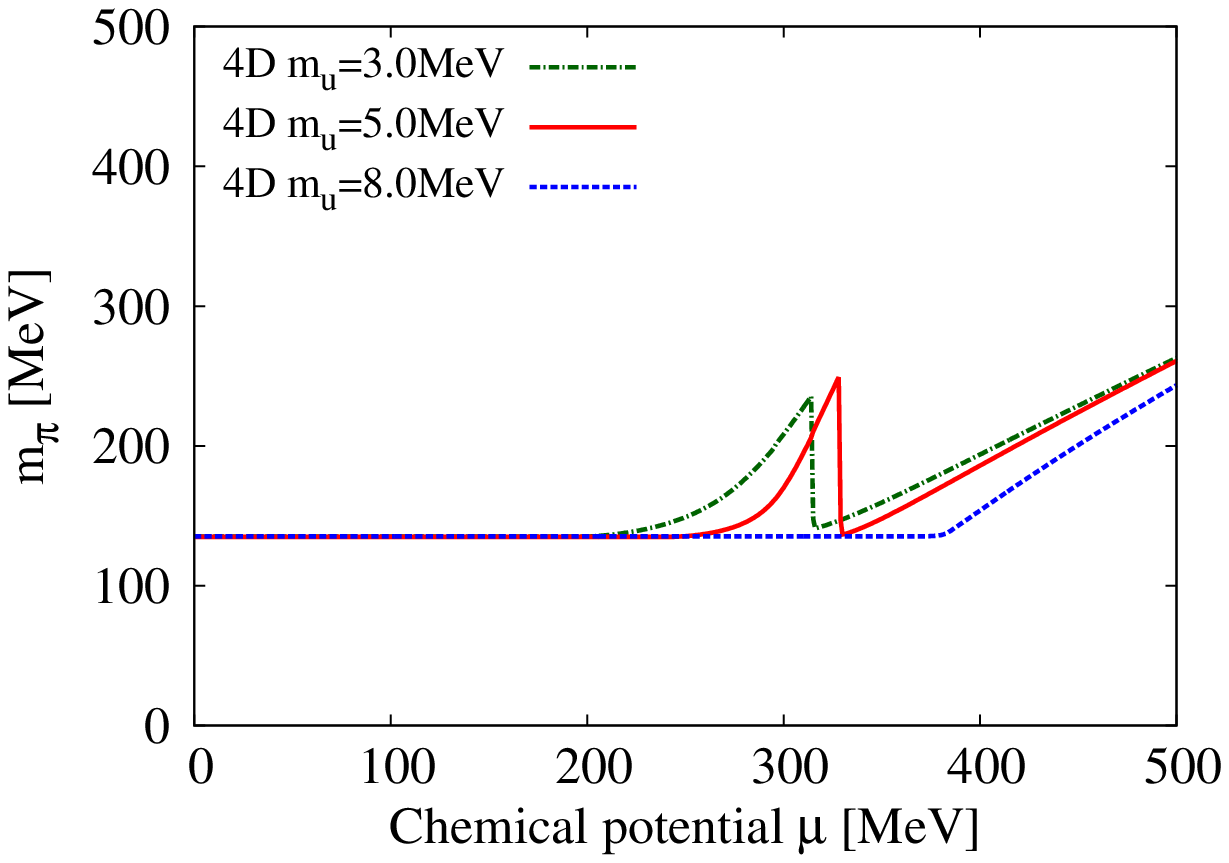} 
  }  
 %\hspace{1.0cm}
  \subfigure{
    \includegraphics[height=4.6cm,keepaspectratio]{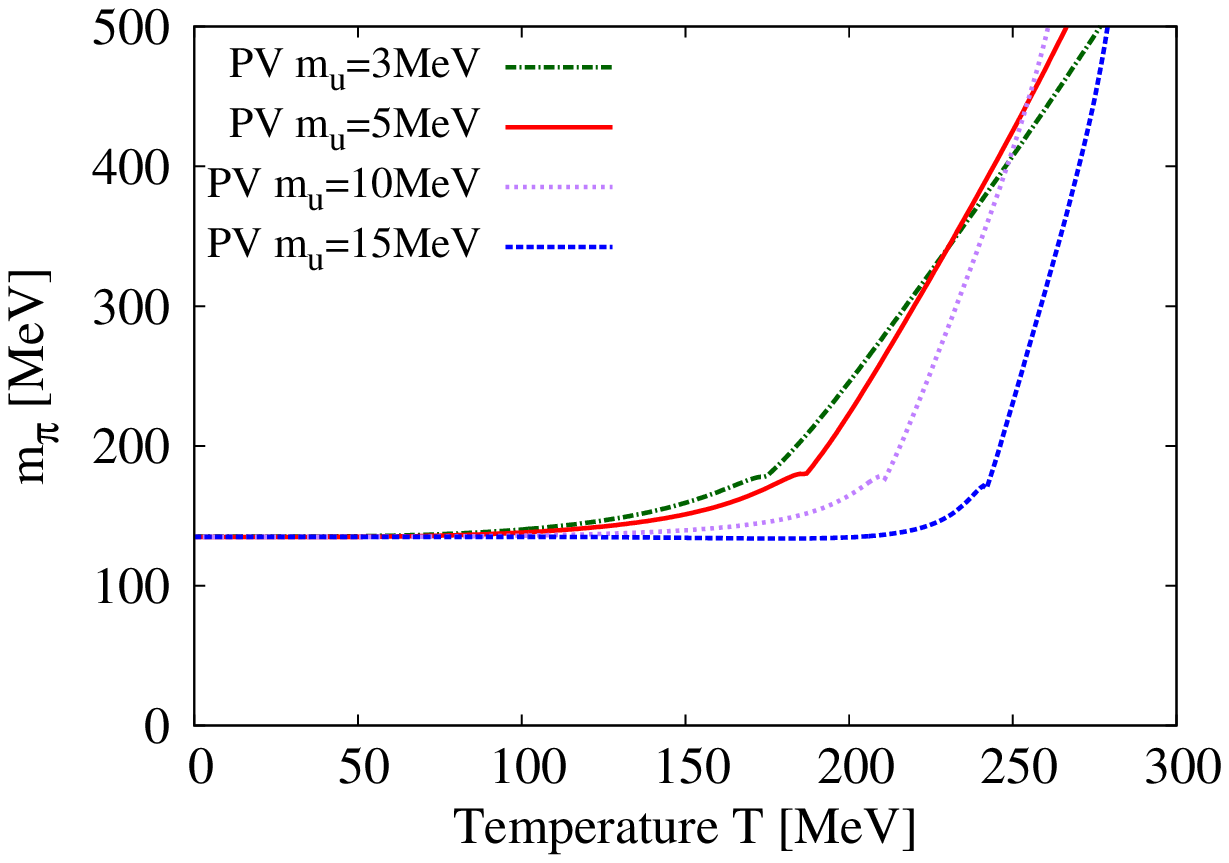} 
  }
  %%\hspace{1.0cm}
  \subfigure{
    \includegraphics[height=4.6cm,keepaspectratio]{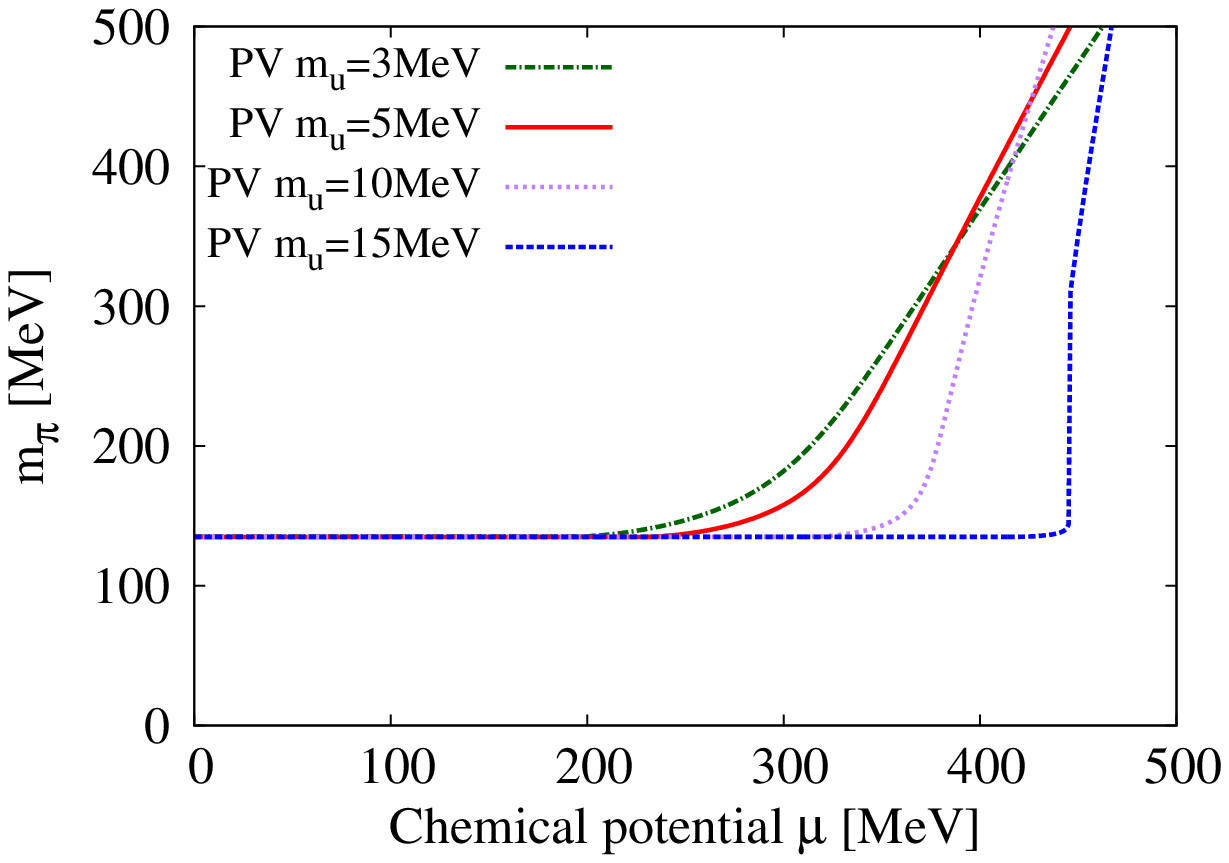} 
  }
  %\hspace{1.0cm}
  \subfigure{
    \includegraphics[height=4.6cm,keepaspectratio]{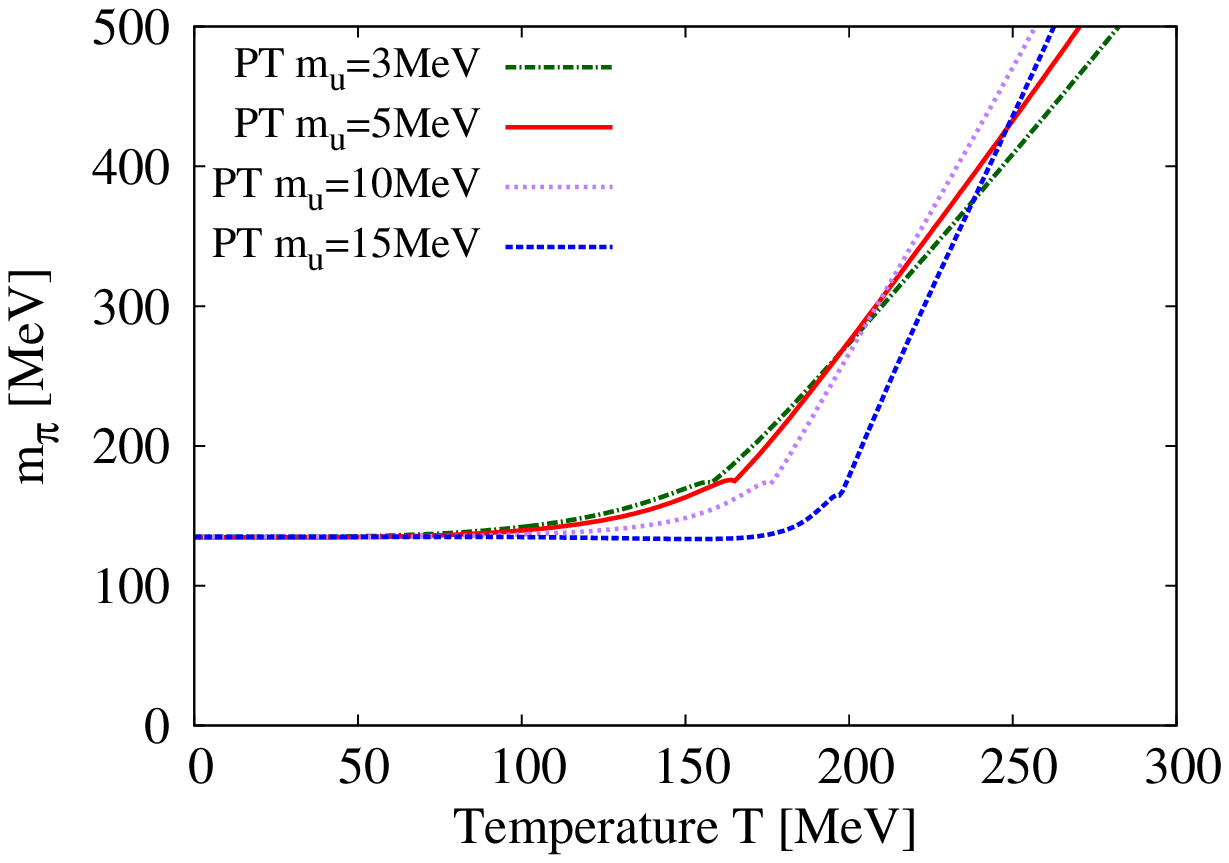} 
  }
  %%\hspace{1.0cm}
  \subfigure{
    \includegraphics[height=4.6cm,keepaspectratio]{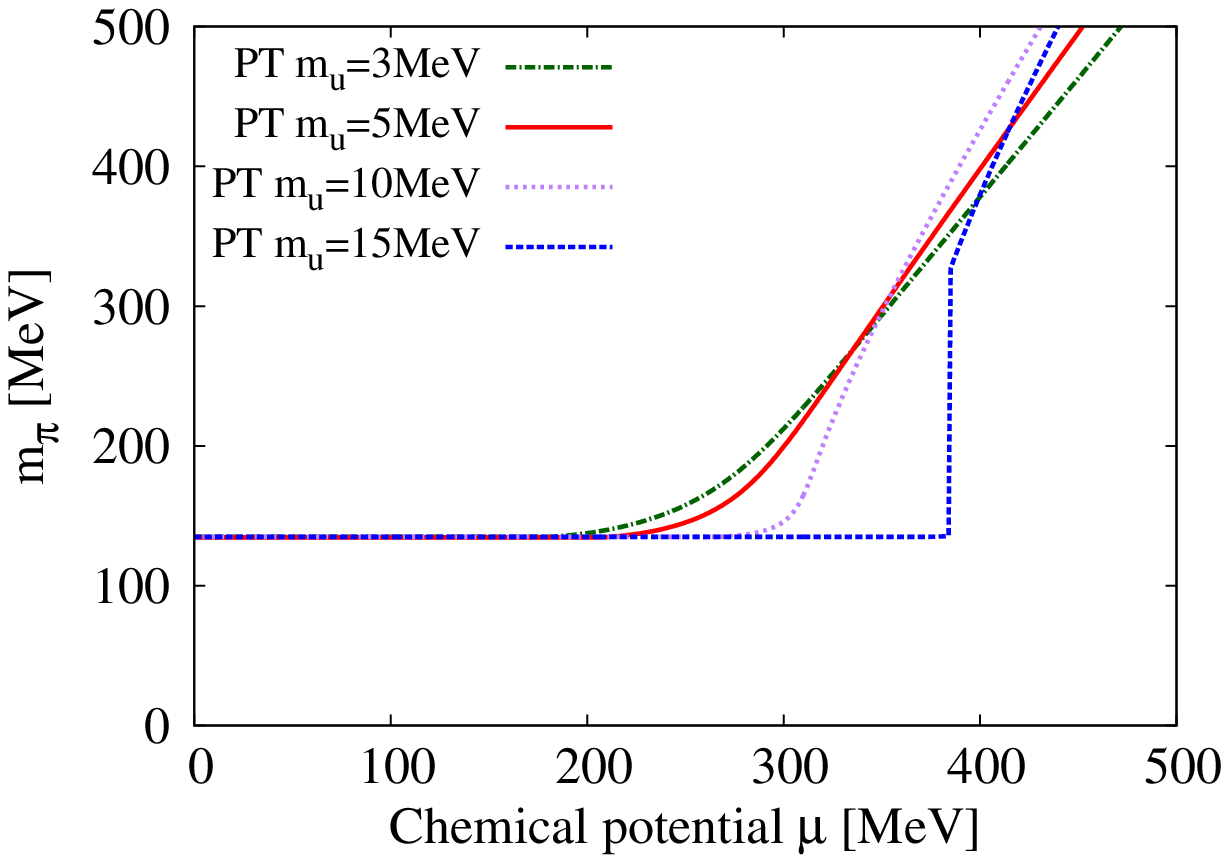} 
  }
  %\hspace{1.0cm}
  \subfigure{
    \includegraphics[height=4.6cm,keepaspectratio]{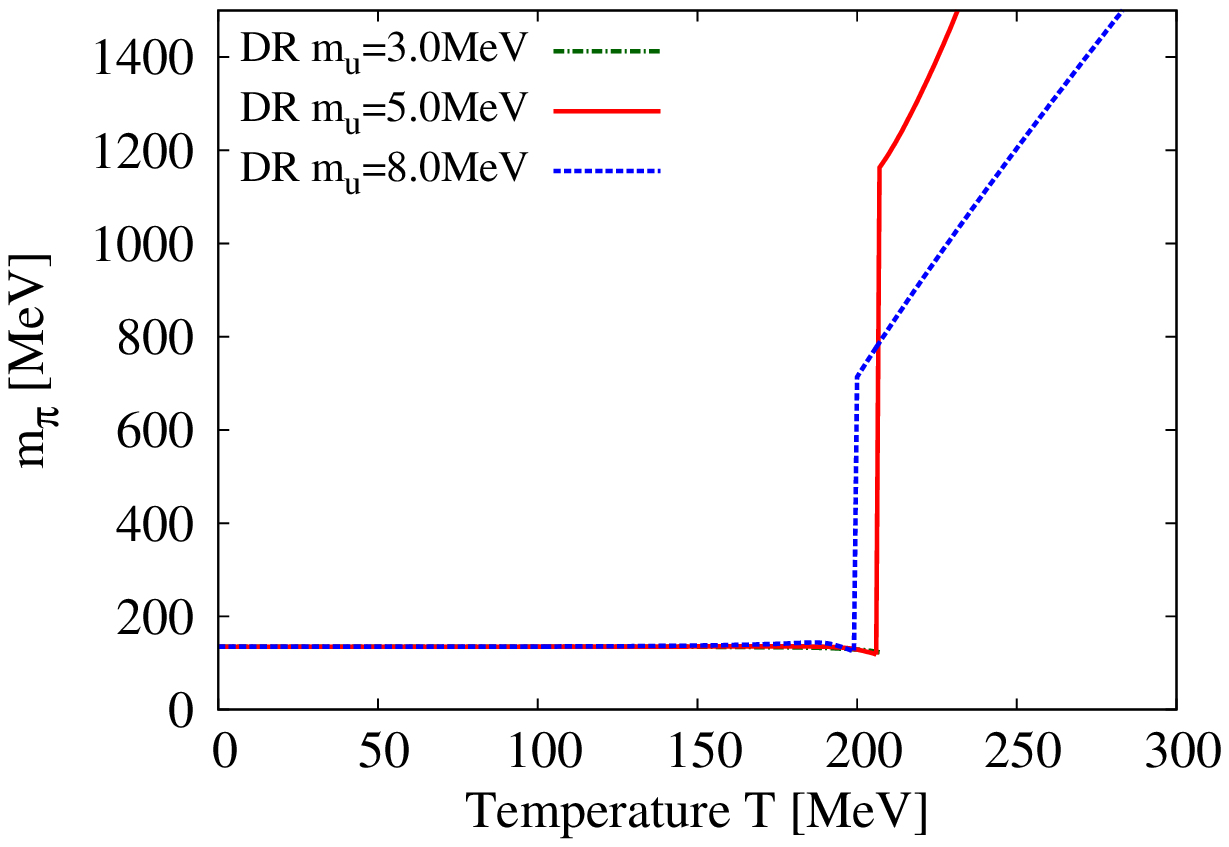} 
  }
  %%\hspace{1.0cm}
  \subfigure{
    \includegraphics[height=4.6cm,keepaspectratio]{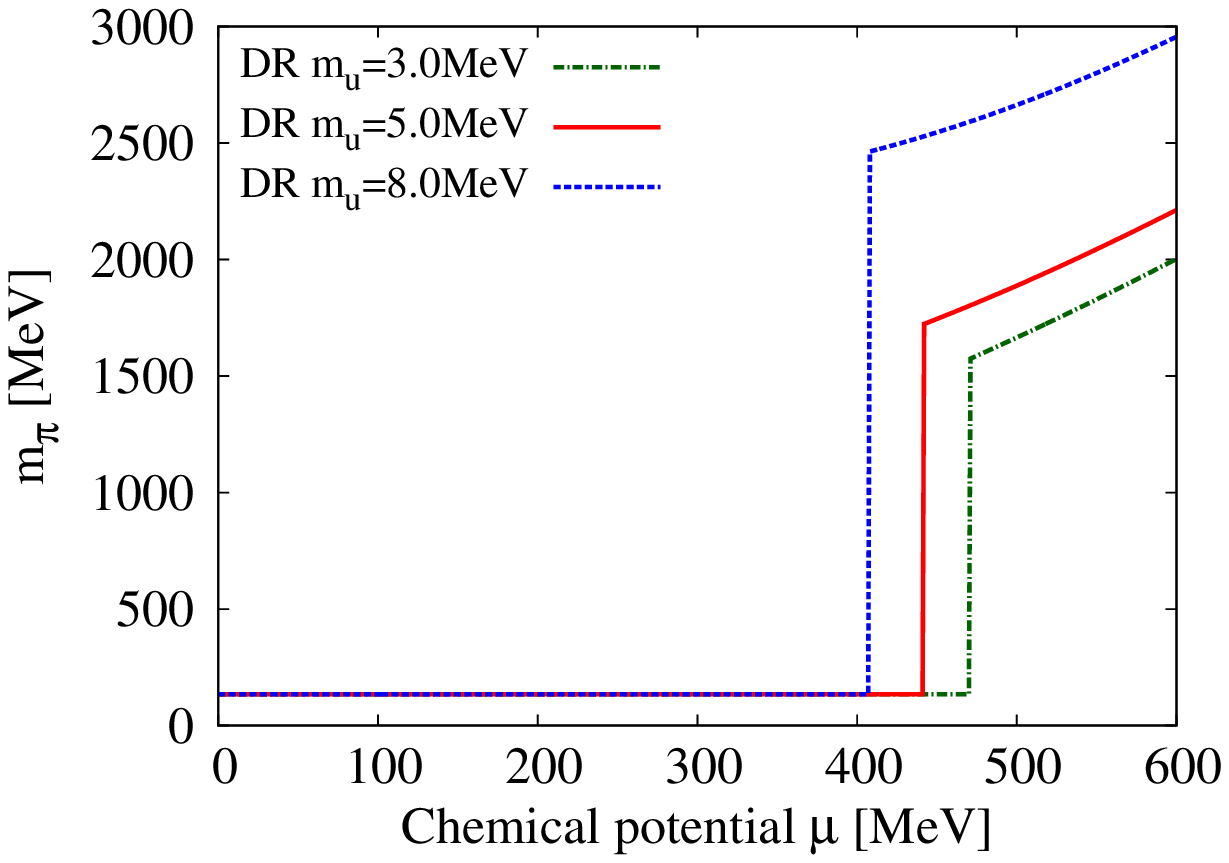} 
  }  
  %\vspace{0.7cm}
  \caption{Pion mass. Left: $\mu=0$. Right: $T=0$.
                }
  \end{center}
\label{fig_pion}
\end{figure}
% --- figure --- %
%%%%%%%%%%%%%%%%%%%%%%%%%%%%
%Figure \ref{fig_pion} 
%%%%%%
Figure 2
%%%%%%
shows how the pion mass changes with respect
to $T$ and $\mu$ for various parameter sets in the previous section.
It should be noted that, for some parameter sets, no real solution exists
at high temperature as seen in the case with the DR and 
$m_u=3.0$MeV~\cite{Inagaki:2011uj}. We observe the similar behavior in each
regularization; the pion mass remains almost
constant for low $T$ and $\mu$, then raises up for higher $T$ and
$\mu$. This comes from the fact that the chiral symmetry is broken
at low $T$ and $\mu$ and 
restores at high $T$ and $\mu$. The pion has smaller
mass when the symmetry is broken due to the Nambu-Goldstone
theorem, while the mass becomes large after symmetry restoration.
We see that the mass starts to increase around $170$MeV which is
comparable to the critical temperature for the chiral symmetry breaking. 
We see that the
temperature and chemical potential where the pion mass glows up
become larger with respect to $m_u$ for the 3D cutoff, 4D cutoff, PV
and PT, while they become smaller for the DR. We also see that the
discontinuity seen around the transition temperature is considerably
larger in the DR case compare to the other regularizations.
Then we expect that the tendency of the first order phase transition
is strong for the DR comparing the other 
regularizations. We will discuss it in the next section.

% Results for the Pion Decay
%%%%%%%%%%%%%%%%%%%%%%%%%%%%
%\input fig_decay.tex
% --- figure --- %
\begin{figure}[h!]
 \begin{center}
 %\vspace{-0.5cm}
 %\hspace{1.0cm}
  \subfigure{
    \includegraphics[height=4.6cm,keepaspectratio]{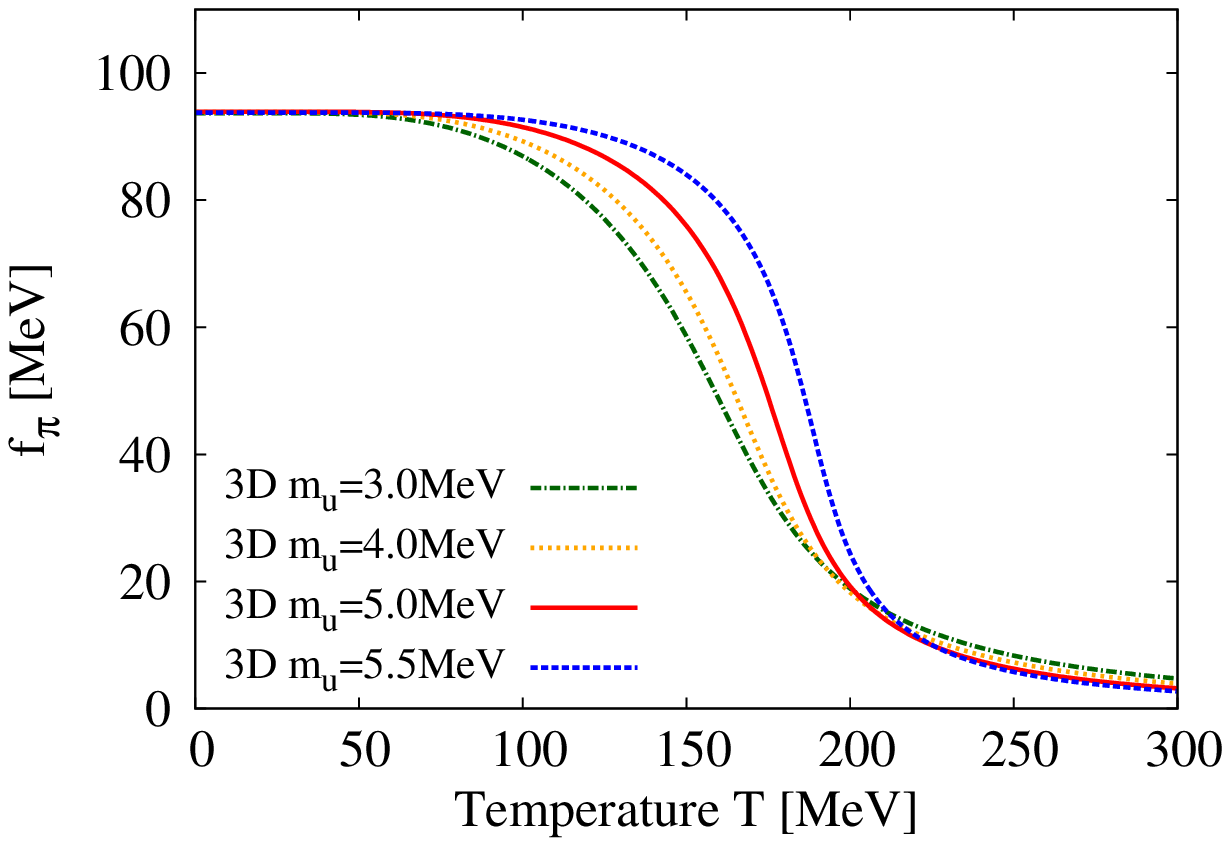} 
  }
  %%\hspace{1.0cm}
  \subfigure{
    \includegraphics[height=4.6cm,keepaspectratio]{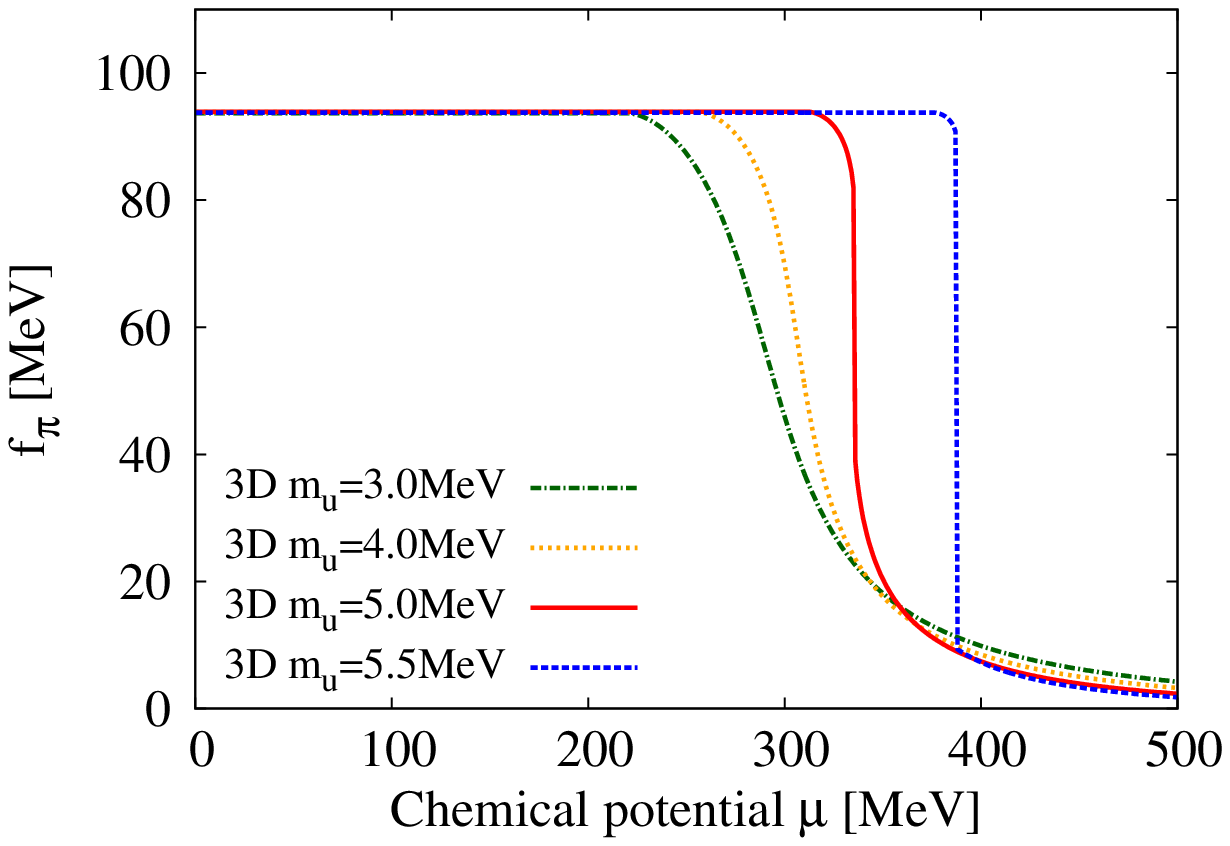} 
  }
  %\hspace{1.0cm}
  \subfigure{
    \includegraphics[height=4.6cm,keepaspectratio]{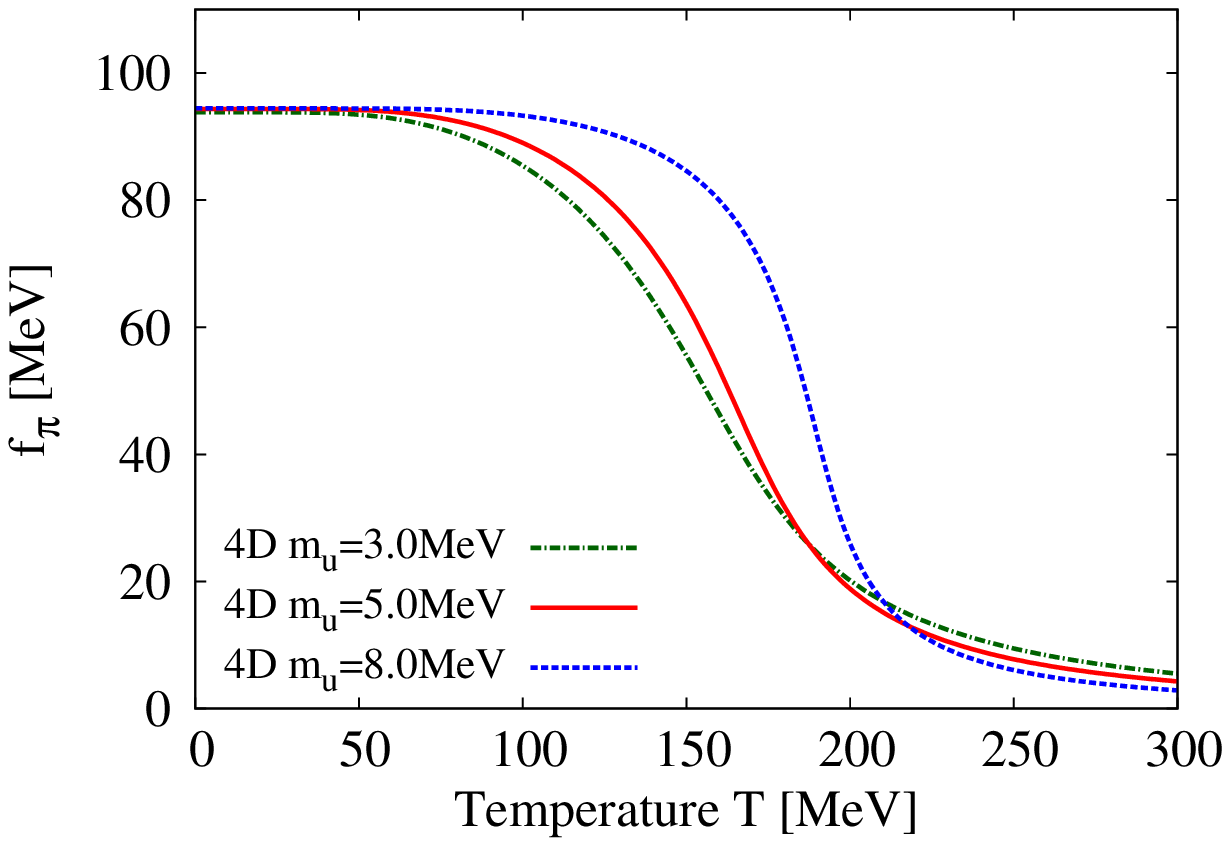} 
  }
  %%\hspace{1.0cm}
  \subfigure{
    \includegraphics[height=4.6cm,keepaspectratio]{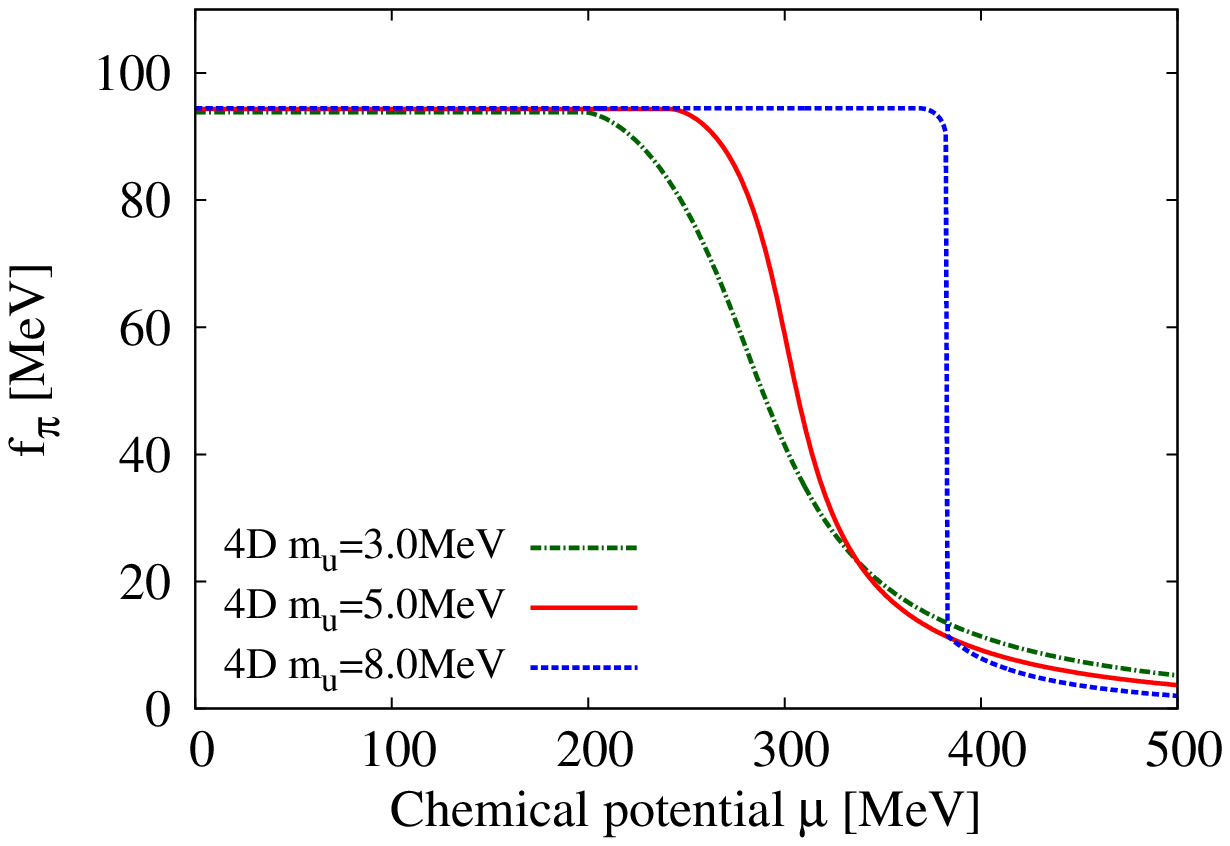} 
  }  
 %\hspace{1.0cm}
  \subfigure{
    \includegraphics[height=4.6cm,keepaspectratio]{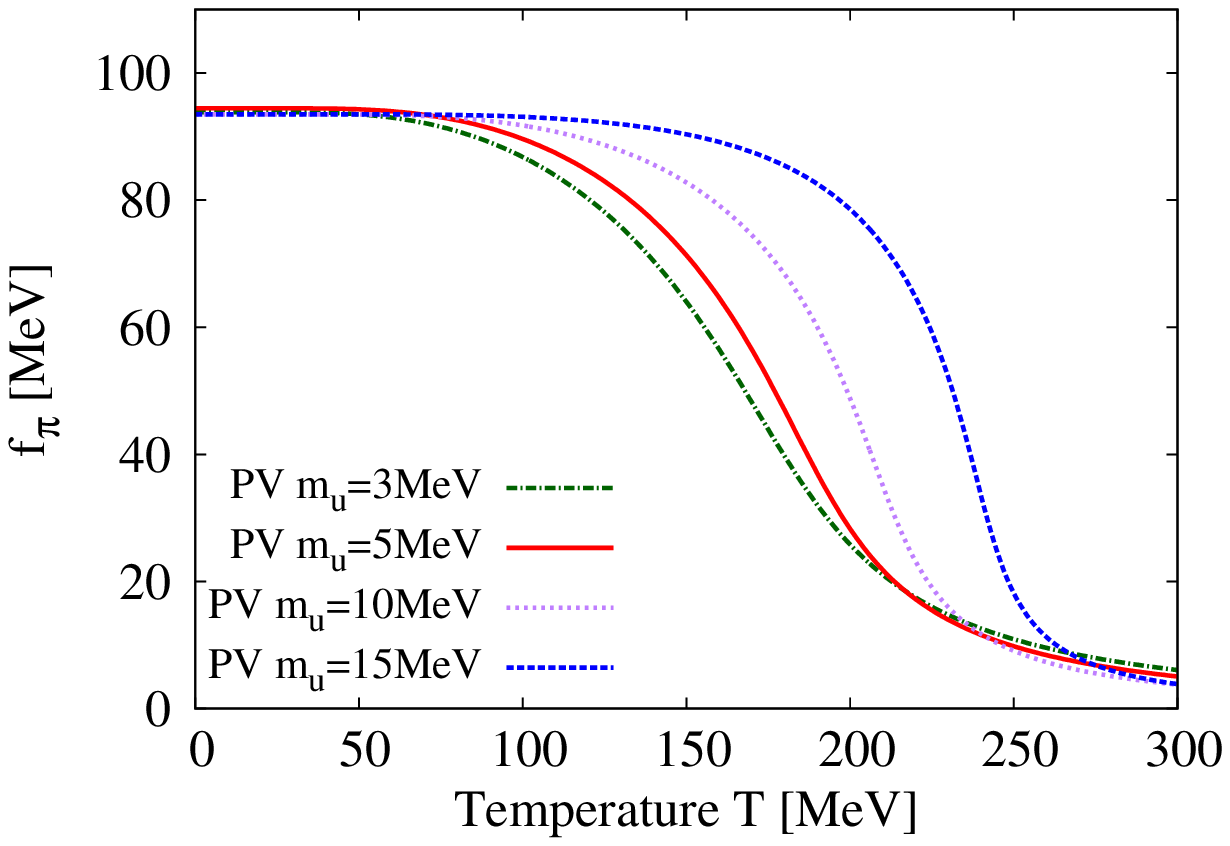} 
  }
  %%\hspace{1.0cm}
  \subfigure{
    \includegraphics[height=4.6cm,keepaspectratio]{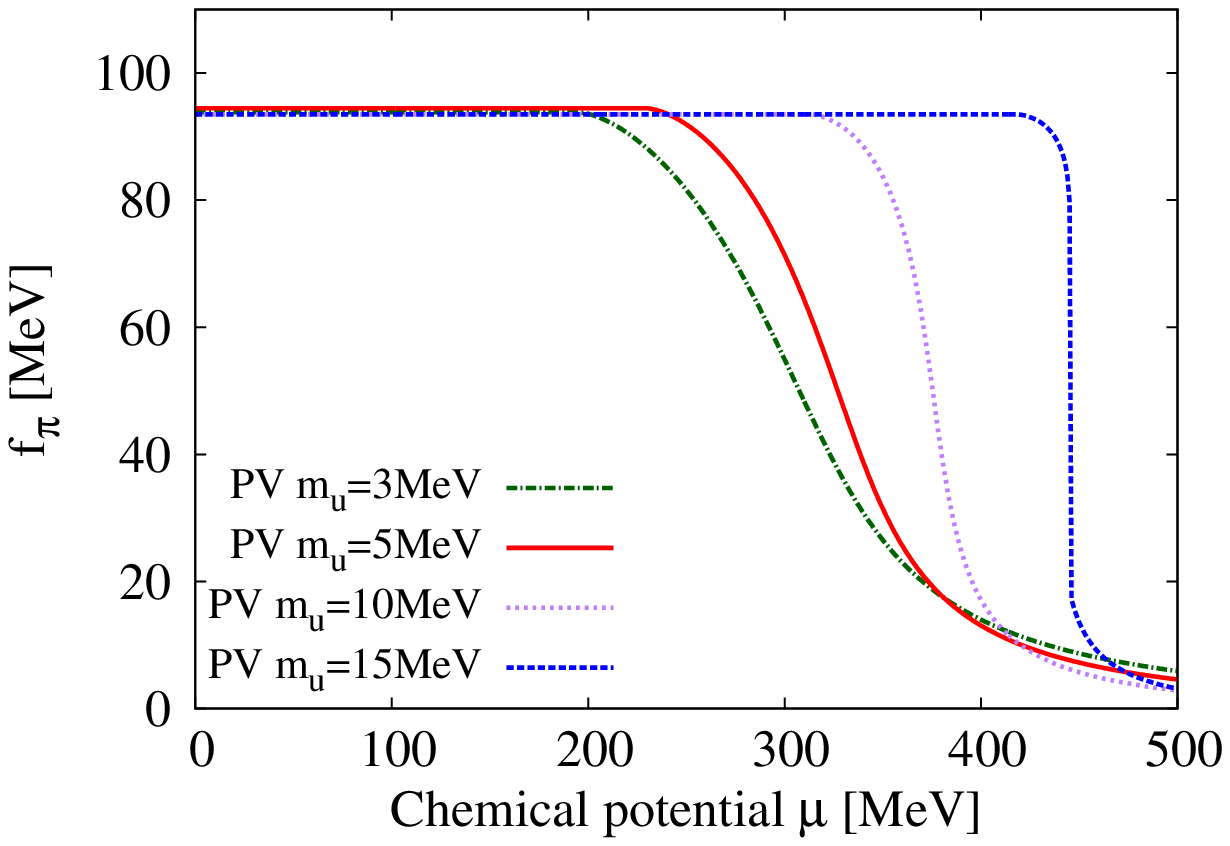} 
  }
  %\hspace{1.0cm}
  \subfigure{
    \includegraphics[height=4.6cm,keepaspectratio]{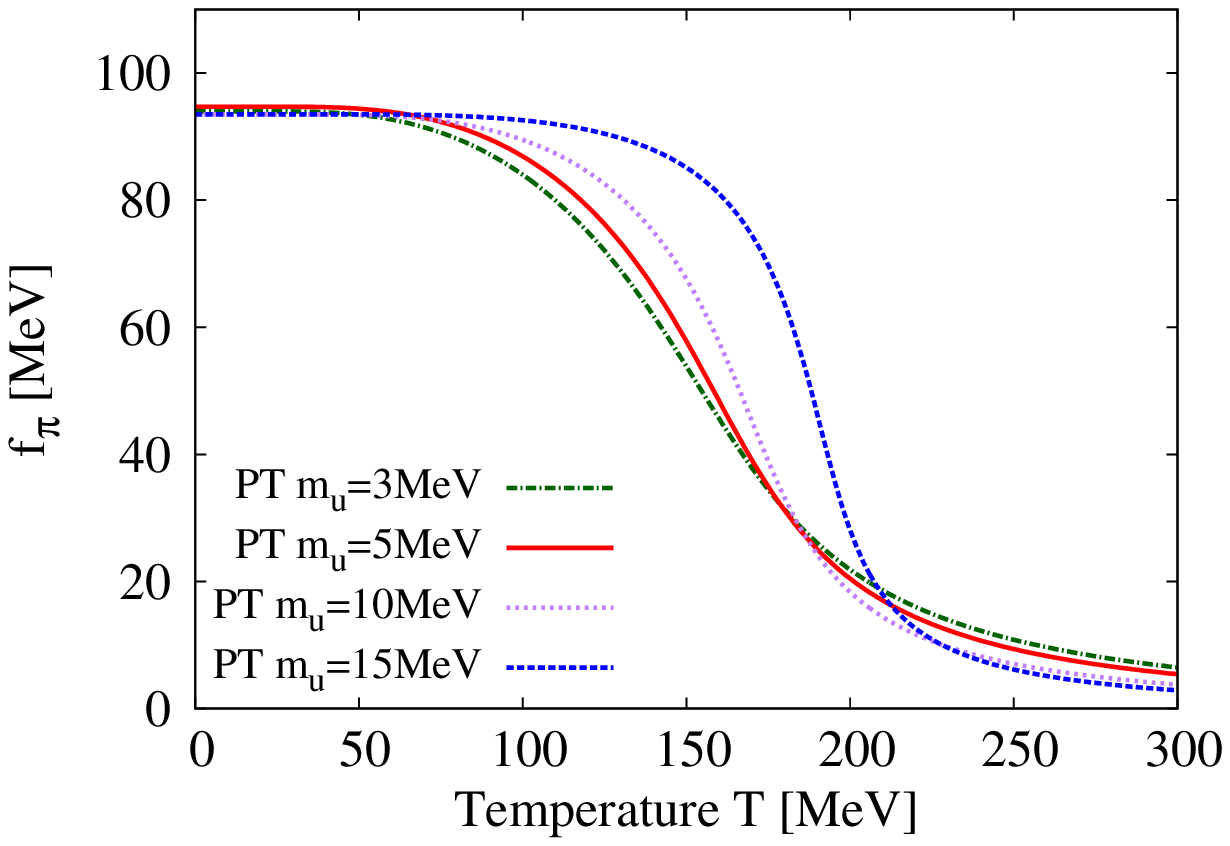} 
  }
  %%\hspace{1.0cm}
  \subfigure{
    \includegraphics[height=4.6cm,keepaspectratio]{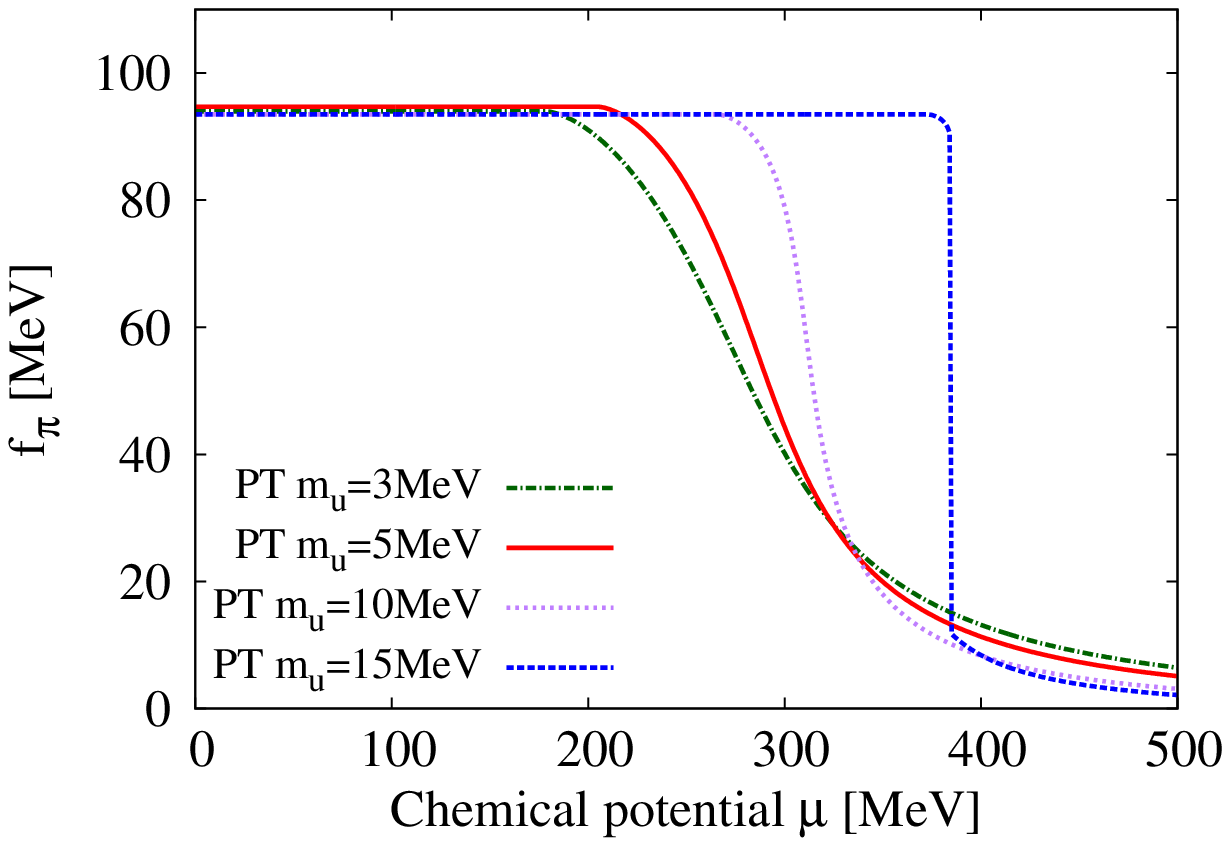} 
  }
 % \hspace{1.0cm}
  \subfigure{
    \includegraphics[height=4.6cm,keepaspectratio]{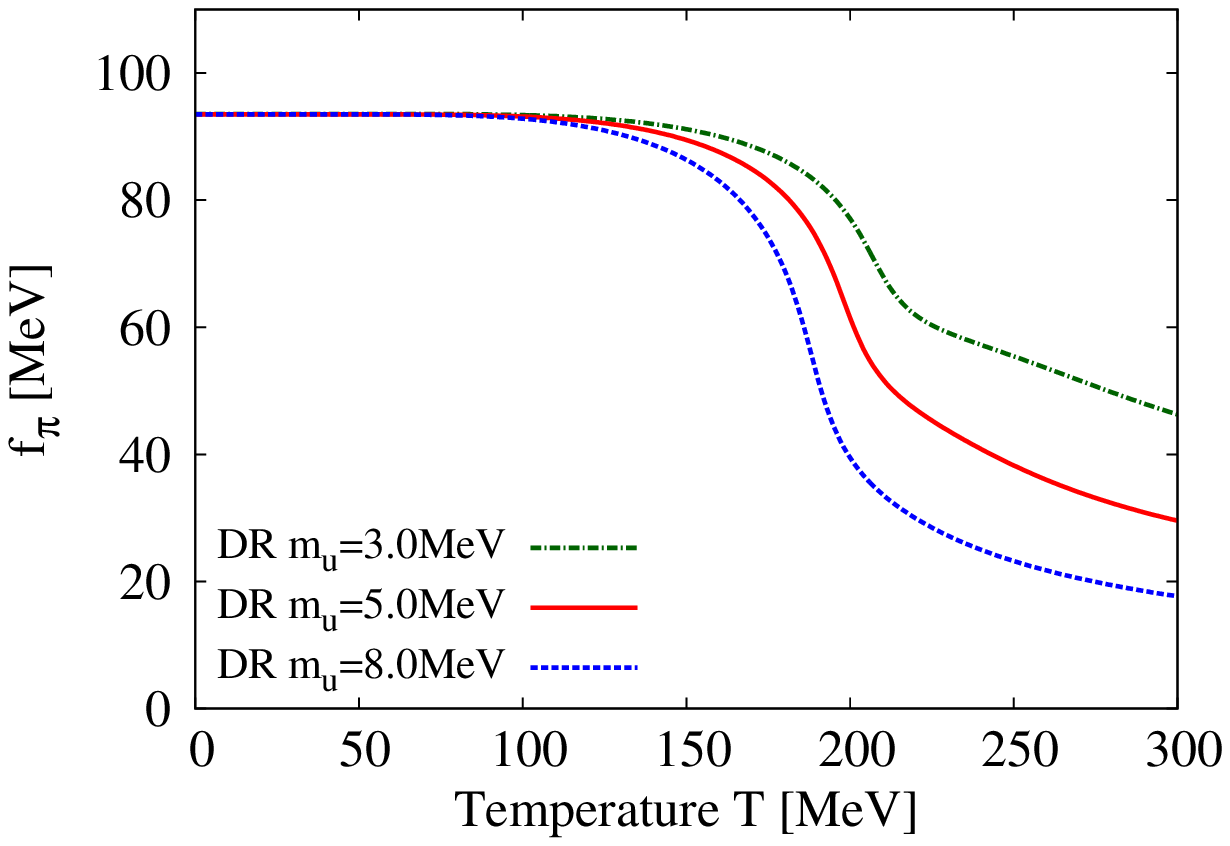} 
  }
  %%\hspace{1.0cm}
  \subfigure{
    \includegraphics[height=4.6cm,keepaspectratio]{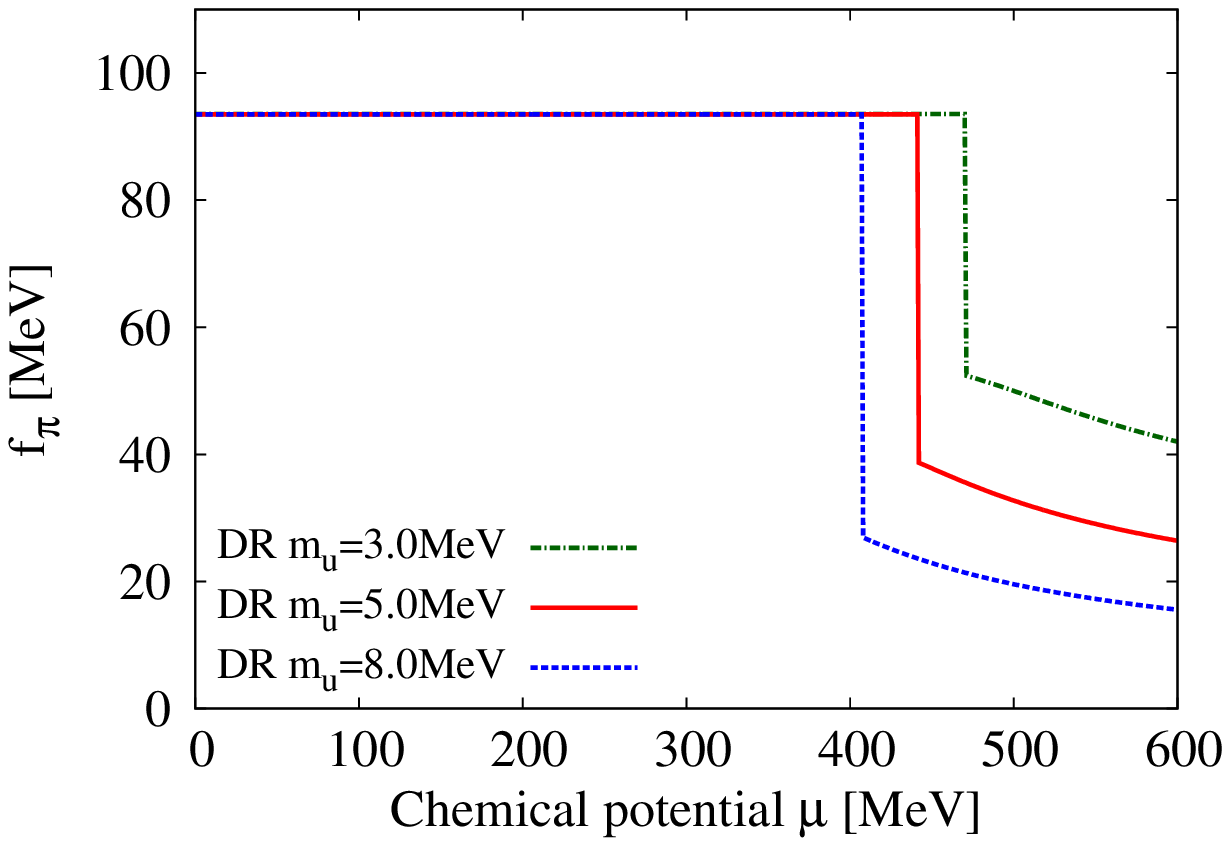} 
  }  
  %\vspace{0.7cm}
  \caption{Pion decay constant. Left: $\mu=0$.
                Right: $T=0$.
                }
  \end{center}
%\vspace{-0.5cm}
\label{fig_decay}
\end{figure}
% --- figure --- %
%%%%%%%%%%%%%%%%%%%%%%%%%%%%
The results of the pion decay are shown in 
%Fig. \ref{fig_decay}.
%%%%%%
Fig. 3.
%%%%%%
One
sees the similar tendency as well; the decay constant is almost constant
for low $T$ and $\mu$, and it decreases when $T$ and $\mu$ exceed
certain values which are around $T\simeq 170$MeV and $\mu \simeq
300-400$MeV. It is interesting to note that the decay constant drops
discontinuously at high $\mu$ for some parameter sets in the $3$D,
$4$D, PV, PT regularizations, while the discontinuity is always the case
in the DR. The existence of the gap is the signal of the first order phase
transition, and the tendency becomes stronger with increasing $m_u$.
This is because the coupling strength is larger for higher $m_u$ as
seen from the parameter tables, so quarks have stronger correlations
when the parameter $m_u$ is larger in the $3$D, $4$D, PV, PT cases.

% Results for the Sigma Mass 
%%%%%%%%%%%%%%%%%%%%%%%%%%%%
%\input fig_sigma.tex
% --- figure --- %
\begin{figure}[h!]
 \begin{center}
 %\vspace{-0.5cm}
 %\hspace{1.0cm}
  \subfigure{
    \includegraphics[height=4.6cm,keepaspectratio]{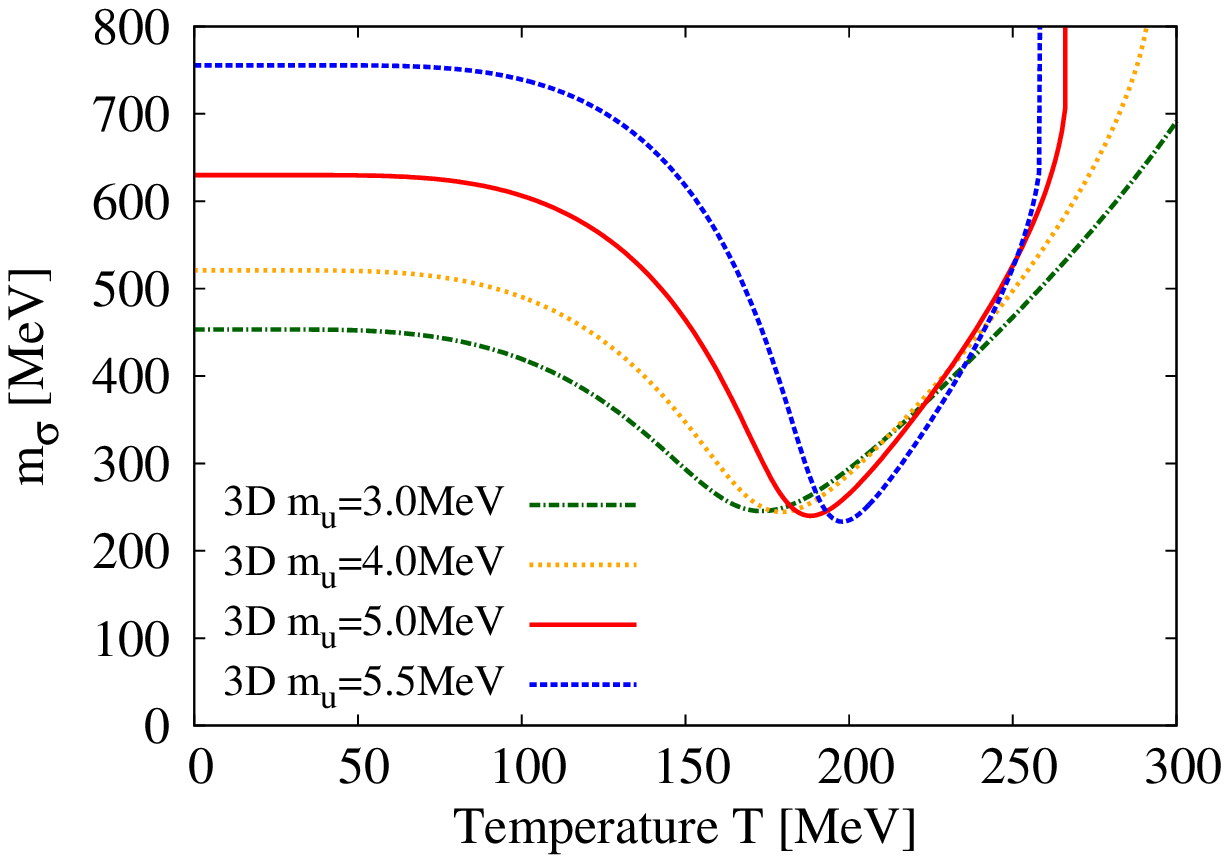} 
  }
  %%\hspace{1.0cm}
  \subfigure{
    \includegraphics[height=4.6cm,keepaspectratio]{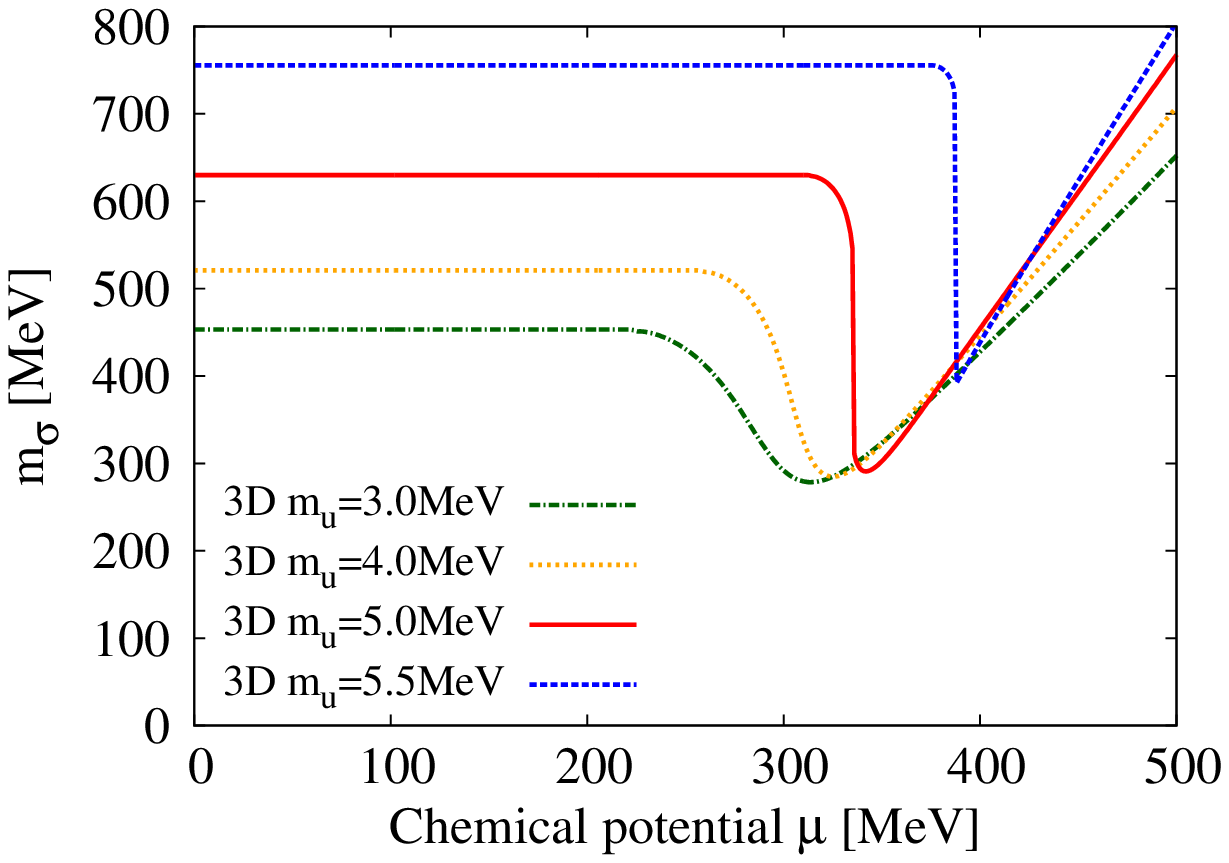} 
  }
  %\hspace{1.0cm}
  \subfigure{
    \includegraphics[height=4.6cm,keepaspectratio]{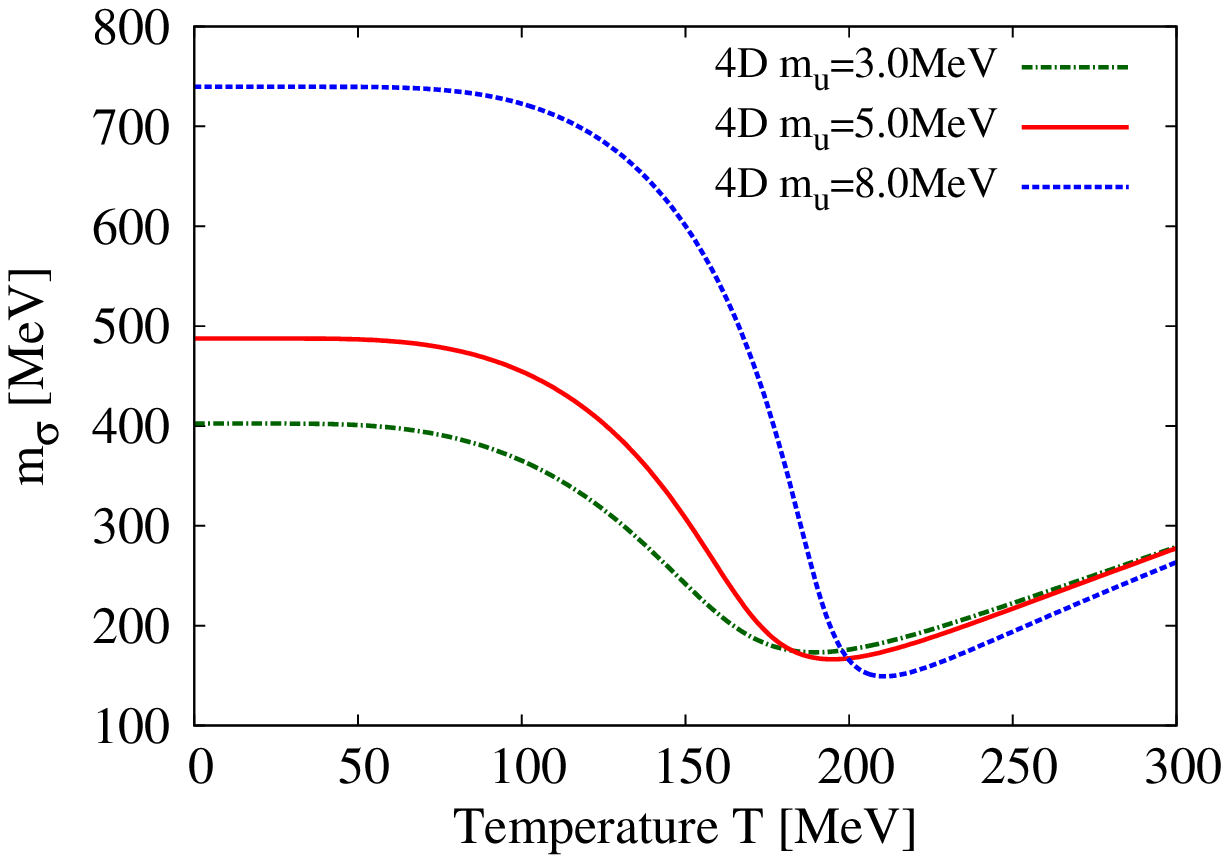} 
  }
  %%\hspace{1.0cm}
  \subfigure{
    \includegraphics[height=4.6cm,keepaspectratio]{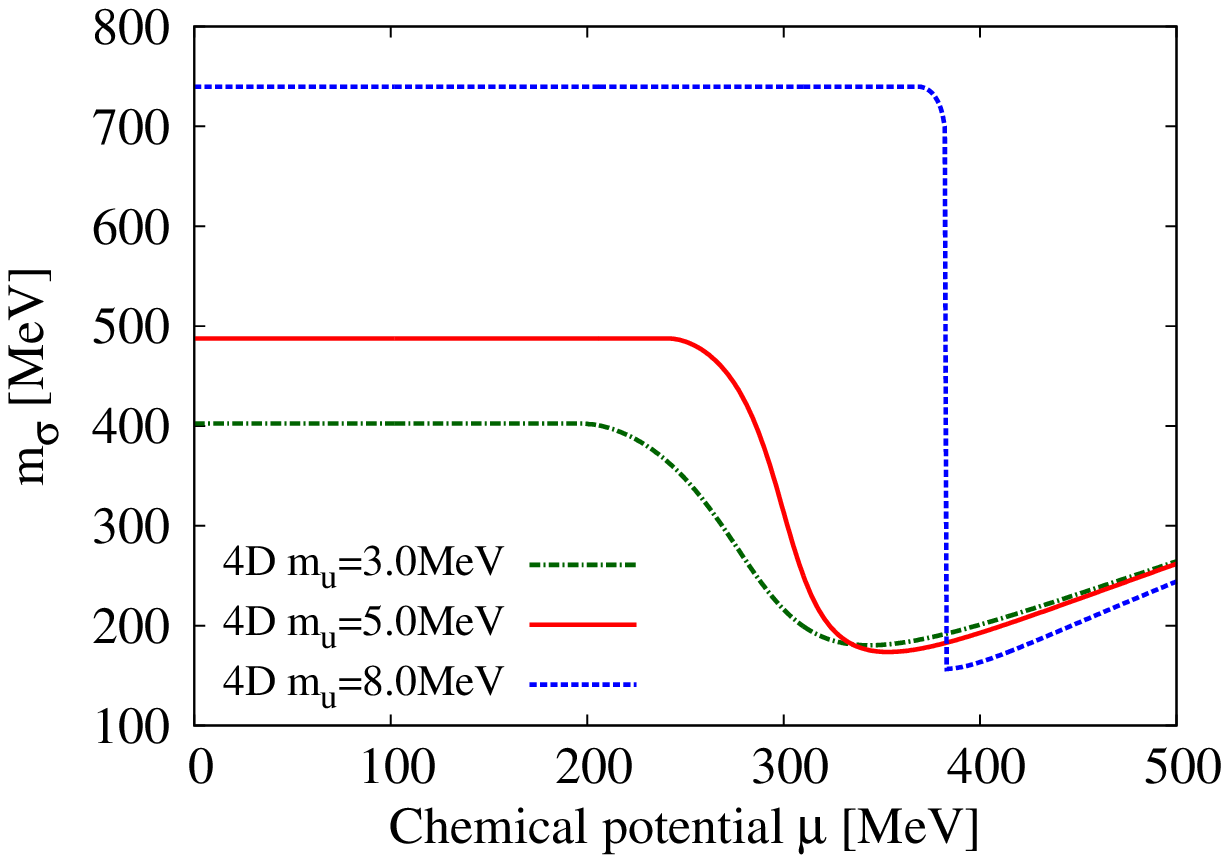} 
  }  
 %\hspace{1.0cm}
  \subfigure{
    \includegraphics[height=4.6cm,keepaspectratio]{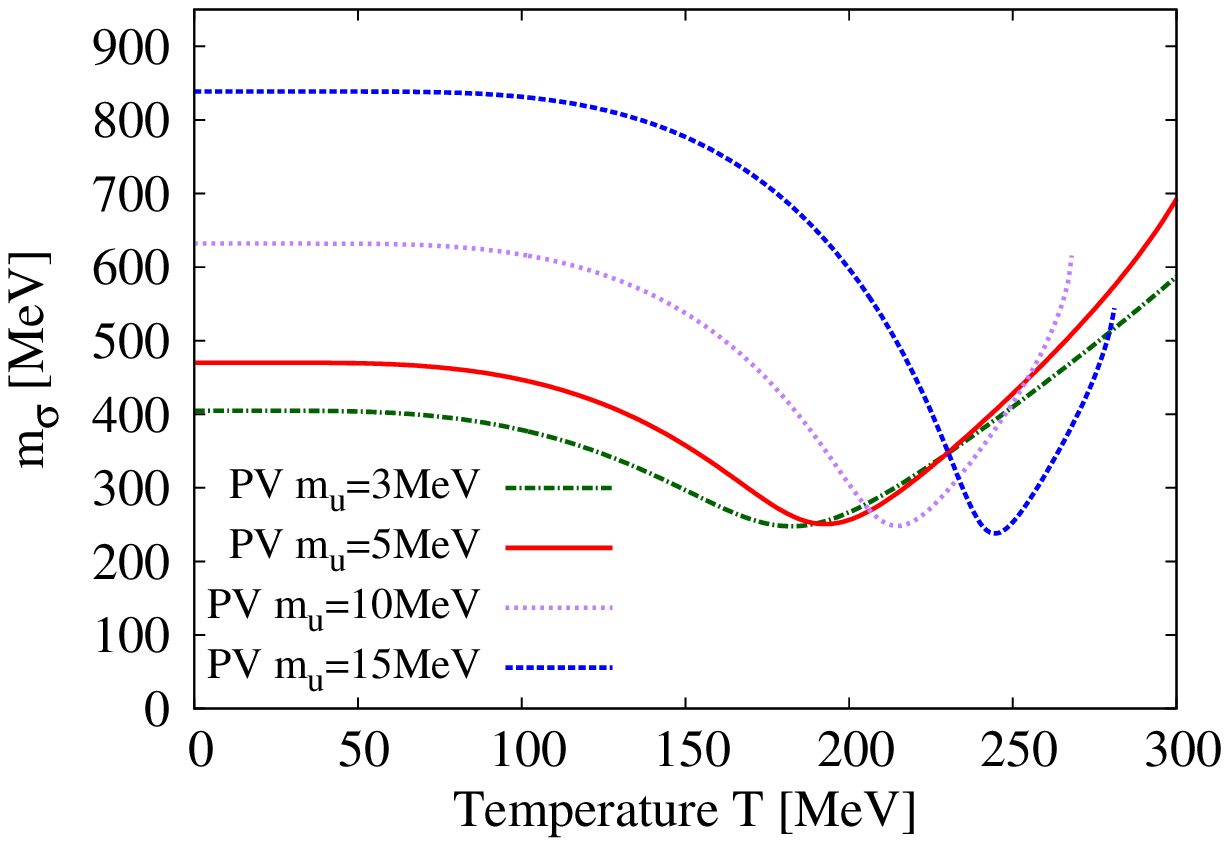} 
  }
  %%\hspace{1.0cm}
  \subfigure{
    \includegraphics[height=4.6cm,keepaspectratio]{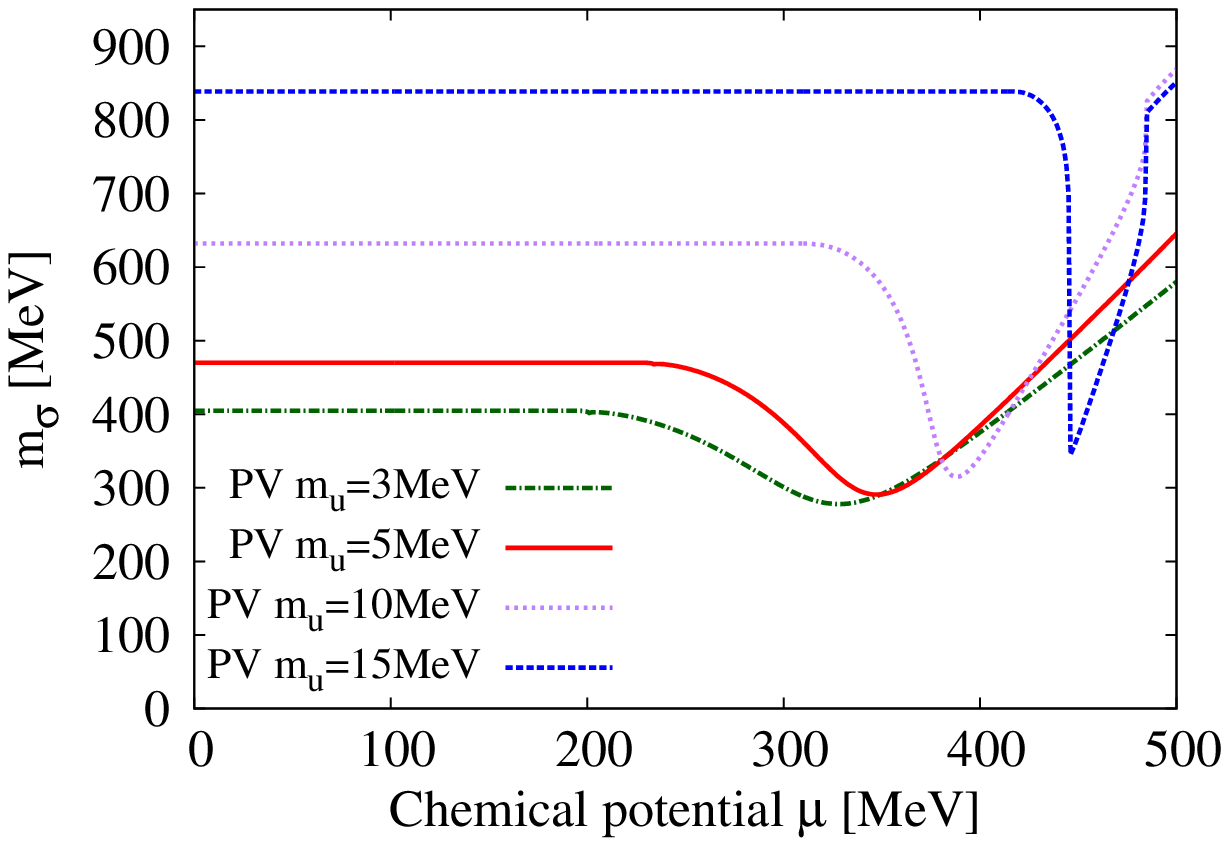} 
  }
  %\hspace{1.0cm}
  \subfigure{
    \includegraphics[height=4.6cm,keepaspectratio]{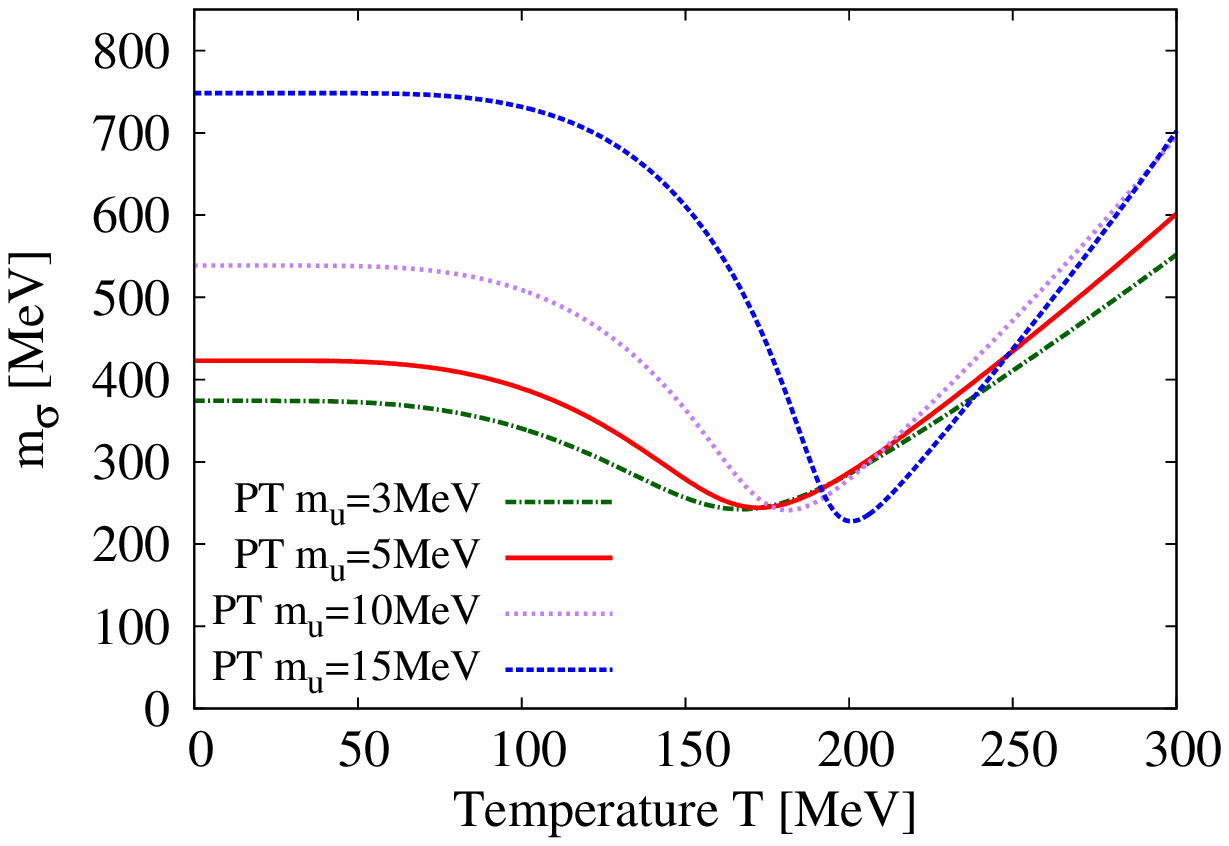} 
  }
  %%\hspace{1.0cm}
  \subfigure{
    \includegraphics[height=4.6cm,keepaspectratio]{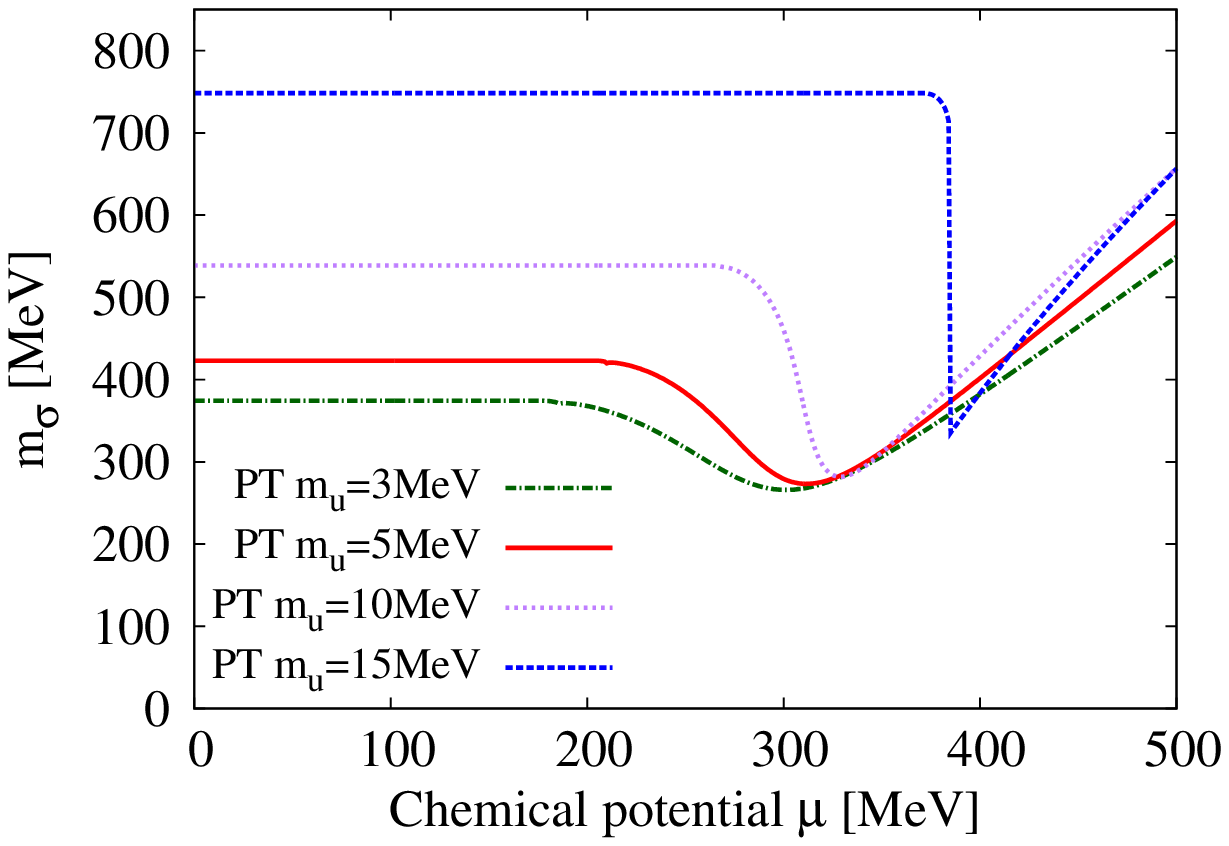} 
  }
  %\hspace{1.0cm}
  \subfigure{
    \includegraphics[height=4.6cm,keepaspectratio]{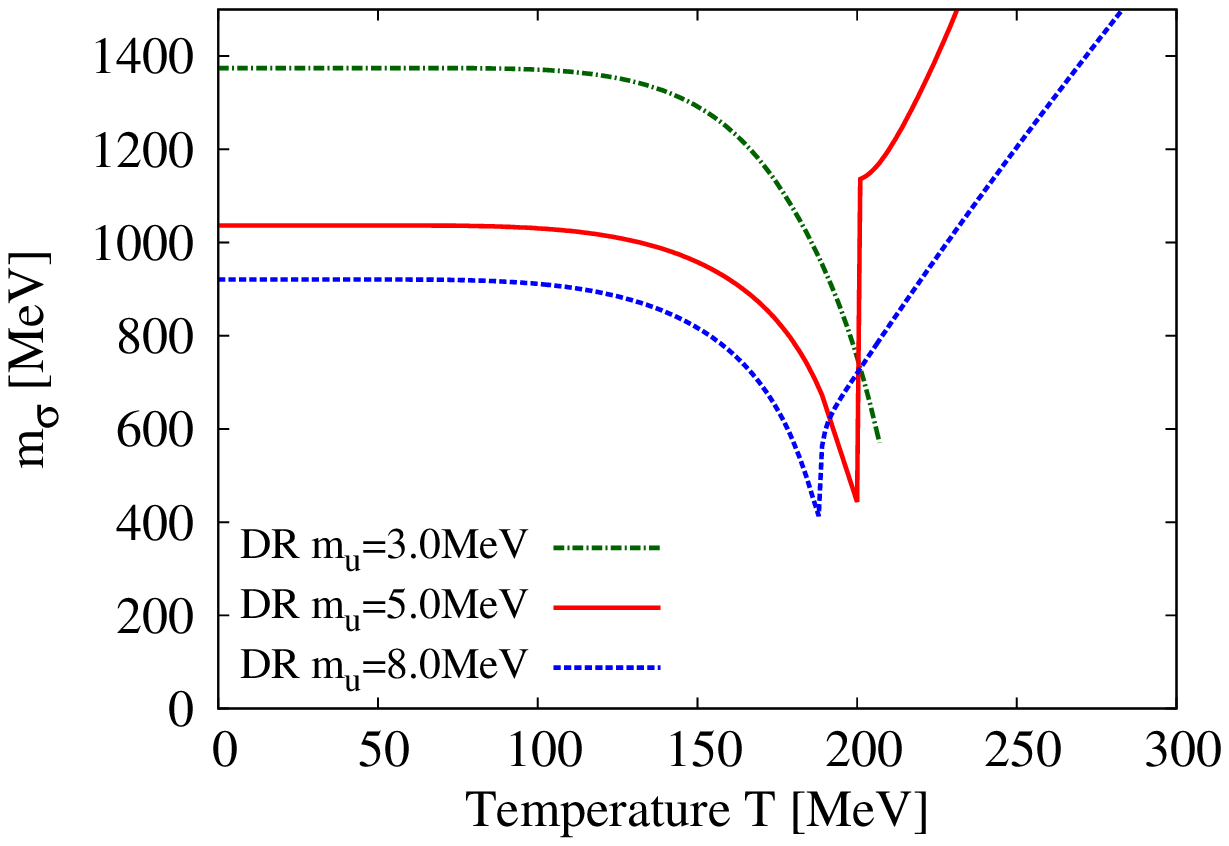} 
  }
  %%\hspace{1.0cm}
  \subfigure{
    \includegraphics[height=4.6cm,keepaspectratio]{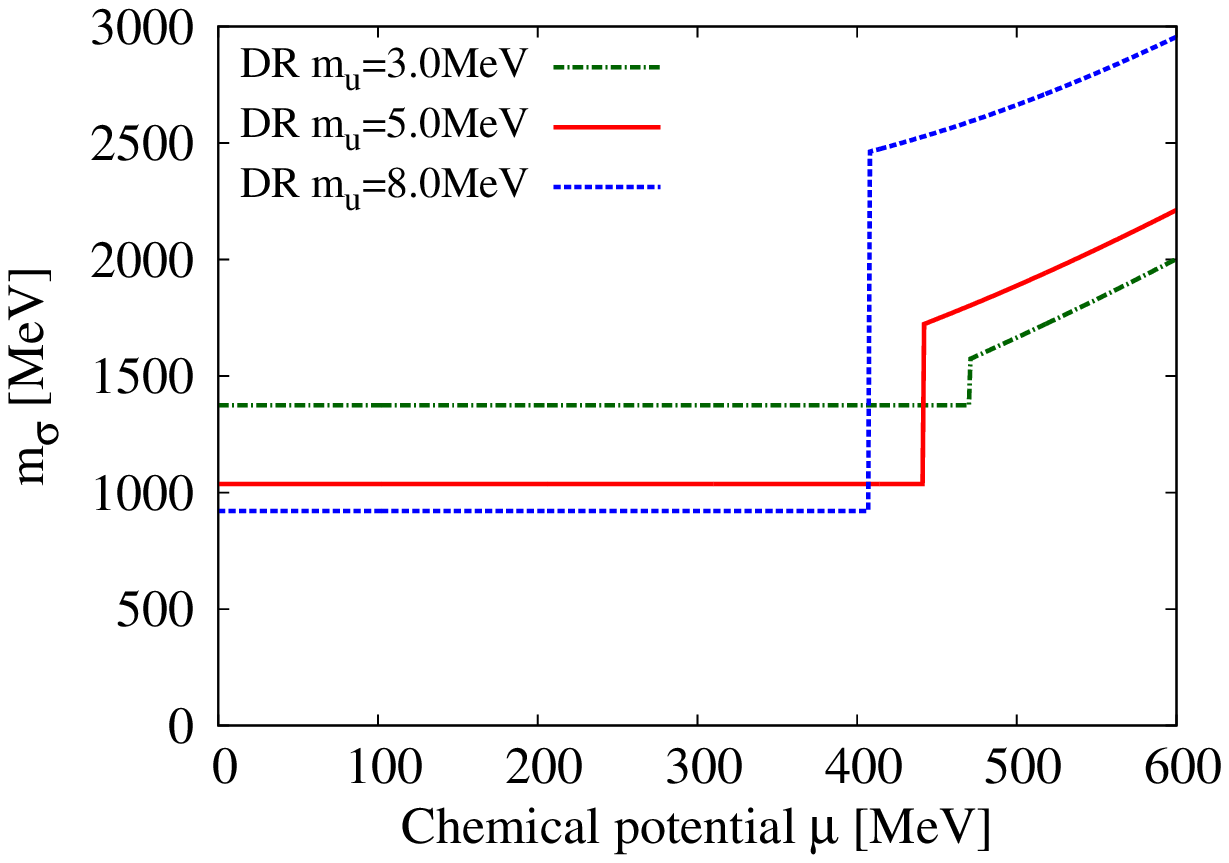} 
  }  
  %\vspace{0.7cm}
  \caption{Sigma mass. Left: $\mu=0$. Right: $T=0$.}
  \end{center}
%\vspace{-0.5cm}
\label{fig_sigma}
\end{figure}
% --- figure --- %
%%%%%%%%%%%%%%%%%%%%%%%%%%%%
Having fixed the parameters with the pion mass and decay constant.
We will calculate the sigma meson mass. It is one of
the predictions of the model. 
%Figure \ref{fig_sigma}
%%%%%%
Figure 4
%%%%%%
shows the numerical results of the sigma mass. 
At $T=0$ and $\mu=0$ we find the band around
$400-900$MeV in $3$D, $4$D, PV,
PT regularizations, $900-1400$MeV in the DR.  Then, in the DR case,
the predicted values are larger than the experimental value
$m_{\sigma} \simeq 500$MeV~\cite{Agashe:2014kda}, in the leading order
of the $1/N_c$ expansion.
As is known we can obtain much smaller sigma meson width in this order.
We should also check the next to leading order contributions for the sigma
meson mass and width.
The features of the curves are the similar to that of the pion; the mass
decreases with increasing $T$ and $\mu$ until some values, then
it increases after exceeding the certain values.  As seen in the pion
mass case, the solution of the sigma mass on the real axis disappears
for some parameter set at high temperature.

%%%%%%%%%%%%%%%%%%%%%%%%%%%%
\subsection{Results with regularizing all contribution}
%%%%%%%%%%%%%%%%%%%%%%%%%%%%
In this subsection, we study the meson properties by applying 
each regularization procedure to both temperature dependent and independent
contributions. It is worth mentioning that the case with $4$D cutoff
method does not give credible results because enough number of
frequency summations is not taken in this method.
The cutoff scale of the $4$D case is around $1$GeV, which means that
at $T=100$MeV only four terms in the Matsubara mode summation,
$n=-2,\pm 1,0$ ($\omega_n=(2n+1)\pi T$), are taken into account.
It is known that finite temperature field theory does not lead
reliable predictions when the number of the frequency summation is
small~\cite{Chen:2010sy}. Therefore, we will not show the results in the
case of $4$D cutoff scheme here, and consider the other four cases $3$D,
PV, PT and DR and call these cases $3$DRT, PVRT, PTRT and DRRT,
respectively. It is also worth mentioning that the calculations technically
become impossible at $T=0$ in the PTRT case as can be read from
Eqs. (\ref{eq_trSpt}) and (\ref{eq_Ipt}). Then, we will show the results
with $T=10{\rm MeV}$ as the representative values on $\mu$ dependence
at low $T$.

% Results for the Pion Mass
%%%%%%%%%%%%%%%%%%%%%%%%%%%%
%\input fig_pion_rt.tex
% --- figure --- %
\begin{figure}[h!]
 \begin{center}
 %\vspace{-0.5cm}
 %\hspace{1.0cm}
  \subfigure{
    \includegraphics[height=4.6cm,keepaspectratio]{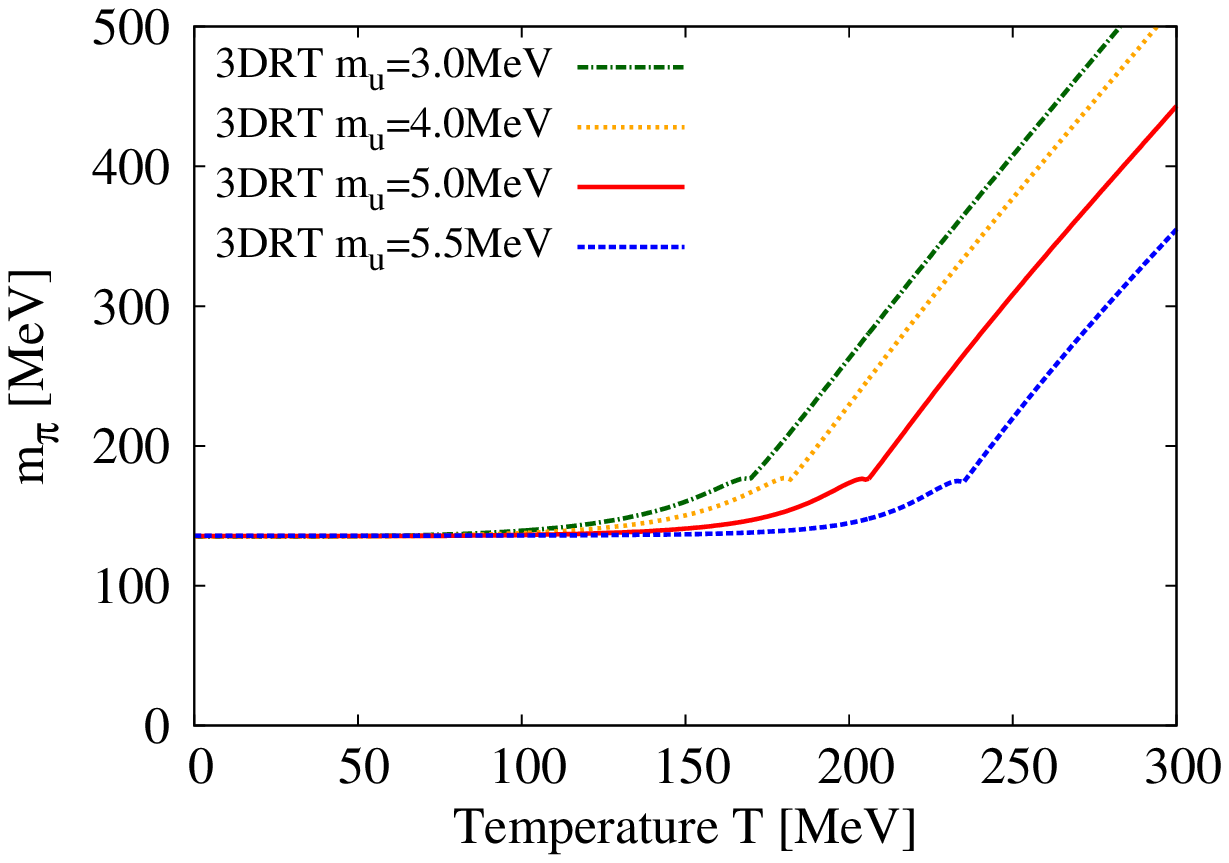} 
  }
  %%\hspace{1.0cm}
  \subfigure{
    \includegraphics[height=4.6cm,keepaspectratio]{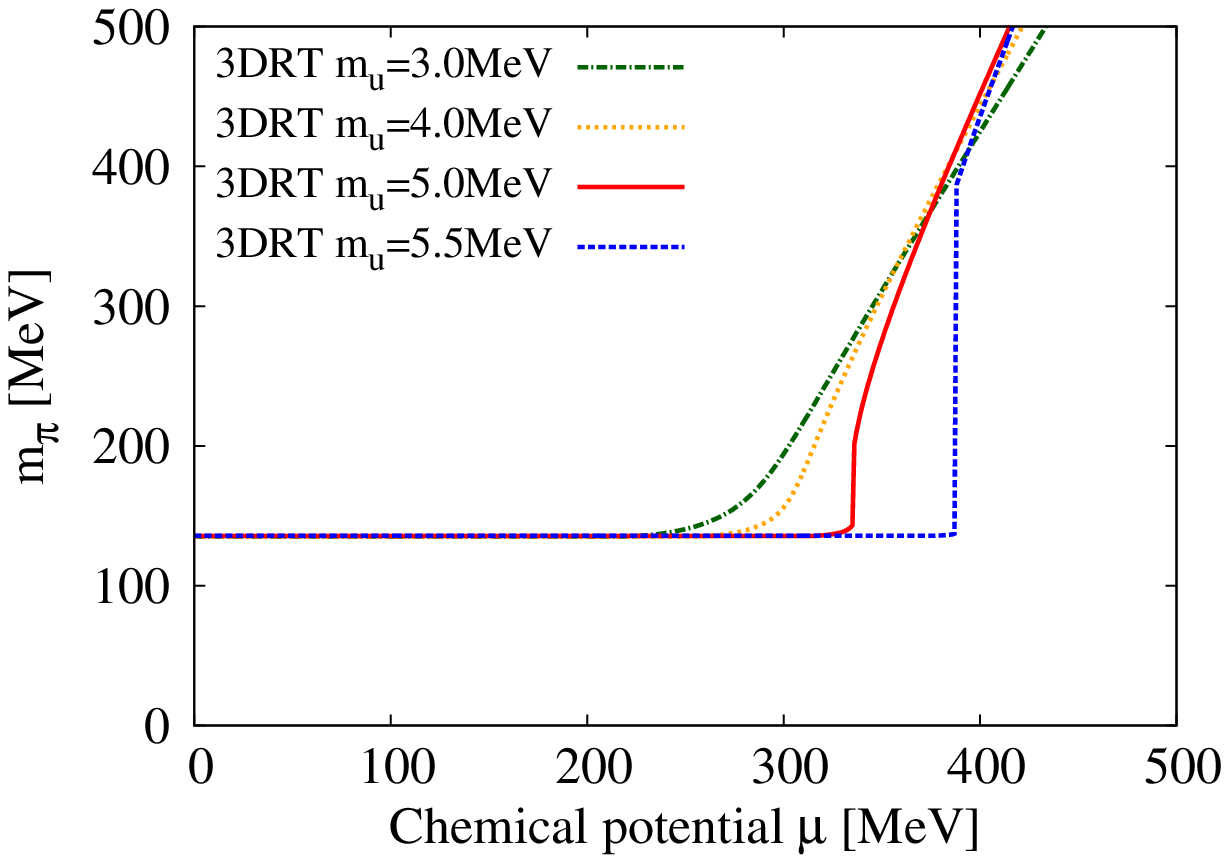} 
  }
  
% \hspace{1.0cm}
  \subfigure{
    \includegraphics[height=4.6cm,keepaspectratio]{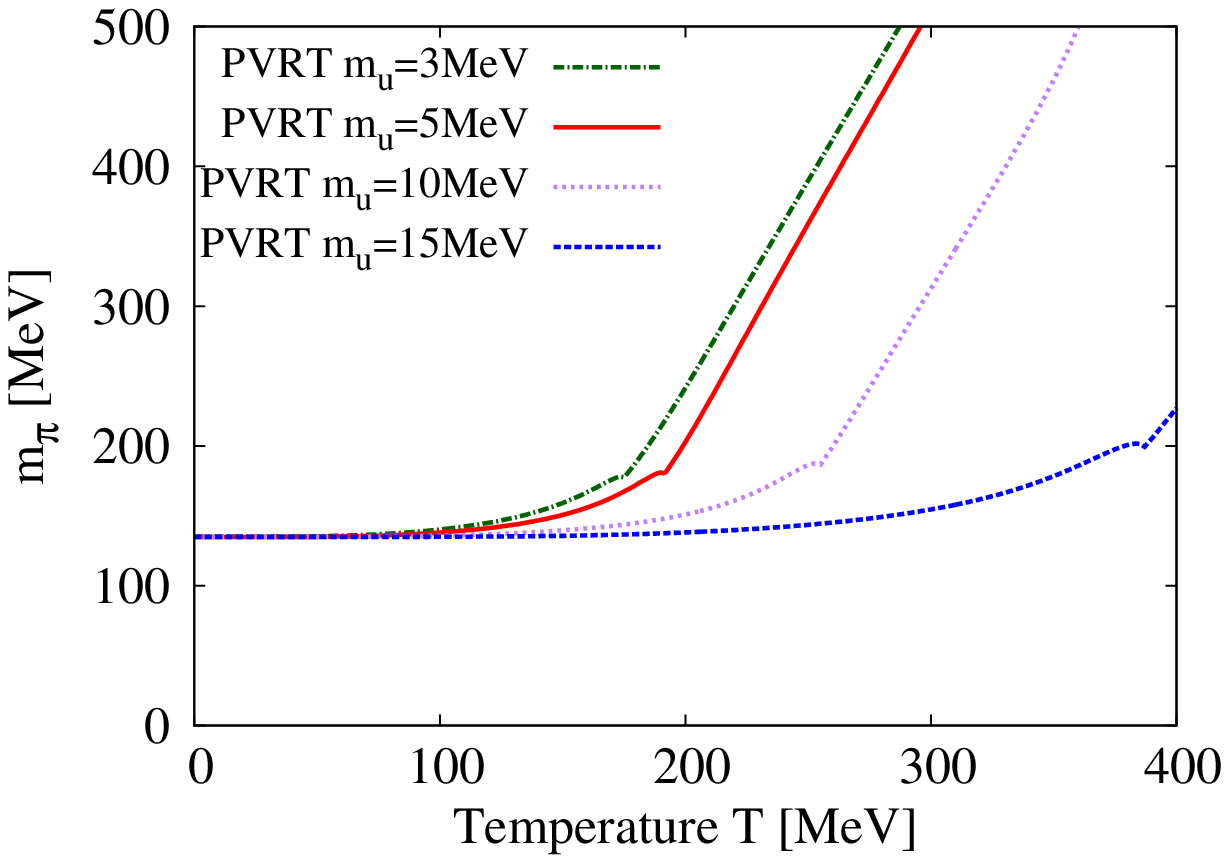} 
  }
  %%\hspace{1.0cm}
  \subfigure{
    \includegraphics[height=4.6cm,keepaspectratio]{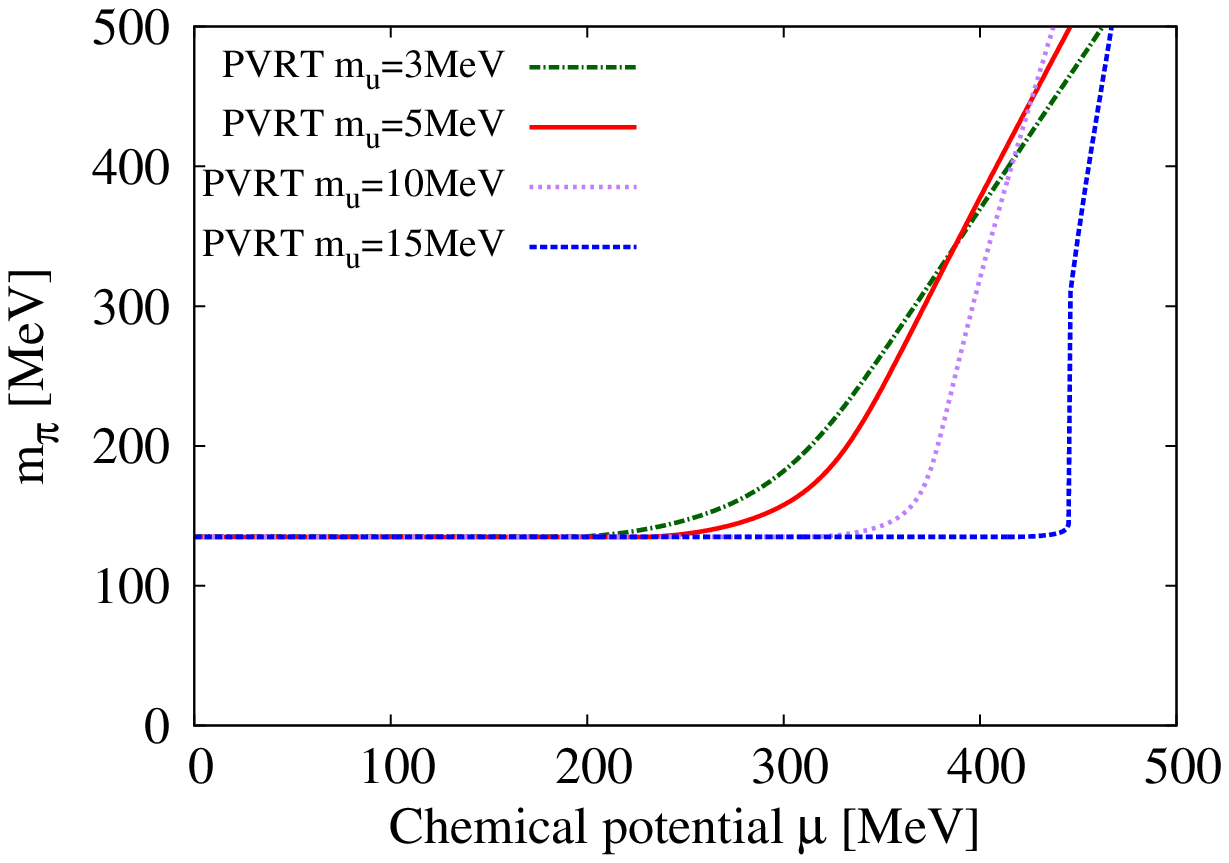} 
  }
  
  %\hspace{1.0cm}
  \subfigure{
    \includegraphics[height=4.6cm,keepaspectratio]{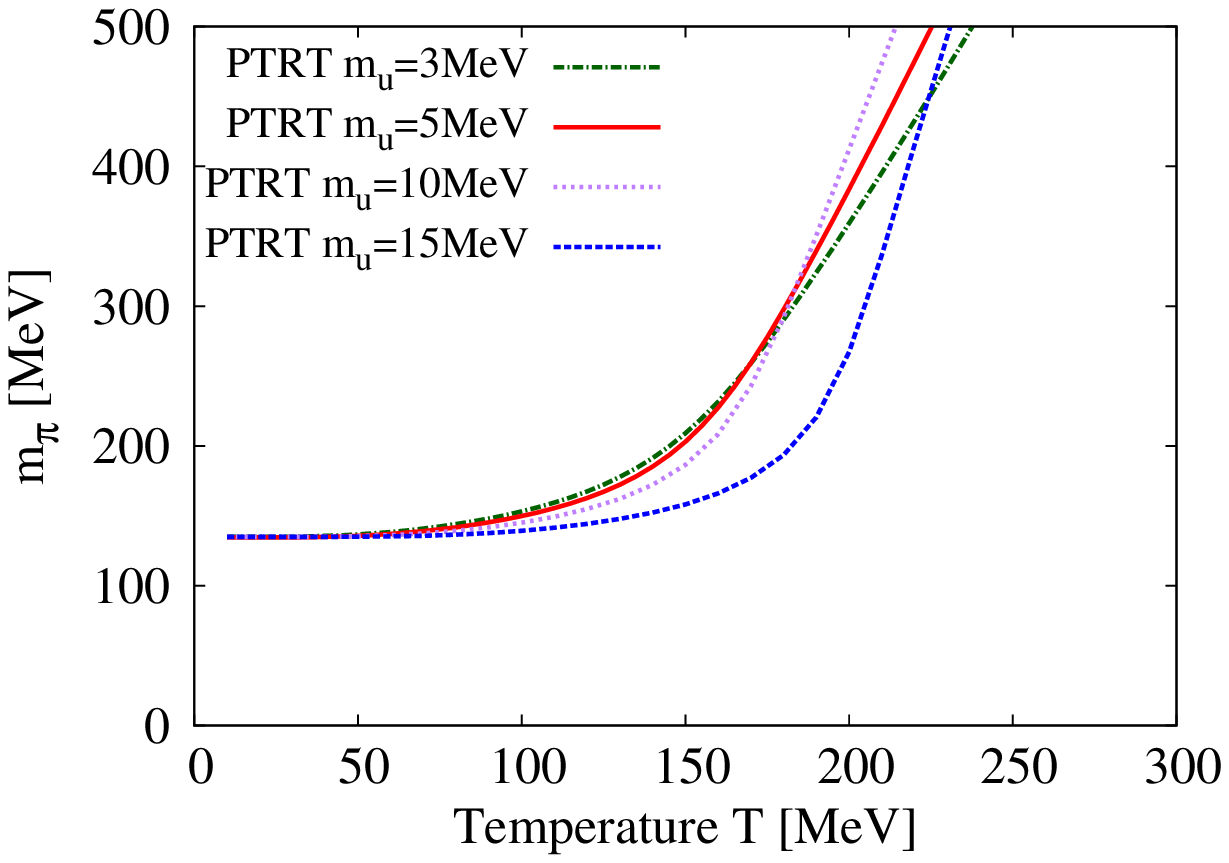} 
  }
  %%\hspace{1.0cm}
  \subfigure{
    \includegraphics[height=4.6cm,keepaspectratio]{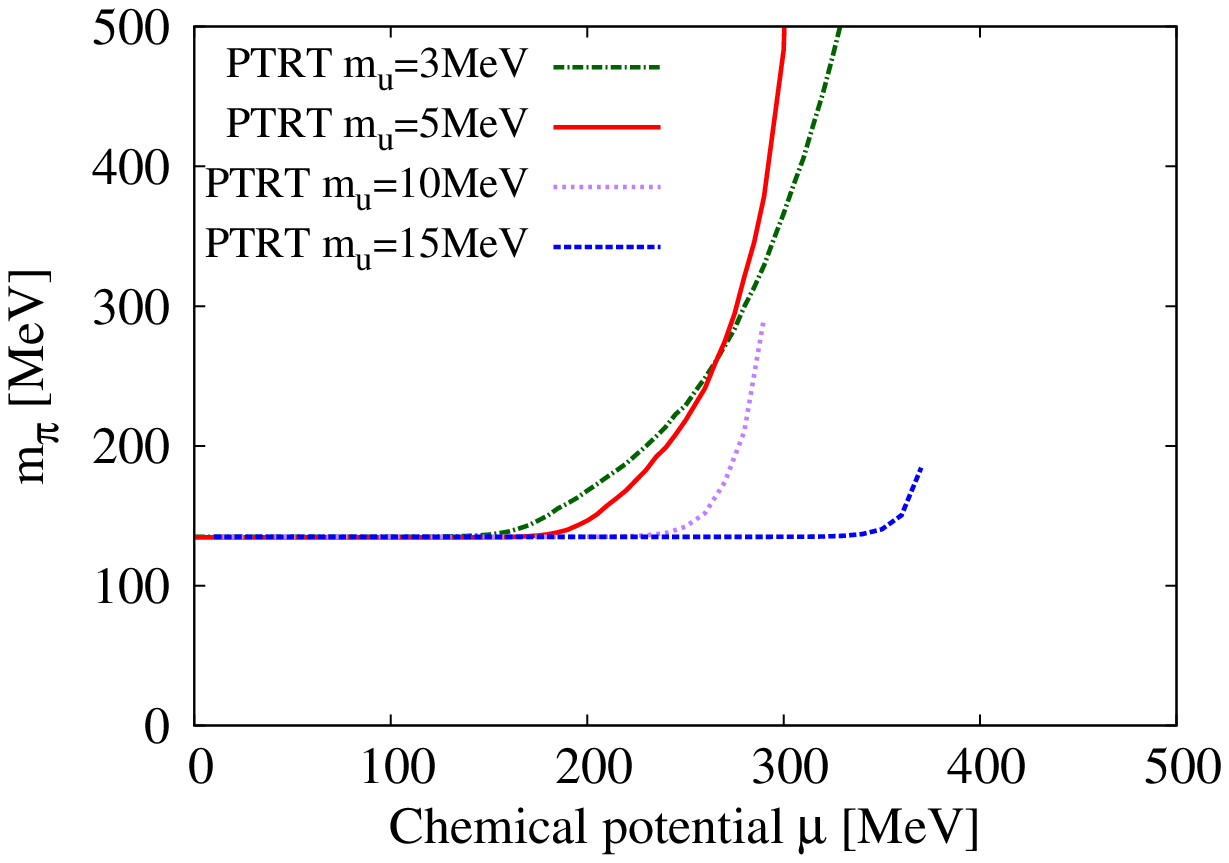} 
  }
  
  %\hspace{1.0cm}
  \subfigure{
    \includegraphics[height=4.6cm,keepaspectratio]{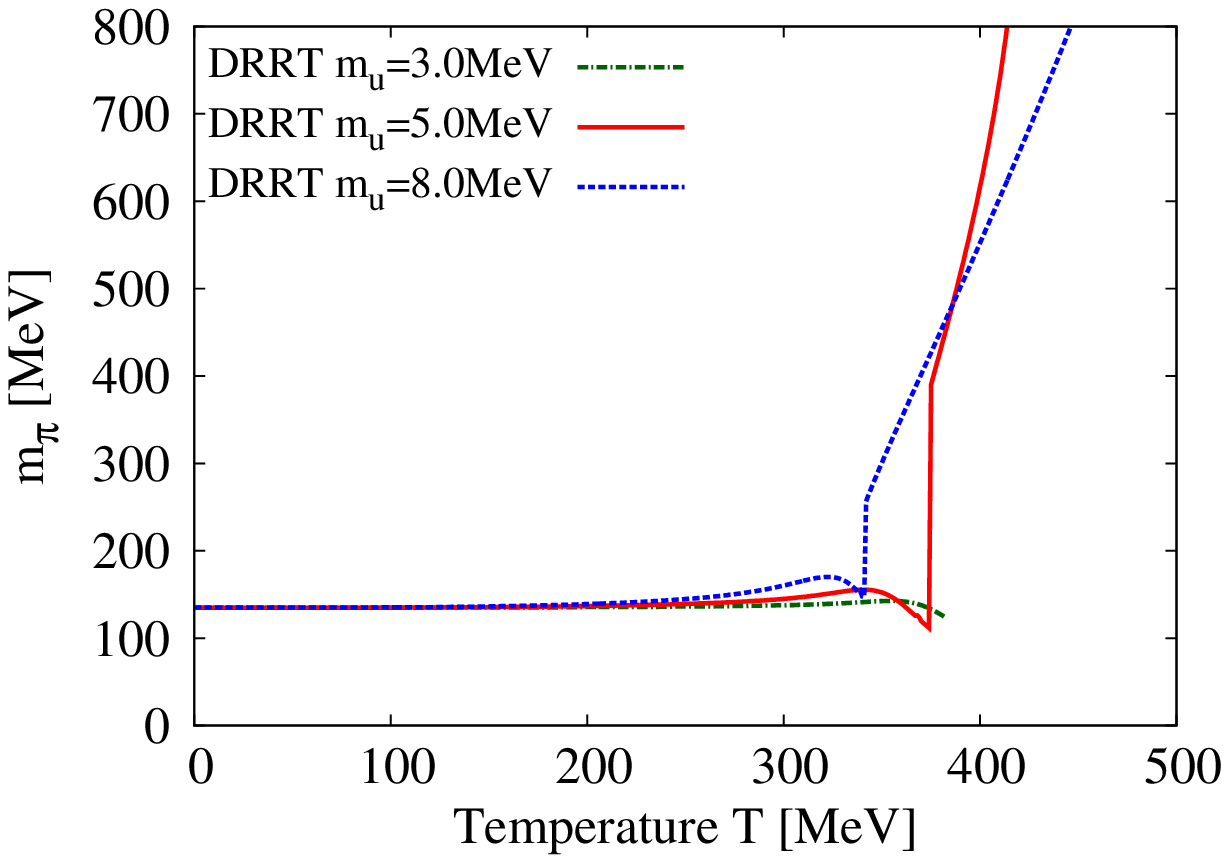} 
  }
  %%\hspace{1.0cm}
  \subfigure{
    \includegraphics[height=4.6cm,keepaspectratio]{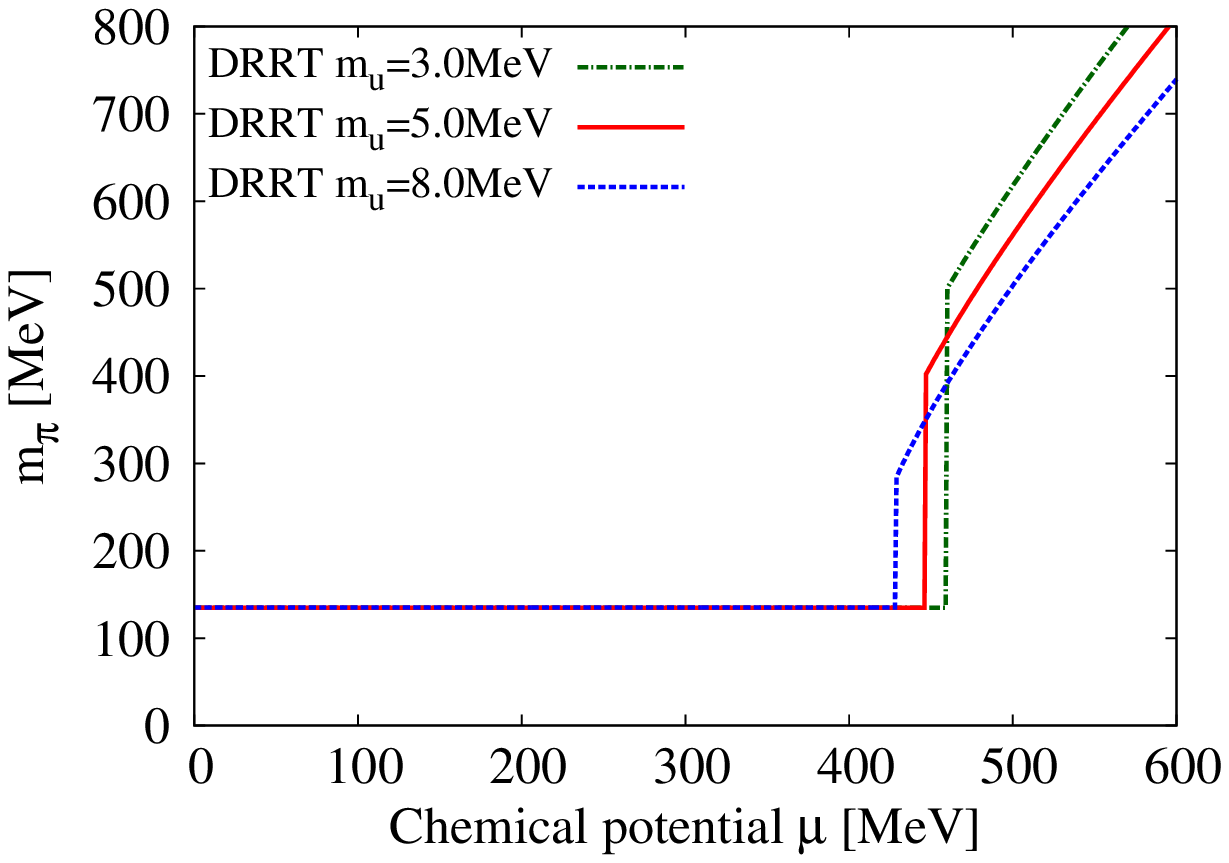} 
  }  
  %\vspace{0.7cm}
  \caption{Pion mass. Left: $\mu=0$.
                Right: $T=0$ for 3DRT, PVRT, DRRT and $T=10$MeV for PTRT.
                }
  \end{center}
%\vspace{-0.5cm}
\label{fig_pion_rt}
\end{figure}
% --- figure --- %
%%%%%%%%%%%%%%%%%%%%%%%%%%%%
%Figure \ref{fig_pion_rt} 
%%%%%%
Figure 5
%%%%%%
displays the results of the pion mass. One sees
that the qualitative feature does not change comparing to the previous
case with regularizing only the temperature independent contributions.
Quantitatively, we note that the changes become smoother at high $T$
and $\mu$. This can easily be understood because the regularization
procedure suppresses the thermal contribution, so the finite temperature
term reduces to give smoother curve with respect to $T$ and $\mu$.

% Results for the Pion Decay
%%%%%%%%%%%%%%%%%%%%%%%%%%%%
%\input fig_decay_rt.tex
% --- figure --- %
\begin{figure}[h!]
 \begin{center}
 %\vspace{-0.5cm}
 %\hspace{1.0cm}
  \subfigure{
    \includegraphics[height=4.6cm,keepaspectratio]{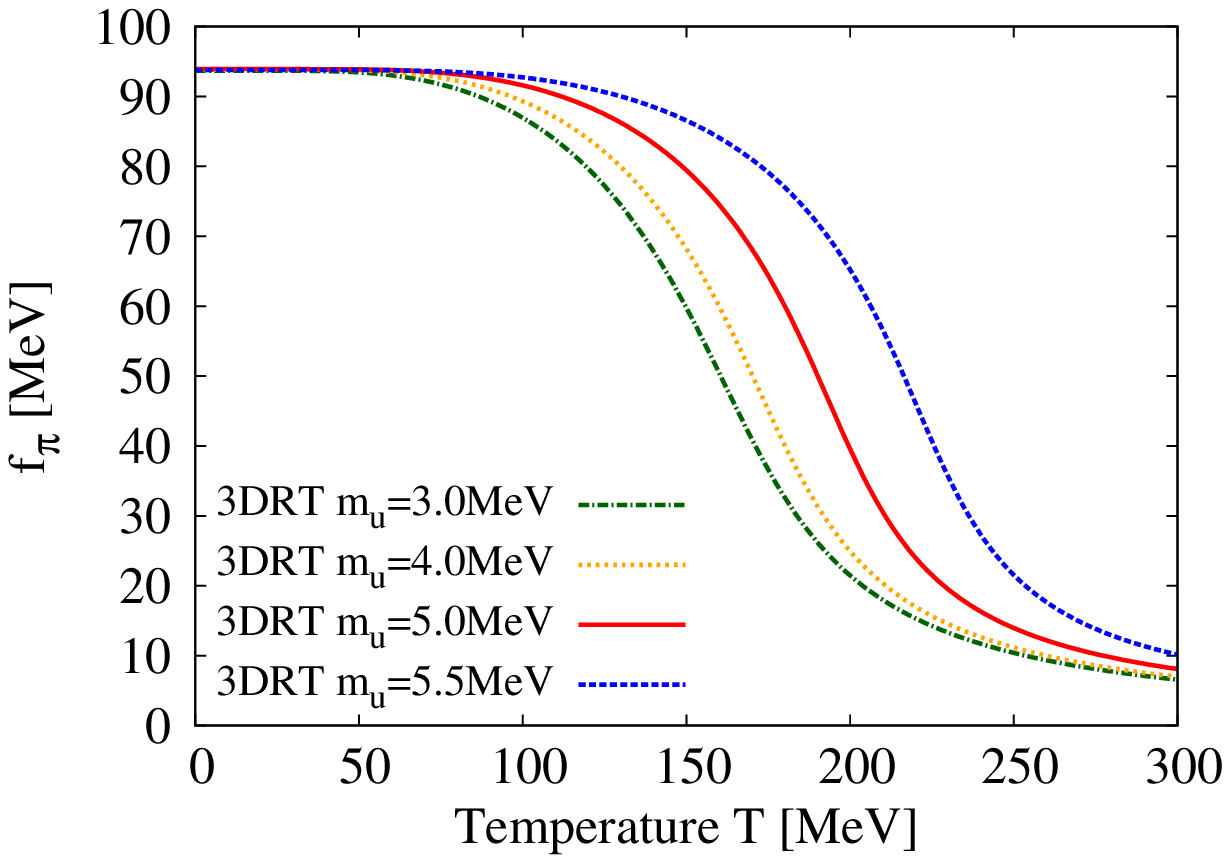} 
  }
  %%\hspace{1.0cm}
  \subfigure{
    \includegraphics[height=4.6cm,keepaspectratio]{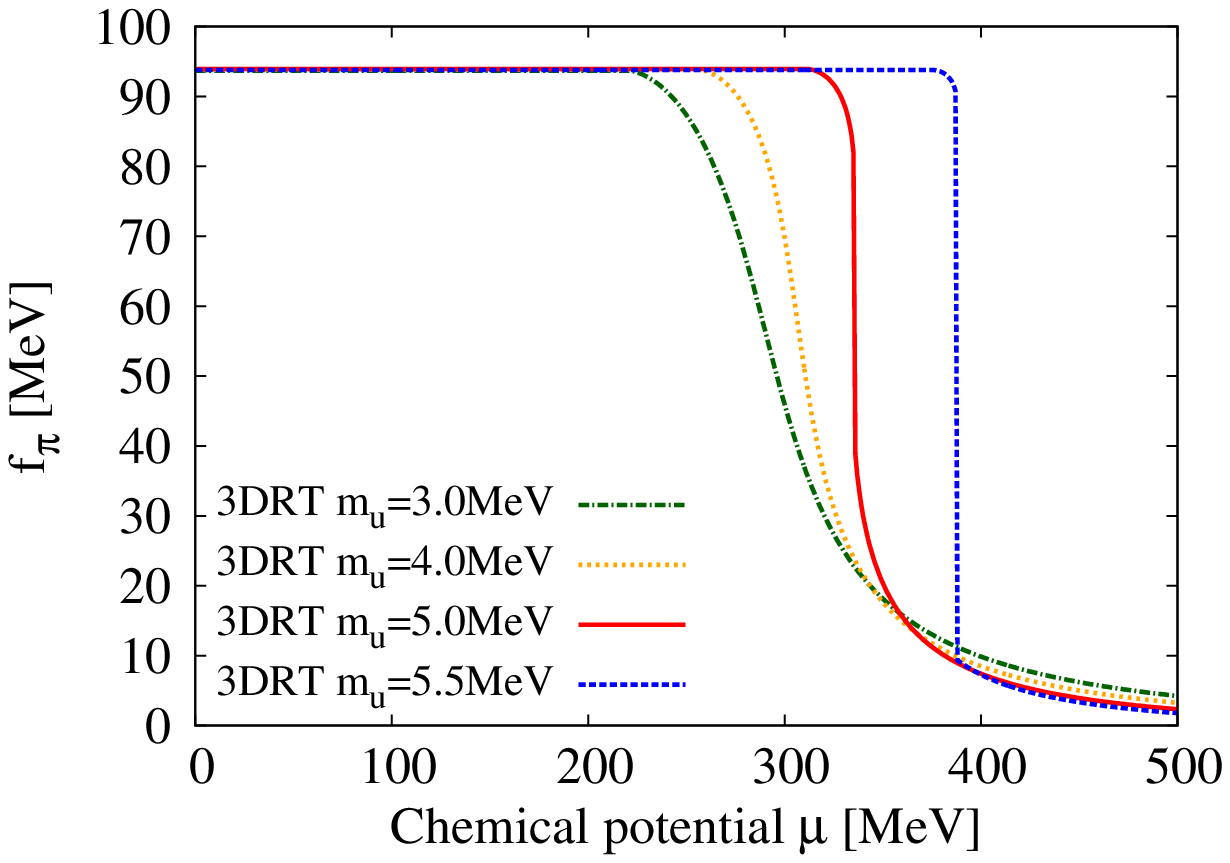} 
  }
   
 %\hspace{1.0cm}
  \subfigure{
    \includegraphics[height=4.6cm,keepaspectratio]{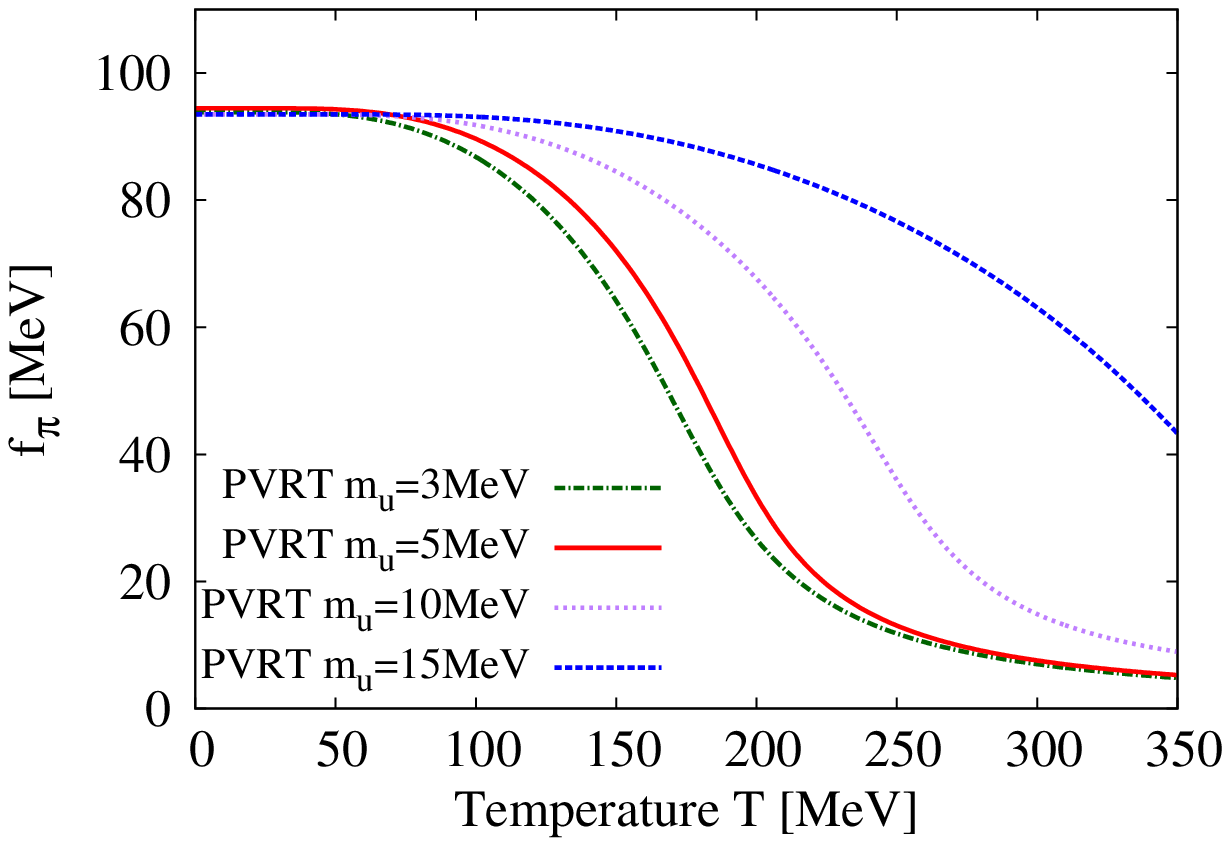} 
  }
  %%\hspace{1.0cm}
  \subfigure{
    \includegraphics[height=4.6cm,keepaspectratio]{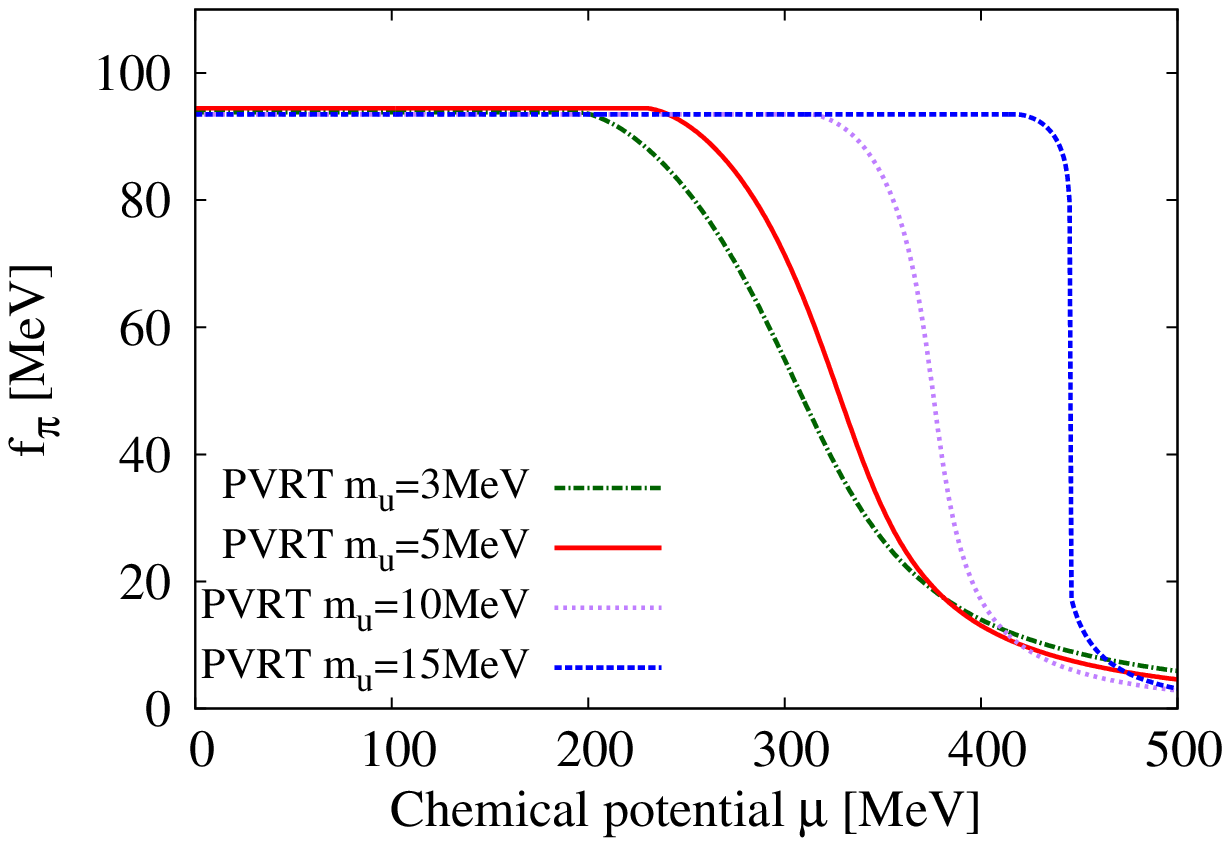} 
  }
  
  %\hspace{1.0cm}
  \subfigure{
    \includegraphics[height=4.6cm,keepaspectratio]{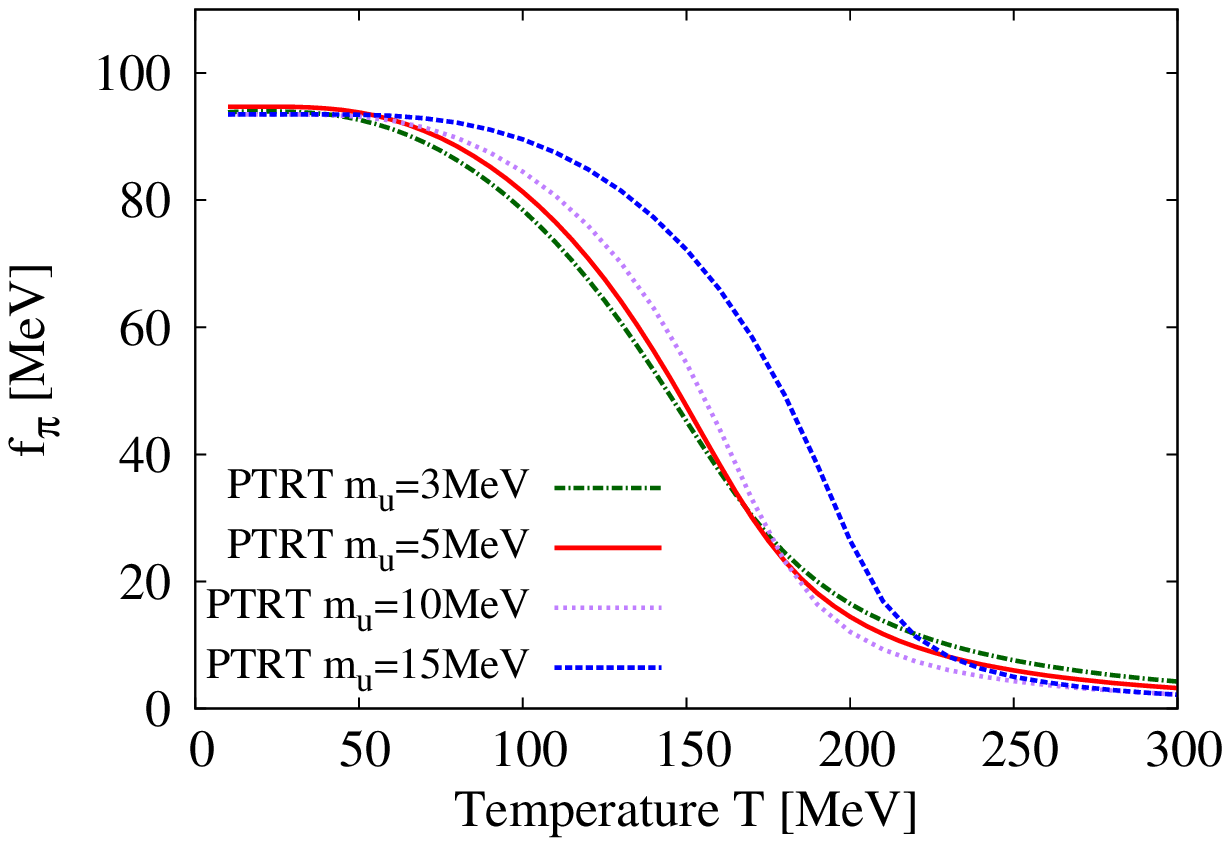} 
  }
  %%\hspace{1.0cm}
  \subfigure{
    \includegraphics[height=4.6cm,keepaspectratio]{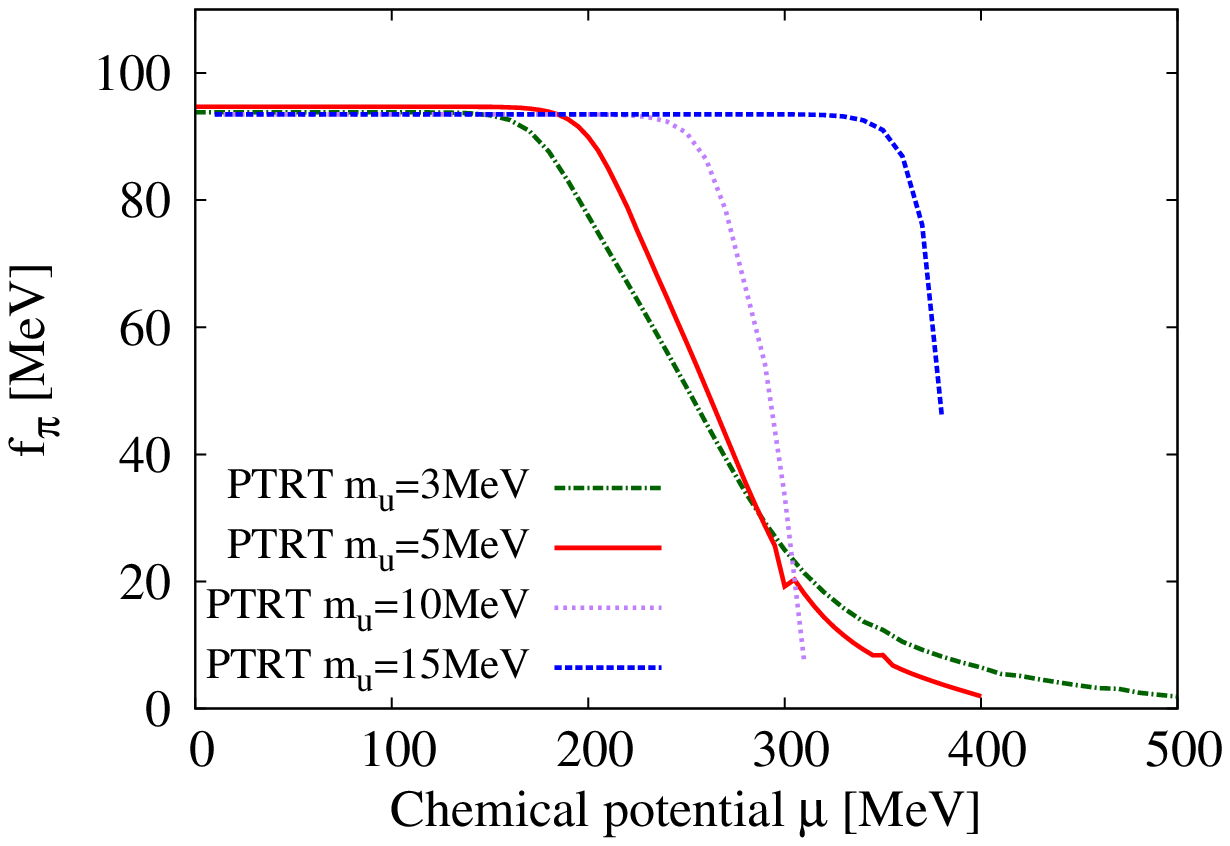} 
  }
  
  %\hspace{1.0cm}
  \subfigure{
    \includegraphics[height=4.6cm,keepaspectratio]{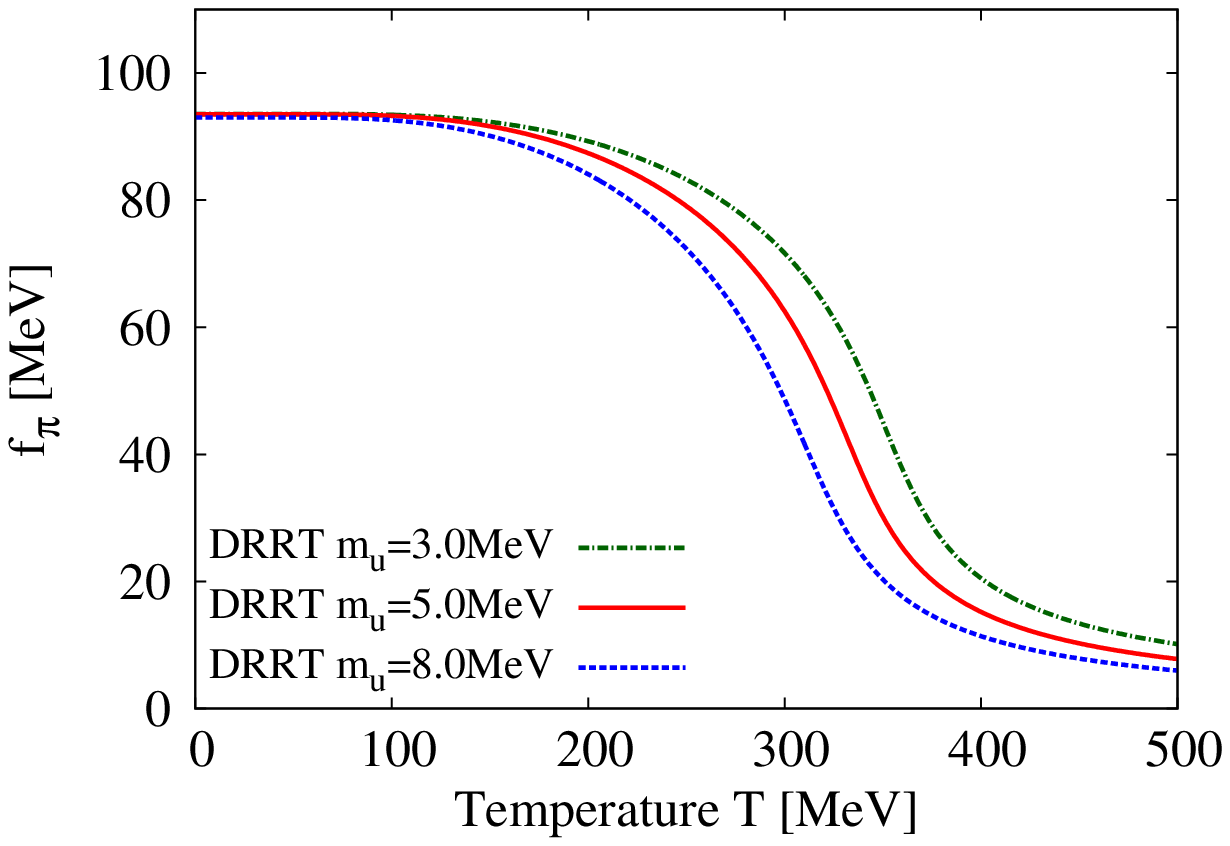} 
  }
  %%\hspace{1.0cm}
  \subfigure{
    \includegraphics[height=4.6cm,keepaspectratio]{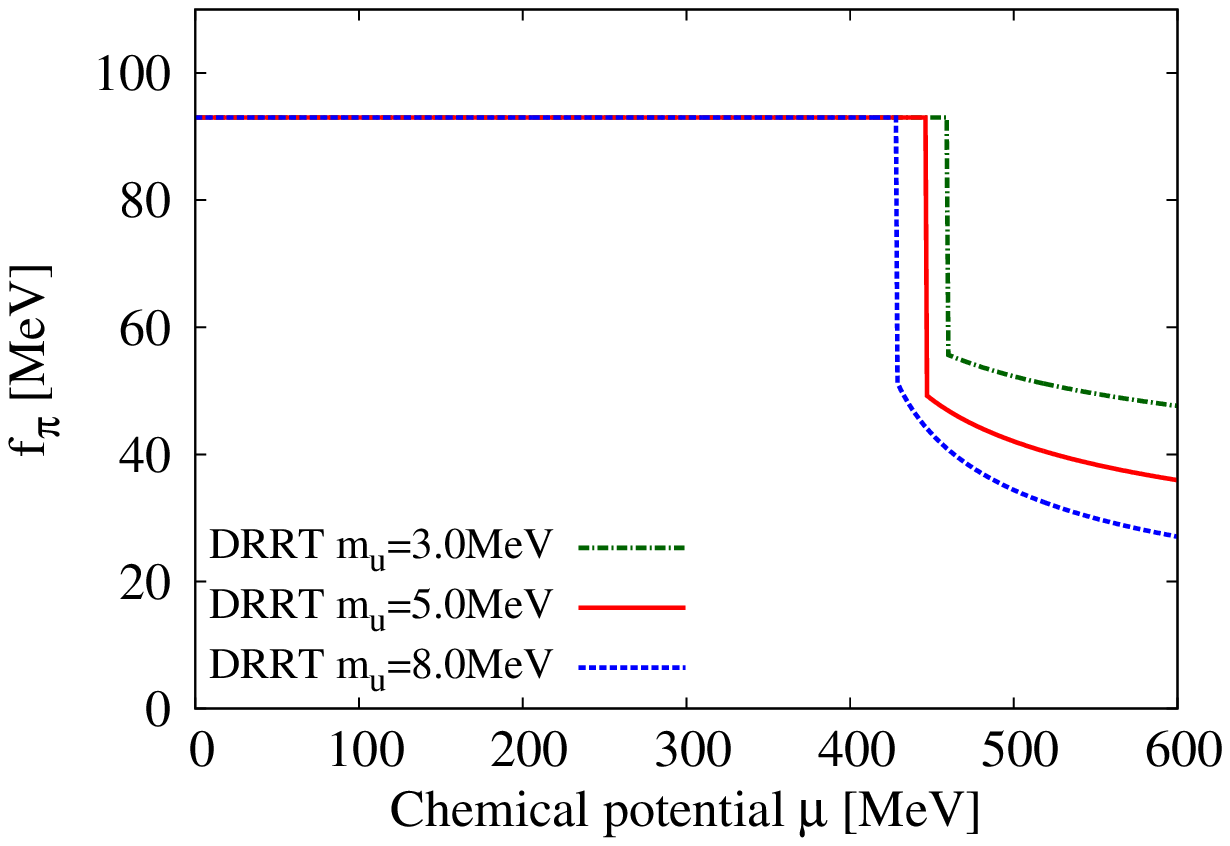} 
  }  
  %\vspace{0.7cm}
  \caption{Pion decay constant. Left: $\mu=0$.
                Right: $T=0$ for 3DRT, PVRT, DRRT and $T=10$MeV for PTRT.
                }
  \end{center}
%\vspace{-0.5cm}
\label{fig_decay_rt}
\end{figure}
% --- figure --- %
%%%%%%%%%%%%%%%%%%%%%%%%%%%%
We aligned the results of the pion decay constant in 
%Fig. \ref{fig_decay_rt}.
%%%%%%
Fig. 6.
%%%%%%
As seen in the pion mass case, the numerical results do not alter
qualitatively, the curves become smooth. Note that, although the $T$
dependence becomes considerably smother, the transition chemical
potential does not change. This is due to the fact that finite temperature
contributions become proportional to the step function $\theta(\mu-m^*)$
for $T=0$, then the transition chemical potential is not affected by the
regularization procedure in this model treatment.

% Results for the Sigma Mass 
%%%%%%%%%%%%%%%%%%%%%%%%%%%%
%\input fig_sigma_rt.tex
% --- figure --- %
\begin{figure}[h!]
 \begin{center}
 %\vspace{-0.5cm}
 %\hspace{1.0cm}
  \subfigure{
    \includegraphics[height=4.6cm,keepaspectratio]{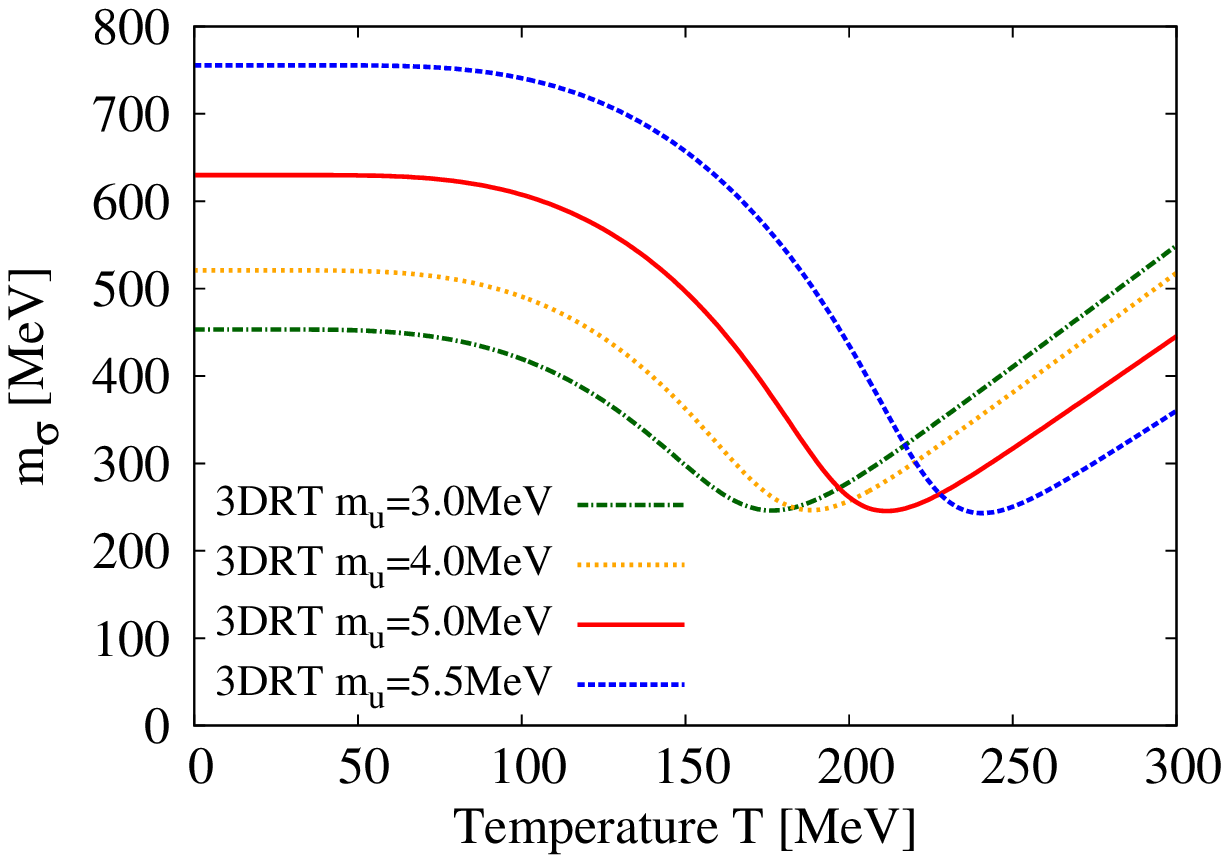} 
  }
  %%\hspace{1.0cm}
  \subfigure{
    \includegraphics[height=4.6cm,keepaspectratio]{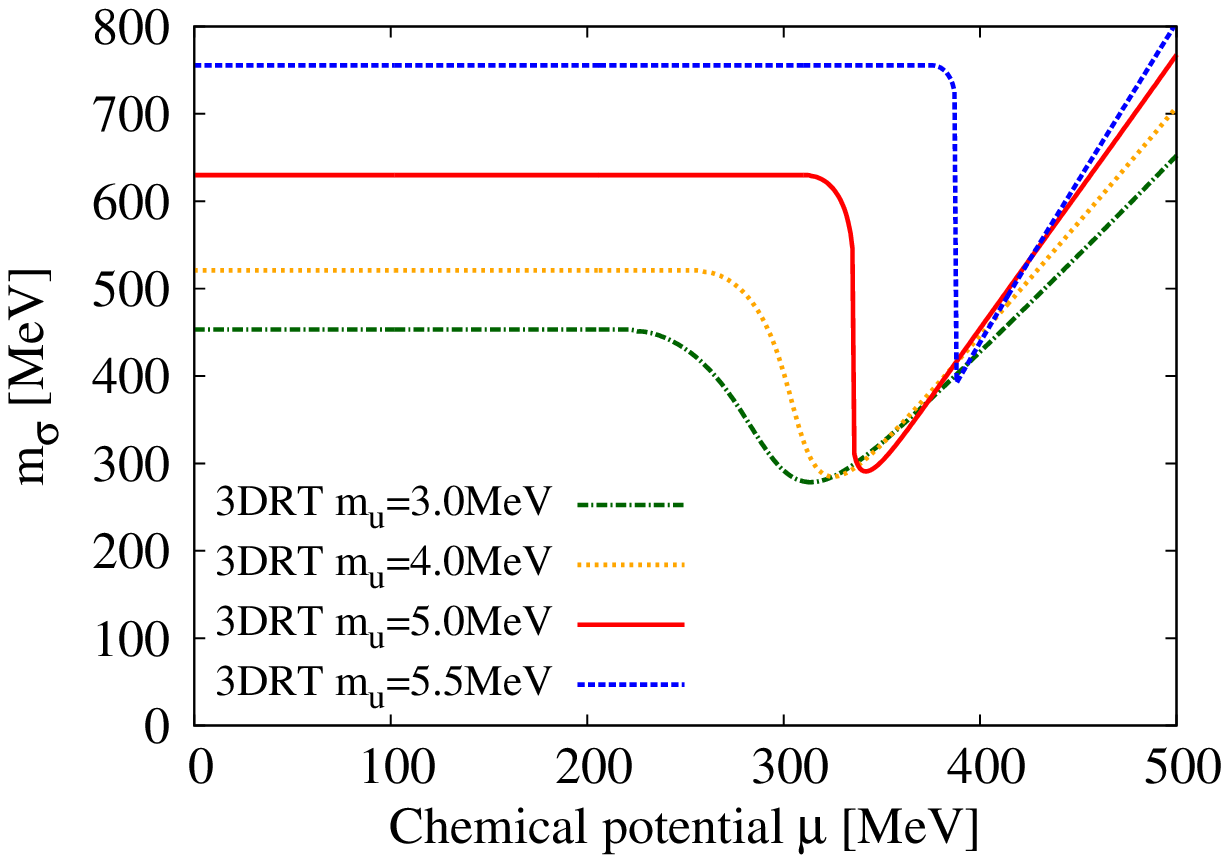} 
  }
   
 %\hspace{1.0cm}
  \subfigure{
    \includegraphics[height=4.6cm,keepaspectratio]{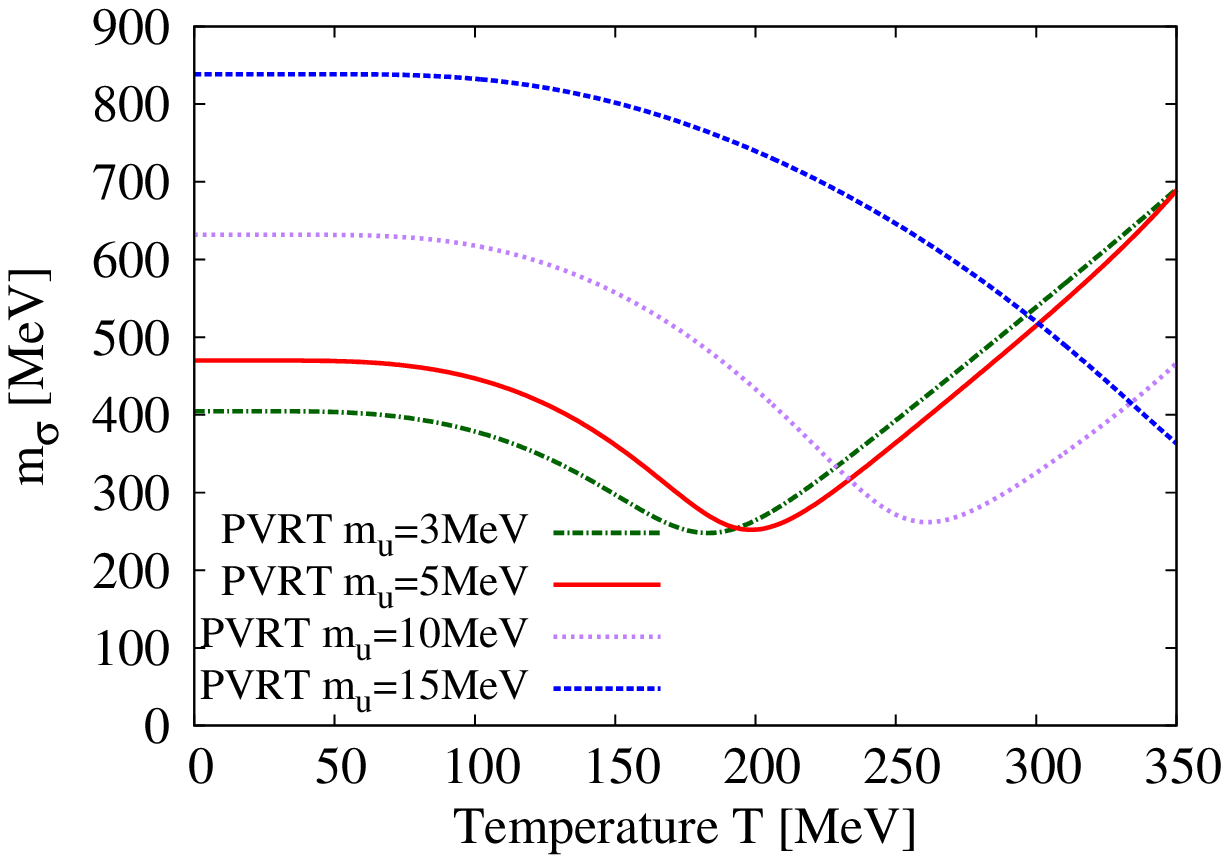} 
  }
  %%\hspace{1.0cm}
  \subfigure{
    \includegraphics[height=4.6cm,keepaspectratio]{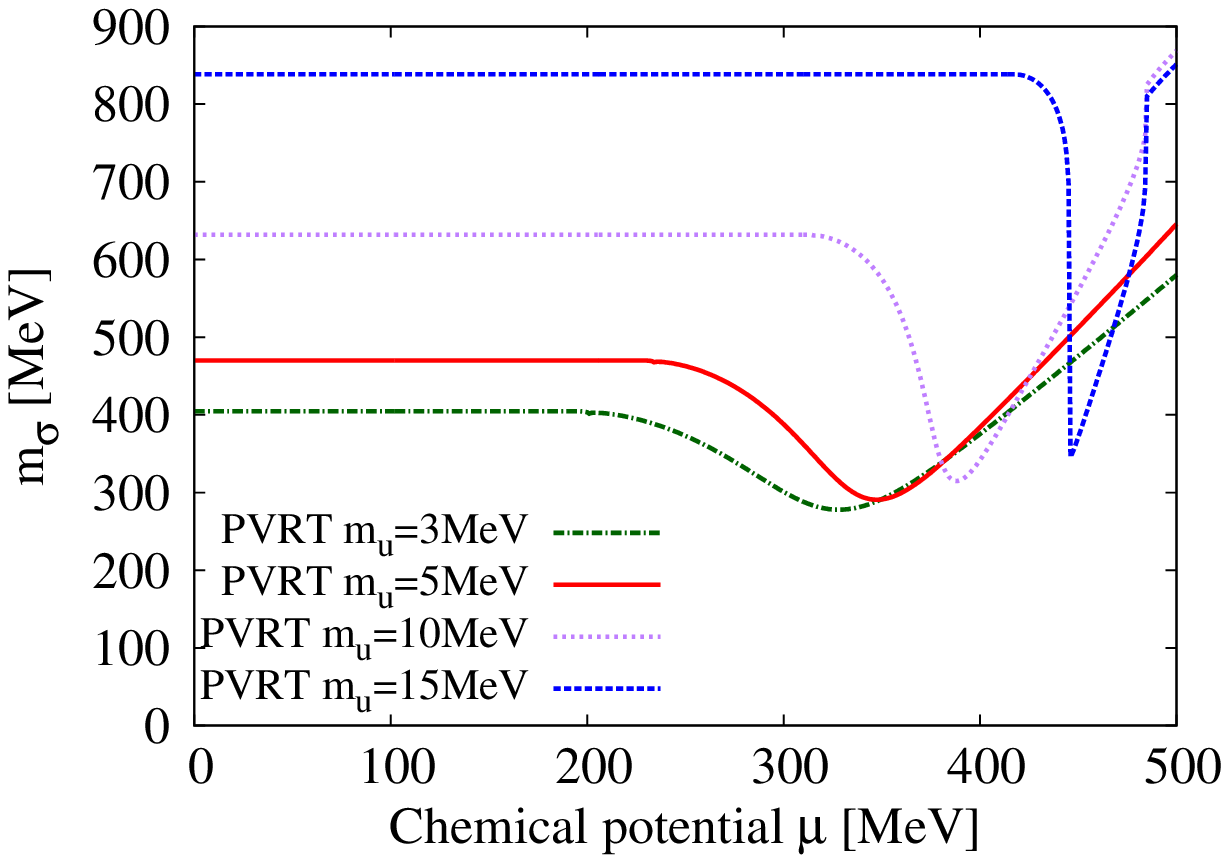} 
  }
  
  %\hspace{1.0cm}
  \subfigure{
    \includegraphics[height=4.6cm,keepaspectratio]{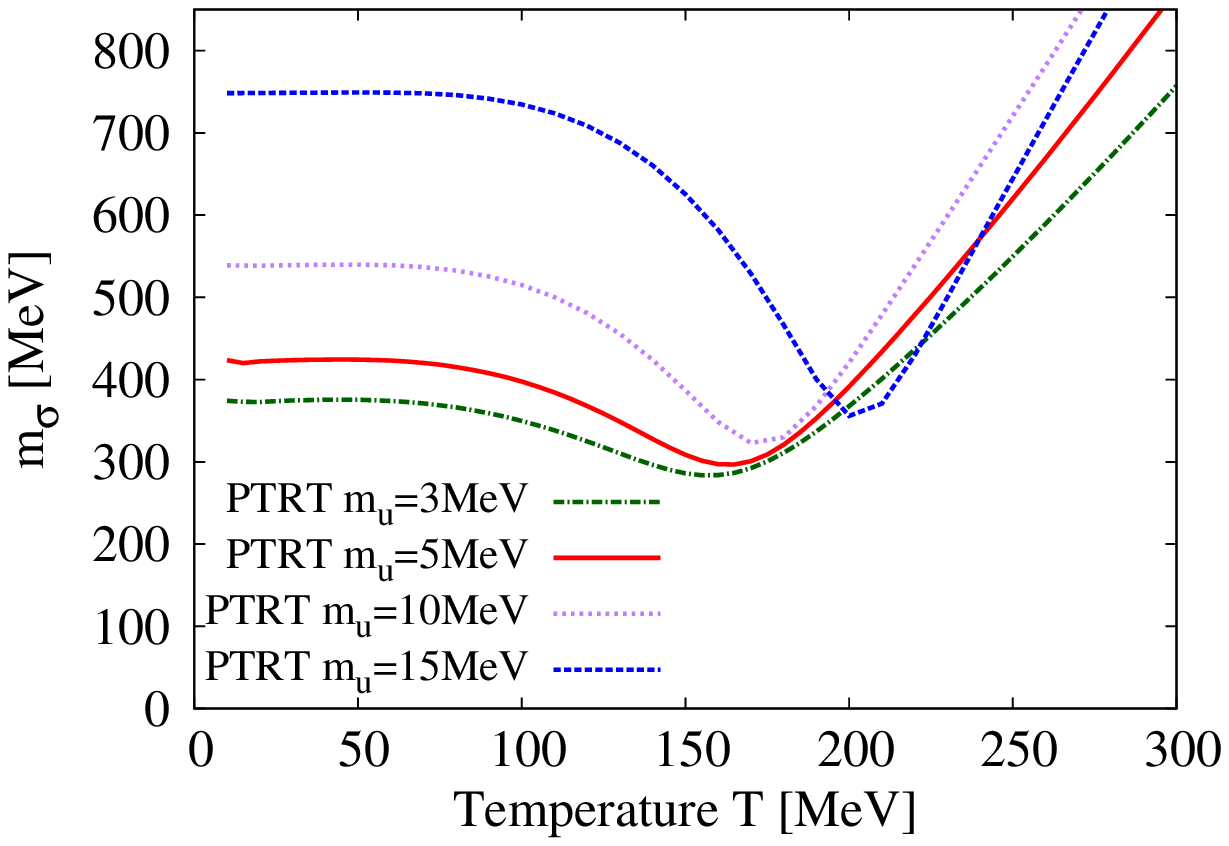} 
  }
  %%\hspace{1.0cm}
  \subfigure{
    \includegraphics[height=4.6cm,keepaspectratio]{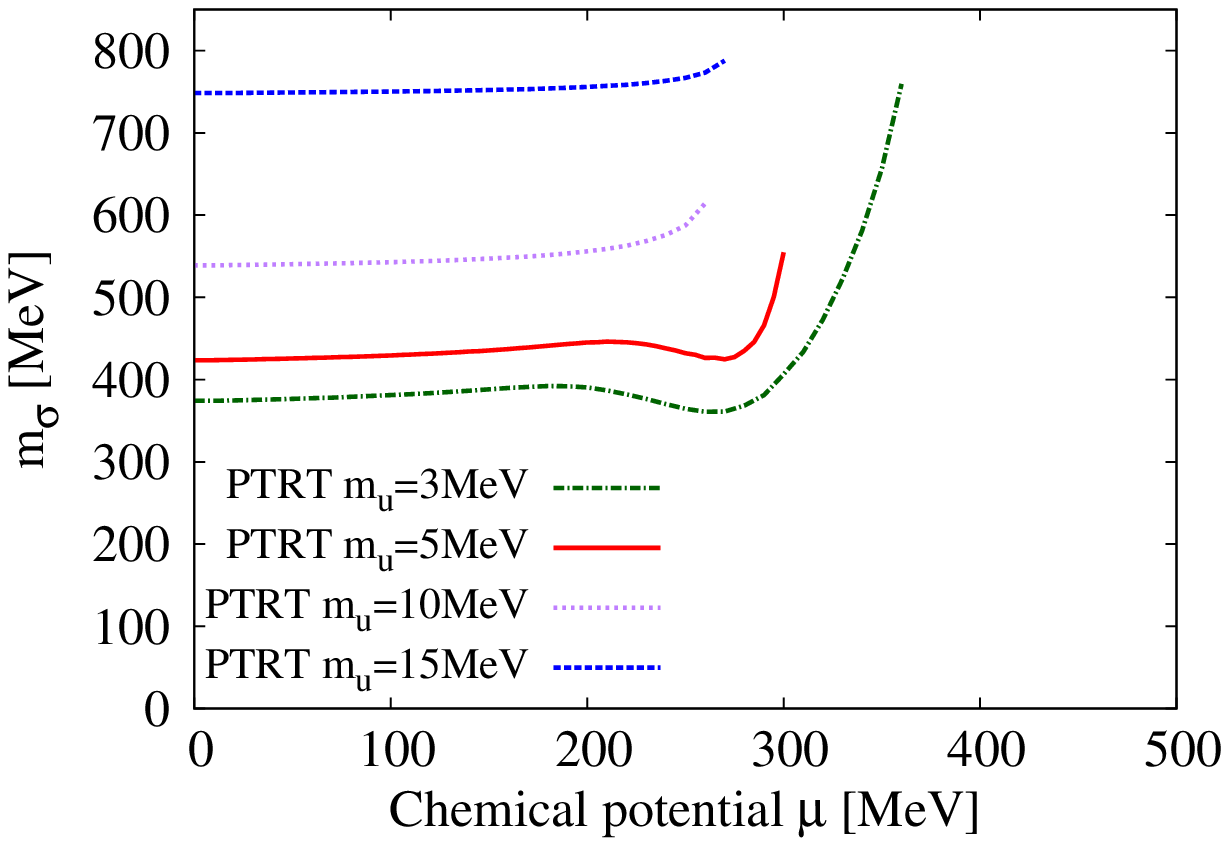} 
  }
  
  %\hspace{1.0cm}
  \subfigure{
    \includegraphics[height=4.6cm,keepaspectratio]{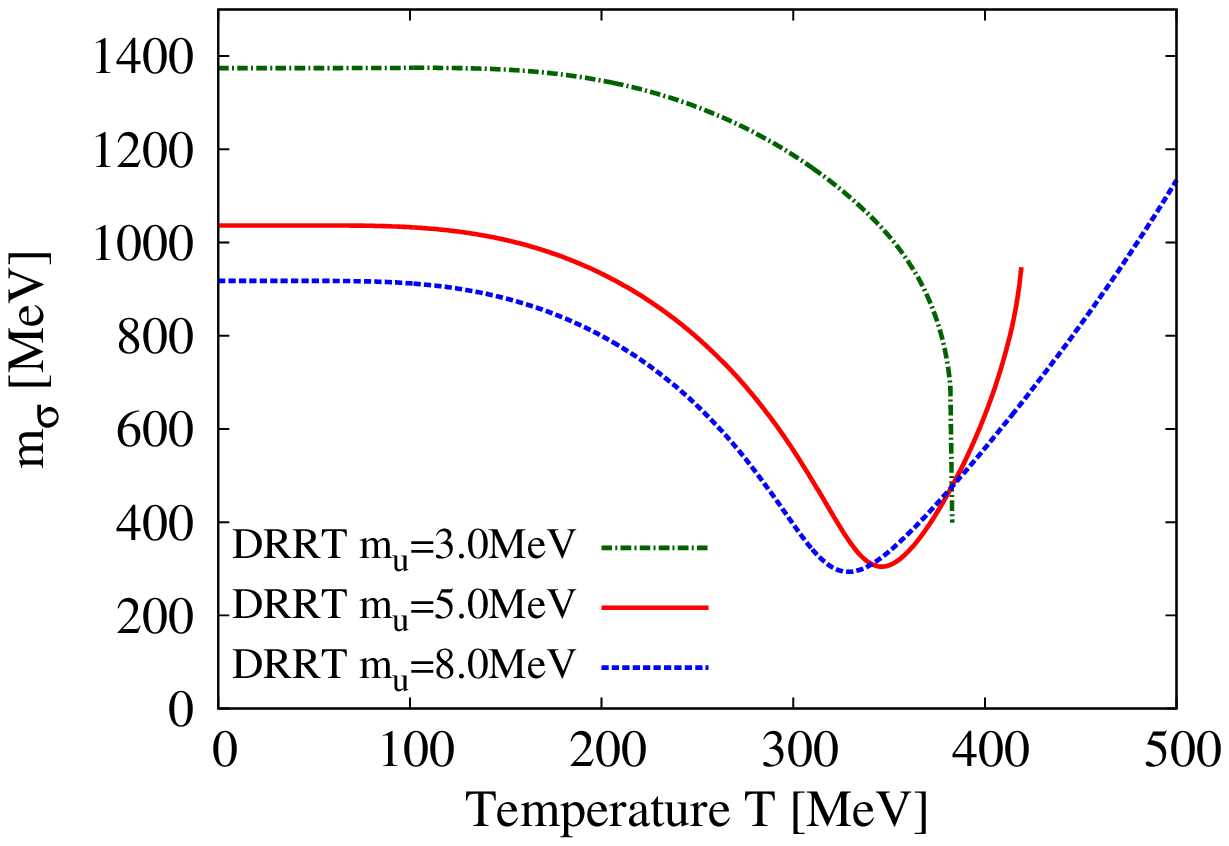} 
  }
  %%\hspace{1.0cm}
  \subfigure{
    \includegraphics[height=4.6cm,keepaspectratio]{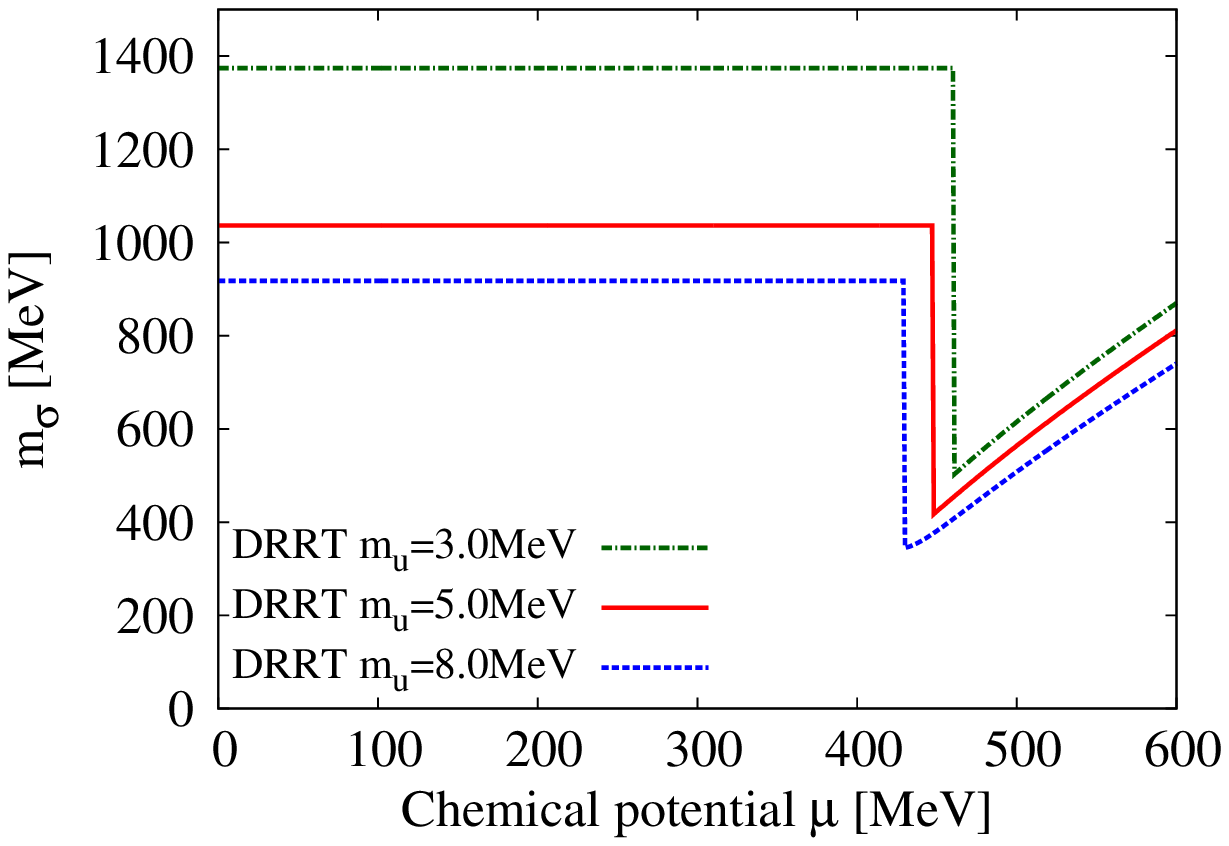} 
  }  
  %\vspace{0.7cm}
  \caption{Sigma mass. Left: $\mu=0$.
                Right: $T=0$ for 3DRT, PVRT, DRRT and $T=10$MeV for PTRT.
                }
  \end{center}
%\vspace{-0.5cm}
\label{fig_sigma_rt}
\end{figure}
% --- figure --- %
%%%%%%%%%%%%%%%%%%%%%%%%%%%%
The predictions on the sigma meson mass are shown in
%Fig. \ref{fig_sigma_rt}. 
%%%%%%
Fig. 7.
%%%%%%
The deviations from the results in the previous subsection in
%Fig. \ref{fig_sigma}
%%%%%%
Fig. 4
%%%%%%
are more or less similar to the deviations on the pion
mass; the curves become flatter with respect to $T$ while $\mu$
dependence does not indicate much difference for the $3$D, PV and PT
cases. However, there appears substantial difference between the cases
of the DR and DRRT where the values of the sigma meson mass at $\mu=500$MeV
are around $1500-2500$MeV for the DR and around $600$MeV for the
DRRT case. This comes from the difference of the integral values between
these two cases, which we will discuss in more detail in
Sec. \ref{sec_discussion}.

We find that the solution on the real axis always disappears for high
$\mu$ and low $T(=10{\rm MeV})$ in the PTRT with $m_u=10$, $15$MeV.
This is because for high $\mu$ some quantities become pure imaginary
number in the calculation due to the complicated counter integral of
$I_{\rm PT}(p^2)$ as seen in Eq. (\ref{eq_Ipt}). Then one can not find
a real solution in that case. This is the numerical reason why the meson
properties behave badly for high $\mu$ in PTRT case.

%%%%%%%%%%%%%%%%%%%%%%%%%%%%
\section{Phase diagram}
\label{sec_pd}
%%%%%%%%%%%%%%%%%%%%%%%%%%%%
We shall draw the phase diagram in this section. We search the phase
transition point by the condition that the maximum change of the chiral 
condensate with respect to $T$ and $\mu$. In more concrete, we
numerically differentiate the condensate with respect to
$r \equiv \sqrt{T^2 +\mu^2}$, then the condition can be written
%%%%%%
\begin{equation}
%  {\rm Maximum \,\,\, change \,\,\, of \,\,\,} \frac{d \phi}{d r}.
\left.\frac{d \langle \bar{u}u \rangle^{1/3}}{dr}\right|_{T=T_c, \mu=\mu_c}
= {\rm Max}\left(\frac{d \langle \bar{u}u \rangle^{1/3}}{dr}\right) .
\label{phasetransition}
\end{equation} 
%%%%%%
We should be careful on the case with the first order transition,
because the condensate has the discontinuous point where we
need to find the minimum of the thermodynamic potential then
determine the chiral condensate. No matter whether the phase
transition is the first order or cross over, we can also use
the above criterion since at the first order point, 
$d \langle \bar{u}u \rangle^{1/3}/dr = \infty$
holds, therefore it is consistent in both cases.

% Results for the Phase Diagram
%%%%%%%%%%%%%%%%%%%%%%%%%%%%
%\input fig_pd.tex
% --- figure --- %
\begin{figure}[h!]
 \begin{center}
 %\vspace{-0.5cm}
 %\hspace{1.0cm}
  \subfigure{
    \includegraphics[height=4.6cm,keepaspectratio]{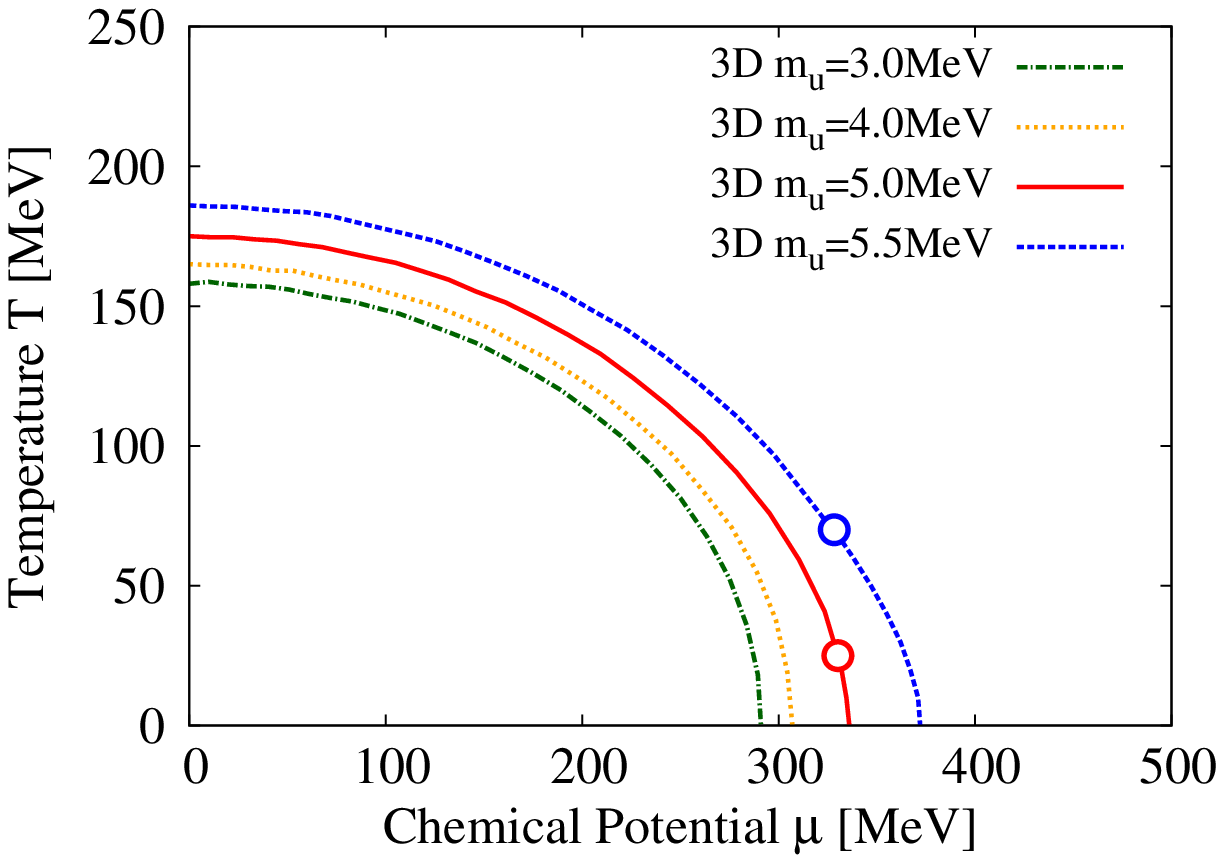} 
  }
  %%\hspace{1.0cm}
  \subfigure{
    \includegraphics[height=4.6cm,keepaspectratio]{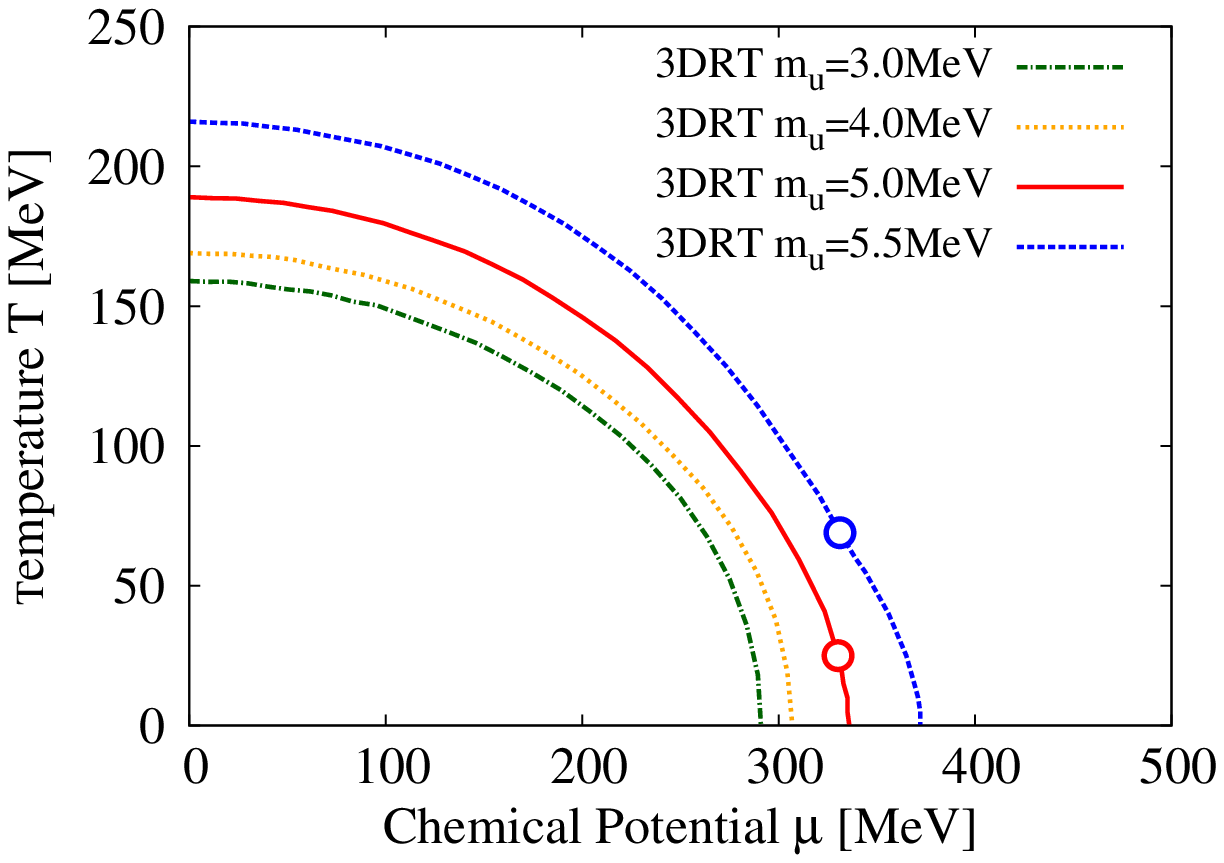} 
  }
  
 %\hspace{1.0cm}
  \subfigure{
    \includegraphics[height=4.6cm,keepaspectratio]{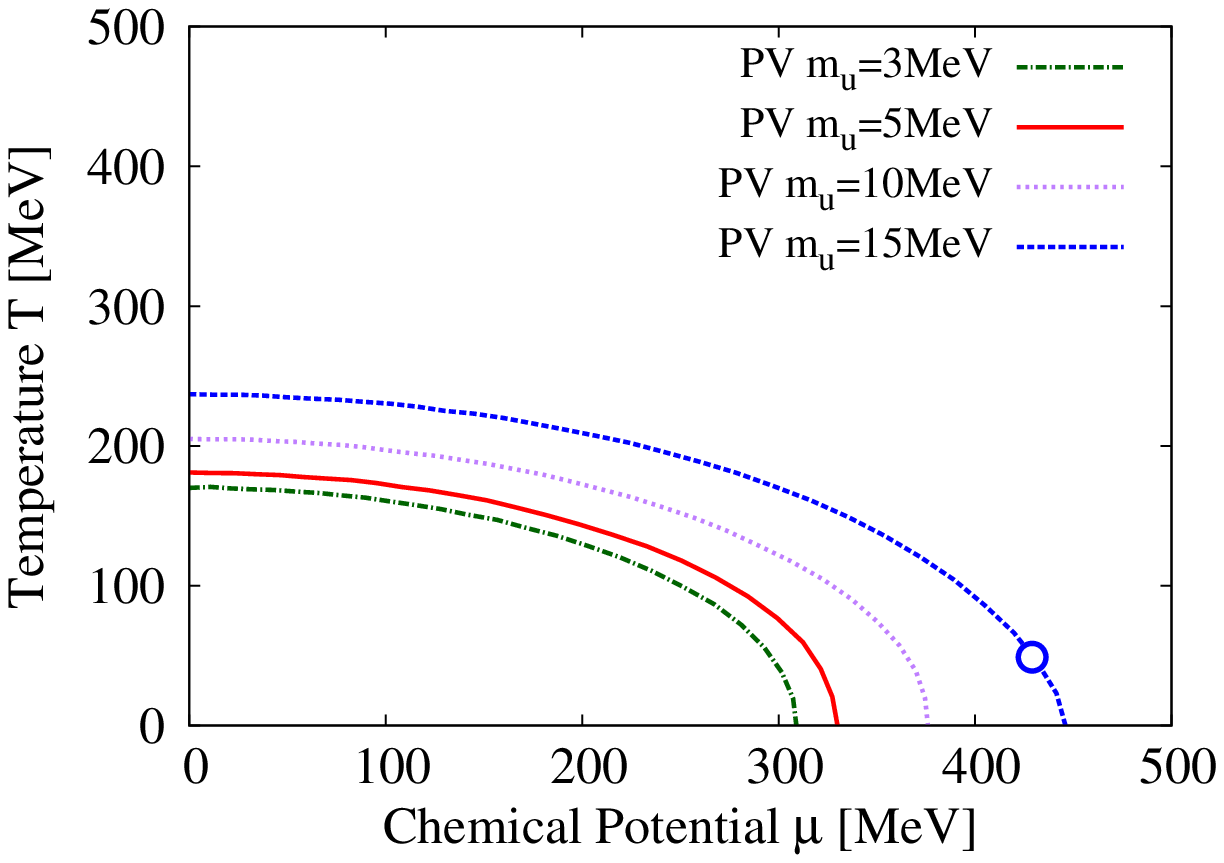} 
  }
  %%\hspace{1.0cm}
  \subfigure{
    \includegraphics[height=4.6cm,keepaspectratio]{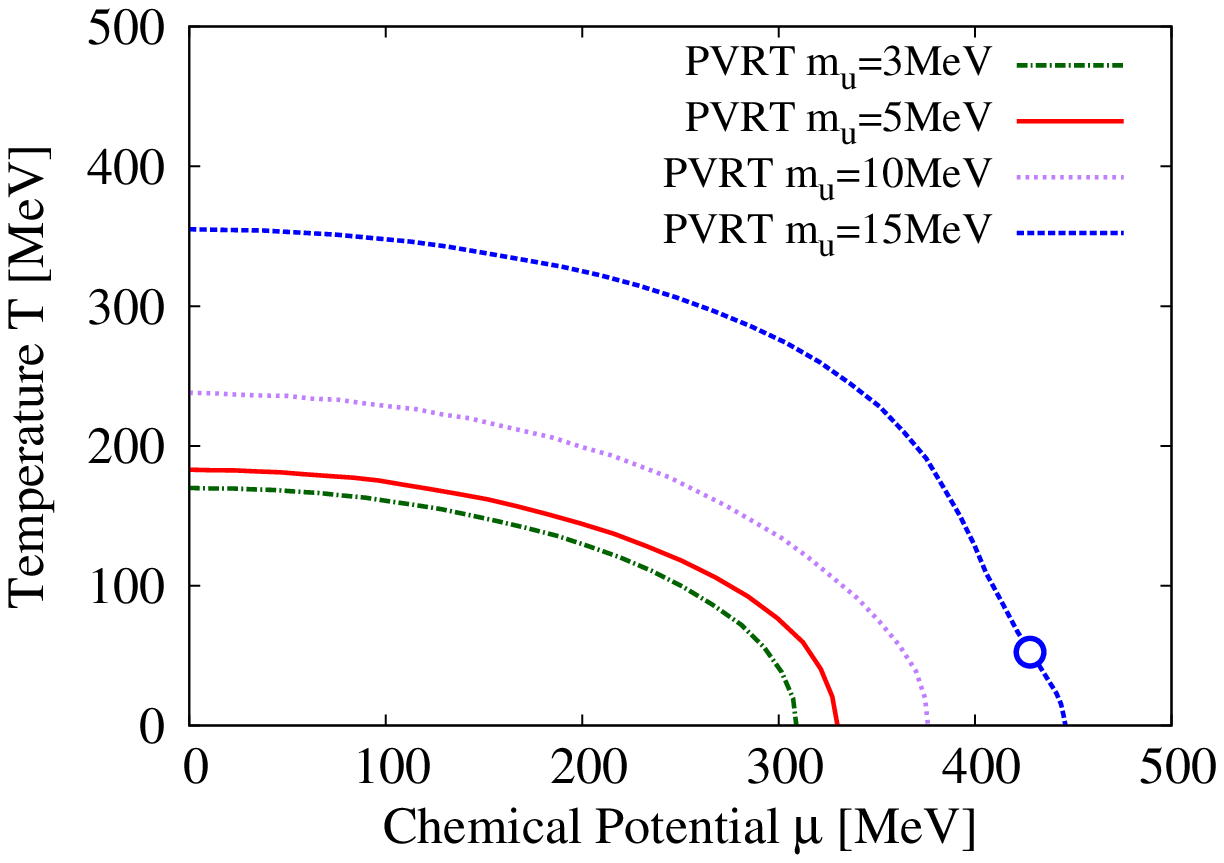} 
  }
  
  %\hspace{1.0cm}
  \subfigure{
    \includegraphics[height=4.6cm,keepaspectratio]{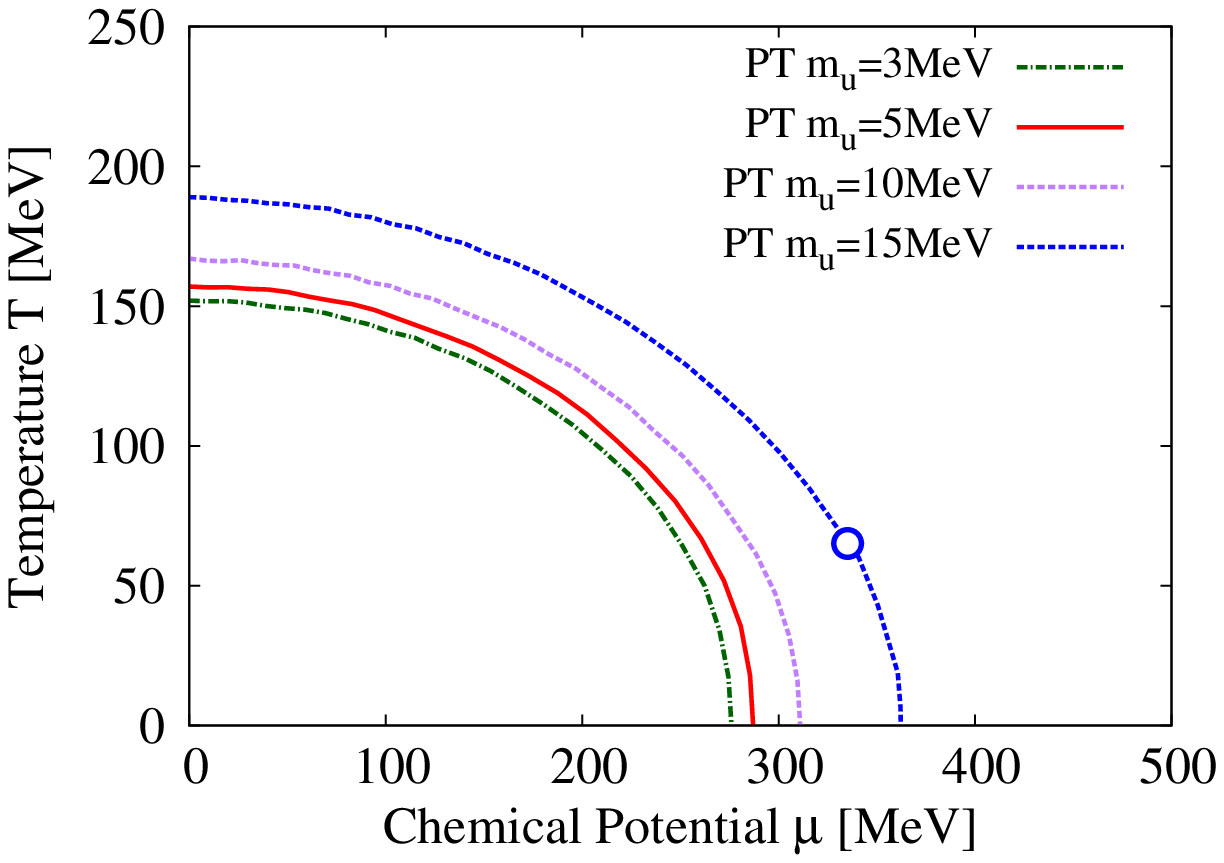} 
  }
  %%\hspace{1.0cm}
  \subfigure{
    \includegraphics[height=4.6cm,keepaspectratio]{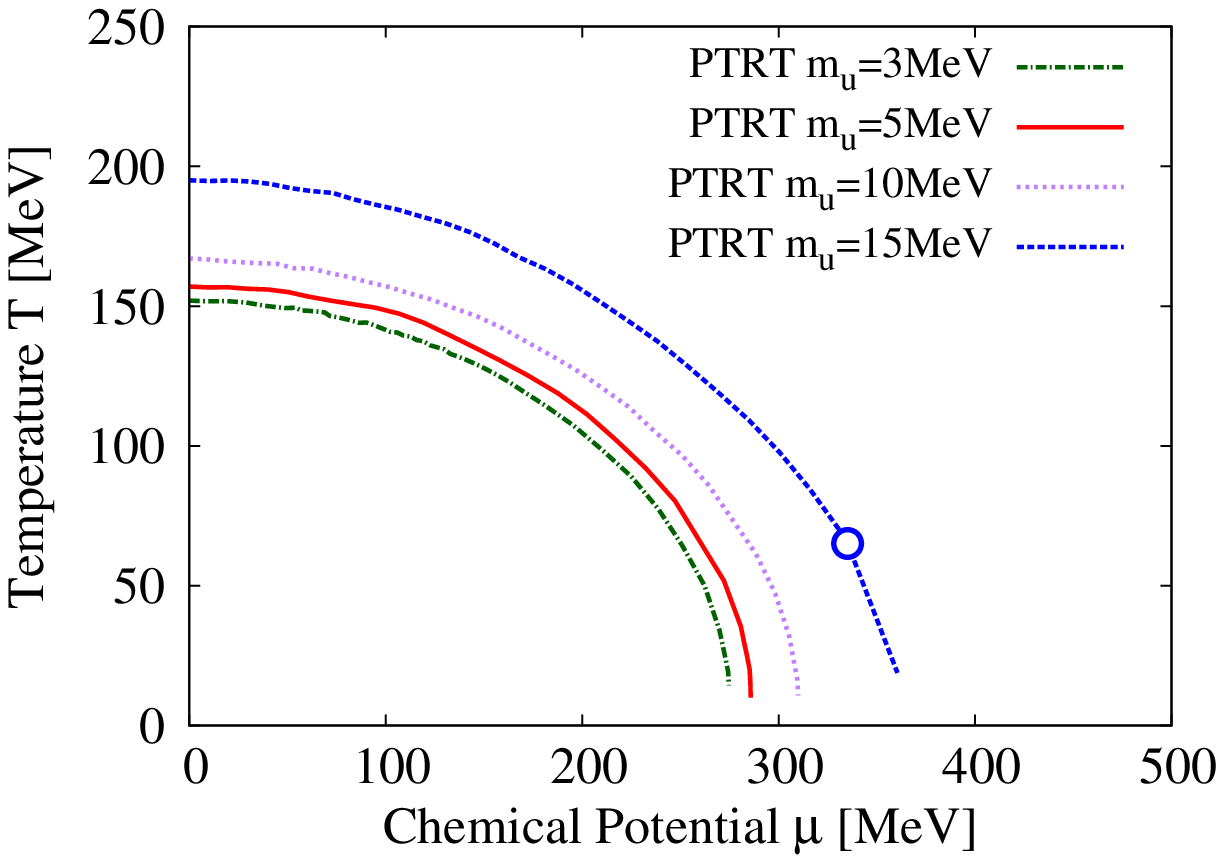} 
  }
  
  %\hspace{1.0cm}
  \subfigure{
    \includegraphics[height=4.6cm,keepaspectratio]{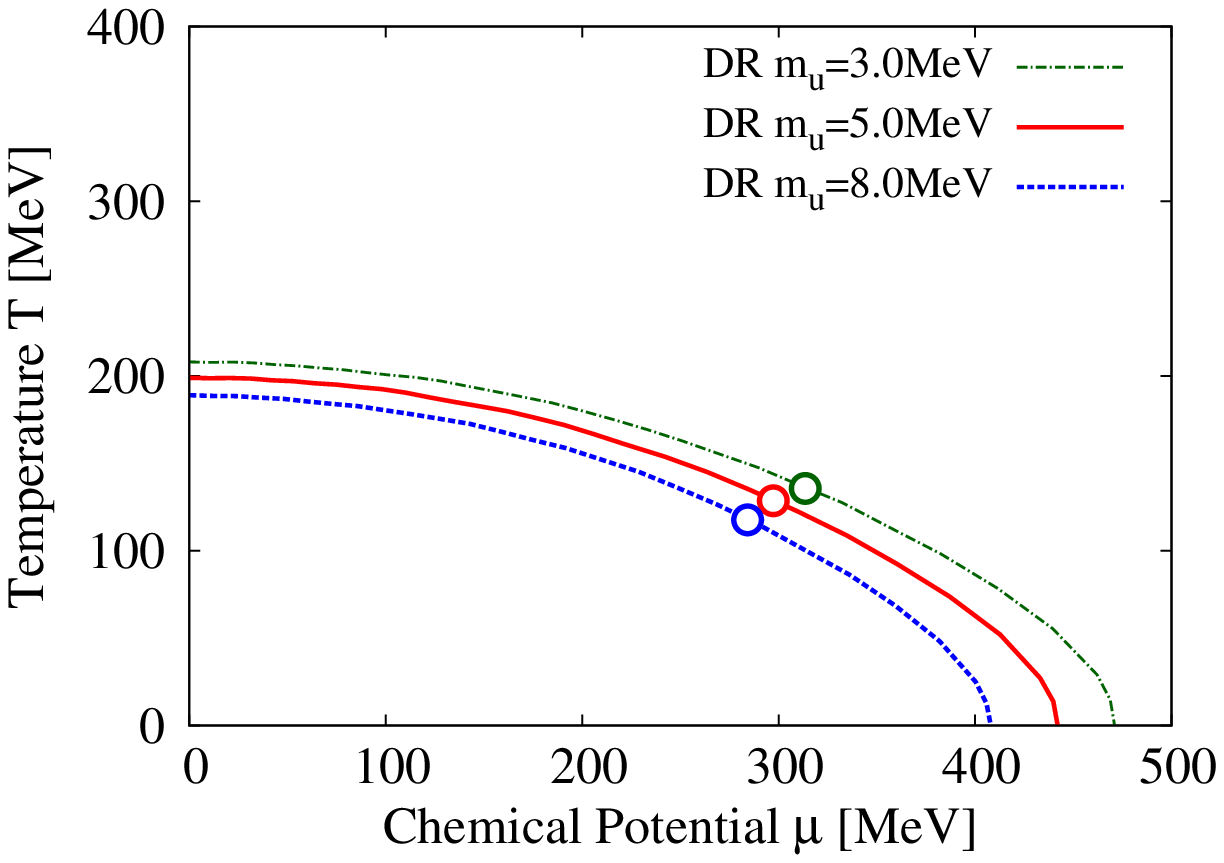} 
  }
  %%\hspace{1.0cm}
  \subfigure{
    \includegraphics[height=4.6cm,keepaspectratio]{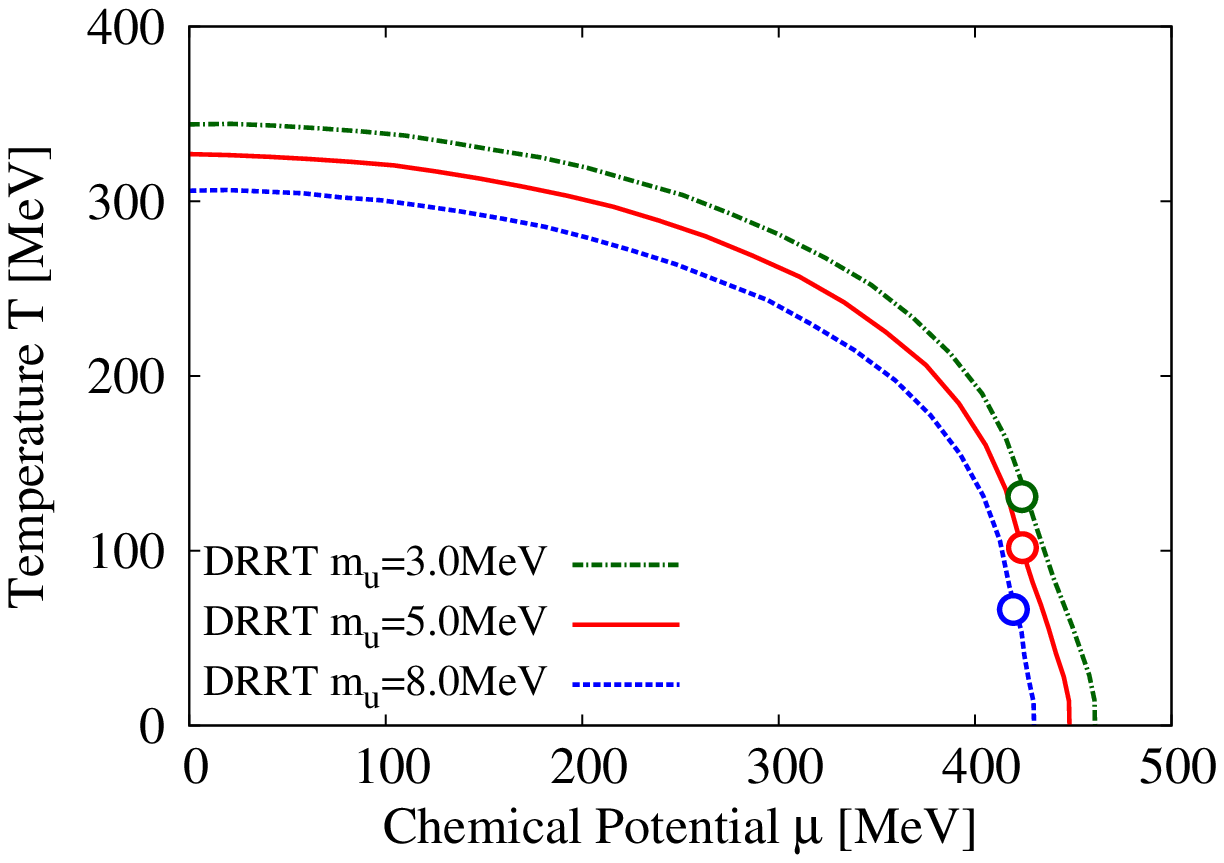} 
  }  
 
  \includegraphics[height=4.6cm,keepaspectratio]{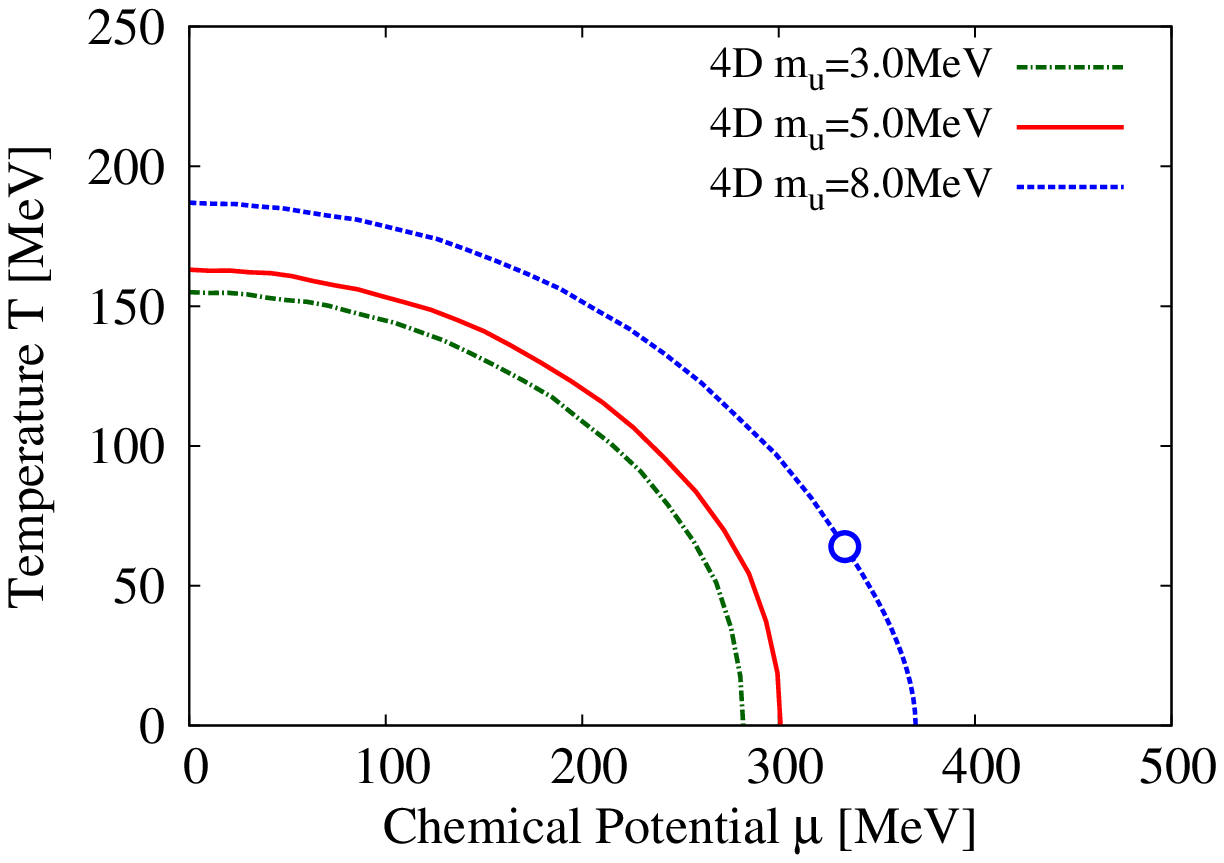}  
  
  \caption{Phase diagram.}
  \end{center}
%\vspace{-0.5cm}
\label{fig_pd}
\end{figure}
% --- figure --- %
%%%%%%%%%%%%%%%%%%%%%%%%%%%%
By searching the maximum number of the differentiate of the chiral
condensate following the condition Eq. (\ref{phasetransition}),
%Evaluating Eq. (\ref{phasetransition}) numerically,
we draw a phase boundary for the chiral symmetry.
%Figure \ref{fig_pd}
%%%%%%
Figure 8
%%%%%%
shows the phase diagrams for the various parameter sets and 
regularizations.
One sees that the critical temperatures, $T_c$ is found
between $150$ and $250$MeV at $\mu=0$ if we regularize
only the temperature independent parts
(left four panels and the bottom panel). 
On the other hand, a higher critical temperature is observed when the regularization
is applied to the temperature dependent and independent parts.
The regularizations $3$DRT and PTRT give a critical temperature around
$T_c \simeq 200$MeV at $\mu=0$, the PVRT induces a
higher critical temperature near $T_c\simeq 400$MeV 
with $m_u=15$MeV, and the DRRT indicates it around $T_c=300-400$MeV. 
One also sees that the critical chemical potential, $\mu_c$,
has no large dependence
on the application of the regularization to
finite temperature term, because the terms are dominated by the step
function $\theta(\mu-m^*)$ as mentioned in the previous section.
Consequently, the area of phase boundary enlarges
in the $T$ direction
when we apply the regularization to the temperature dependent term,
which is numerically confirmed by the figure.

We think the most interesting comparison from the figure is on the 
existence of the critical end point 
%$(T_c,\mu_c)$. 
where the first order phase transition starts on the phase boundary.
For $3$D, $4$D, PV, PT, no critical end point 
appears for the parameter sets with small $m_u$.
While the critical end point always appears in DR.
Then we can numerically conclude
that the DR has stronger tendency of the first order phase transition
comparing with the other regularizations.
The existence of the critical end point
in the other four regularizations can be understood by seeing the value of the parameter $G$. 
Briefly speaking, the critical end point appears when $G$ is large. This is
physically reasonable because $G$ represents the strength of the correlation
between quarks, then the larger $G$ makes the condensation stronger.

%%%%%%%%%%%%%%%%%%%%%%%%%%%%
\section{Phase diagram with fixing $m^*$}
\label{sec_m_fixed}
%%%%%%%%%%%%%%%%%%%%%%%%%%%%
We have seen the phase diagram and how the location of the critical end point
depends on the parameters in the various regularizations. Considering
the fact that the transition $T$ and $\mu$ is essentially determined by the
value of the constituent quark mass since its dependence appears in the 
thermal distribution, $f(E)$, with $E=\sqrt{k^2+m^{*2}}$, we think it may as
well be interesting to compare the phase diagram with the parameter sets
which lead the same value of $m^*$($=311$MeV) at $T=0$ and $\mu=0$ 
instead of fixing the current quark mass, $m_u$.

%%%%%%%%%%%%%%%%%%%%%%%%%%%%
%\input fig_M_C.tex
\begin{figure}[h!]
 \begin{center}
 %\vspace{-0.5cm}
 %\hspace{1.0cm}
  \subfigure{
    \includegraphics[height=4.6cm,keepaspectratio]{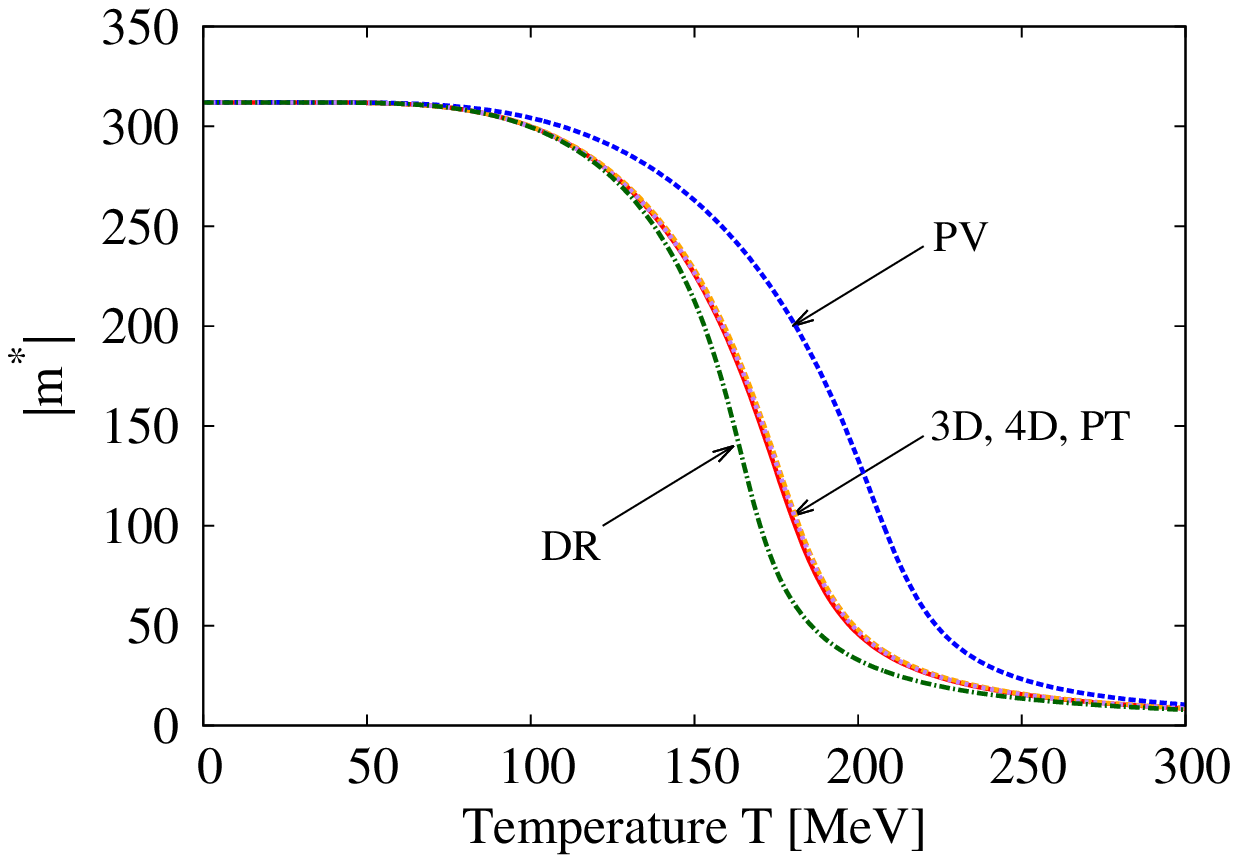} 
  }
   %\hspace{1.0cm} 
    \subfigure{
    \includegraphics[height=4.6cm,keepaspectratio]{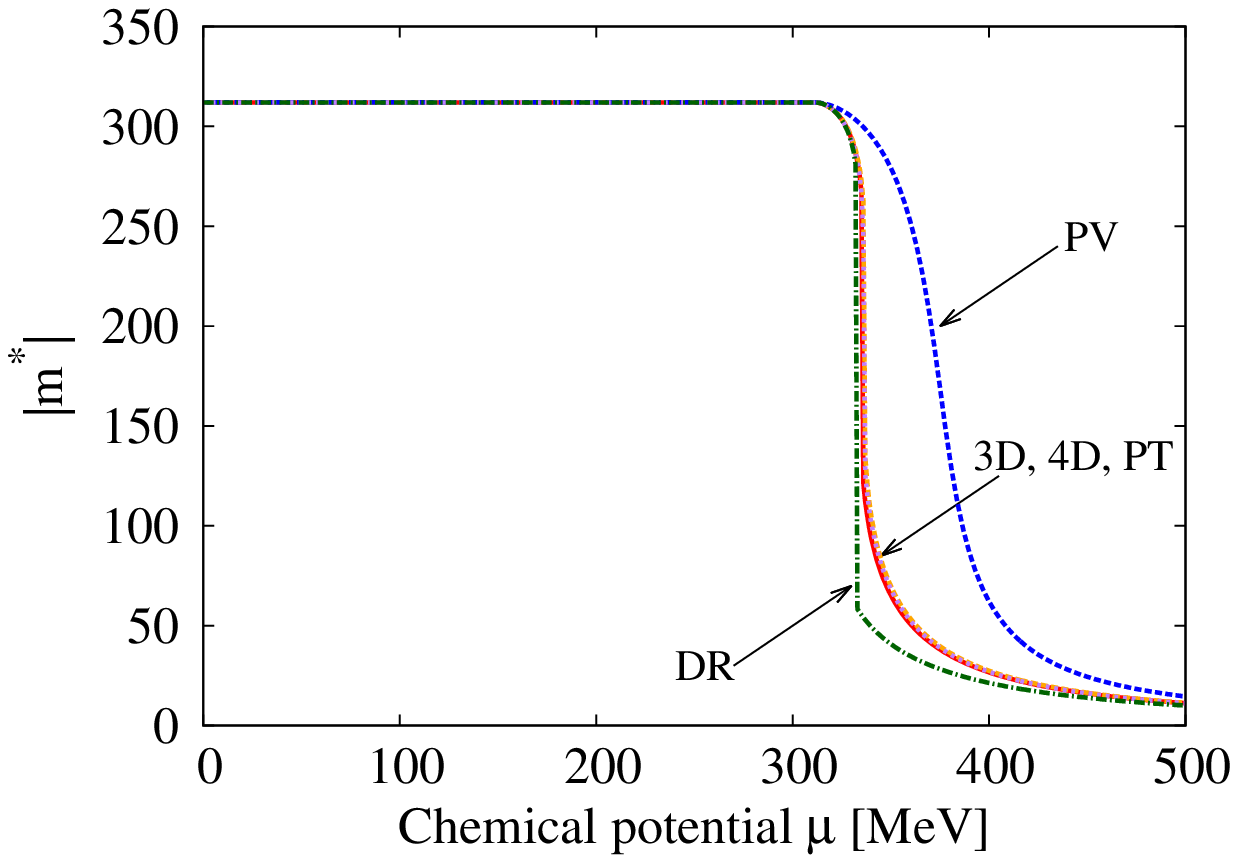} 
  }
  %\vspace{0.7cm}
  \caption{Comparison of the constituent quark mass with fixed parameters
  under $m^*=331$MeV. Left: $\mu=0$.
                Right: $T=0$.}
  \end{center}
%\vspace{-0.5cm}
\label{fig_MC}
\end{figure}
%%%%%%%%%%%%%%%%%%%%%%%%%%%%
First, we see the results based on the one with regularizing
only the temperature independent parts,
3D, 4D, PV, PT and DR. The behaviors of $m^*$ for each regularization
are shown in
%Fig. \ref{fig_MC}.
%%%%%%
Fig. 9.
%%%%%%
All the results have no large difference, 
because the gap equations of the temperature 
independent part for each regularization has the similar behavior. 
The gap equations contain the following form,
%%%%%%
\begin{eqnarray}
{\rm tr}S^0 (m^*) \ni f(m^*,\Lambda) + m^{*2}\ln(g(m^*,\Lambda)) ,
\end{eqnarray}
%%%%%%
in the regularizations, 3D, 4D, PV and PT.
While in DR $m^*$ and $\Lambda$ are replaced by $M_0$ and $D$ 
in some parts (see appendix \ref{app_DR}). In 
%Fig. \ref{fig_MC},
%%%%%%
Fig. 9,
%%%%%% 
we note that the results almost coincide in
three regularizations, 3D, 4D and PT. The behavior in PV shows
a smaller and DR gives %a
steeper slope than the others.

% Results for the Phase Diagram Comparison
%%%%%%%%%%%%%%%%%%%%%%%%%%%%
%\input fig_pd_C.tex
\begin{figure}[h!]
 \begin{center}
 %\vspace{-0.5cm}
 %\hspace{1.0cm}
    \includegraphics[height=4.6cm,keepaspectratio]{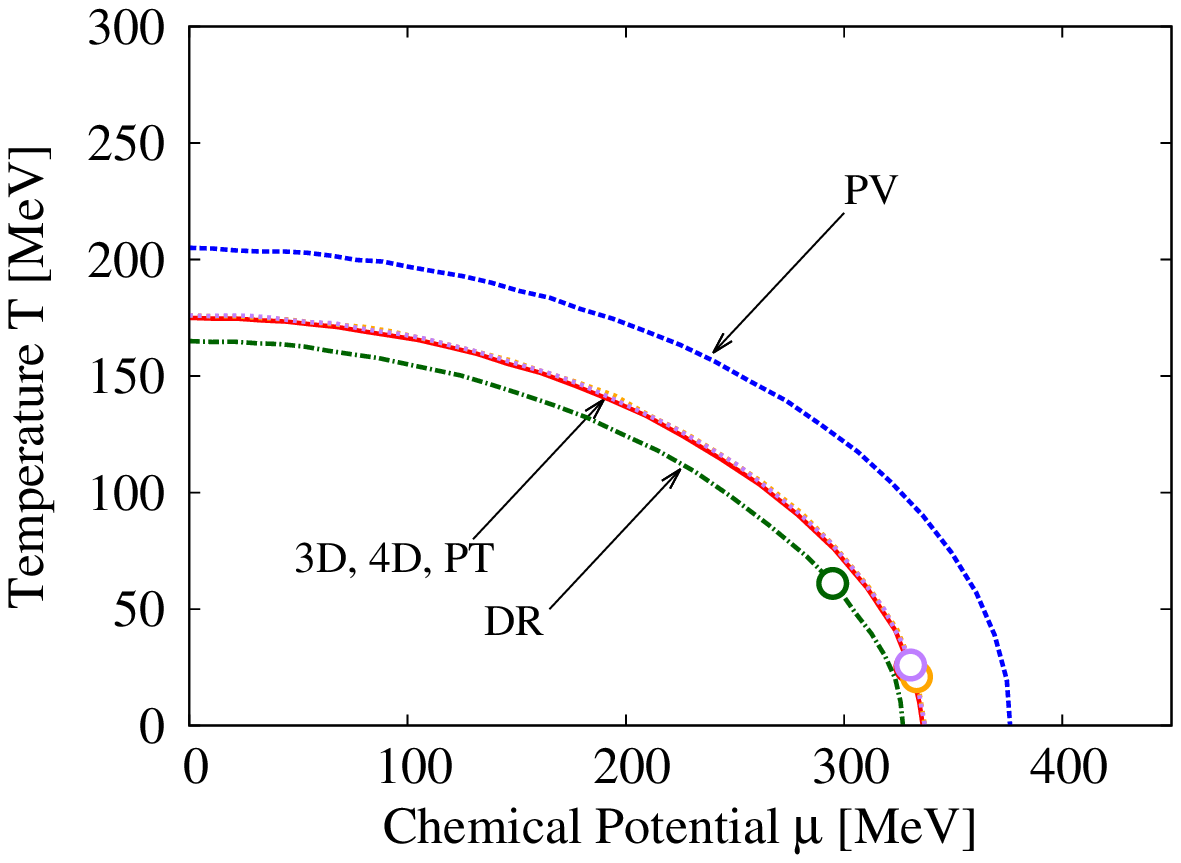} 
  %\vspace{0.7cm}
  \caption{Comparison of the phase diagrams with fixed parameters under
  $m^*=331$MeV.}
  \end{center}
%\vspace{-0.5cm}
\label{fig_pdC}
\end{figure}
%%%%%%%%%%%%%%%%%%%%%%%%%%%%
In
%Fig. \ref{fig_pdC}
%%%%%%
Fig. 10
%%%%%%
the phase diagram is illustrated in each regularization.
The phase boundaries of 3D, 4D and PT 
show almost equivalent behavior. The area
of the chiral symmetry 
broken phase for PV is larger than the others. This tendency comes 
from the behavior of $m^*$ in 
%Fig. \ref{fig_MC}.
%%%%%%
Fig. 9.
%%%%%%
Since the %$\Lpt$
$\Lpv$ 
is entered in the form of the dynamical mass, the 
chiral symmetry breaking contribution is enhanced than the other 
regularizations.  The area for the chiral symmetry broken phase for the
DR is smaller than the others. The critical end point for the DR locates 
higher temperature than the one for 3D, 4D and PT. These tendency
also comes from the behavior of $m^*$ in
%Fig. \ref{fig_MC}.   
%%%%%%
Fig. 9.
%%%%%%

%%%%%%%%%%%%%%%%%%%%%%%%%%%%
%\input fig_M_Crt.tex
\begin{figure}[h!]
 \begin{center}
 %\vspace{-0.5cm}
 %\hspace{1.0cm}
  \subfigure{
    \includegraphics[height=4.6cm,keepaspectratio]{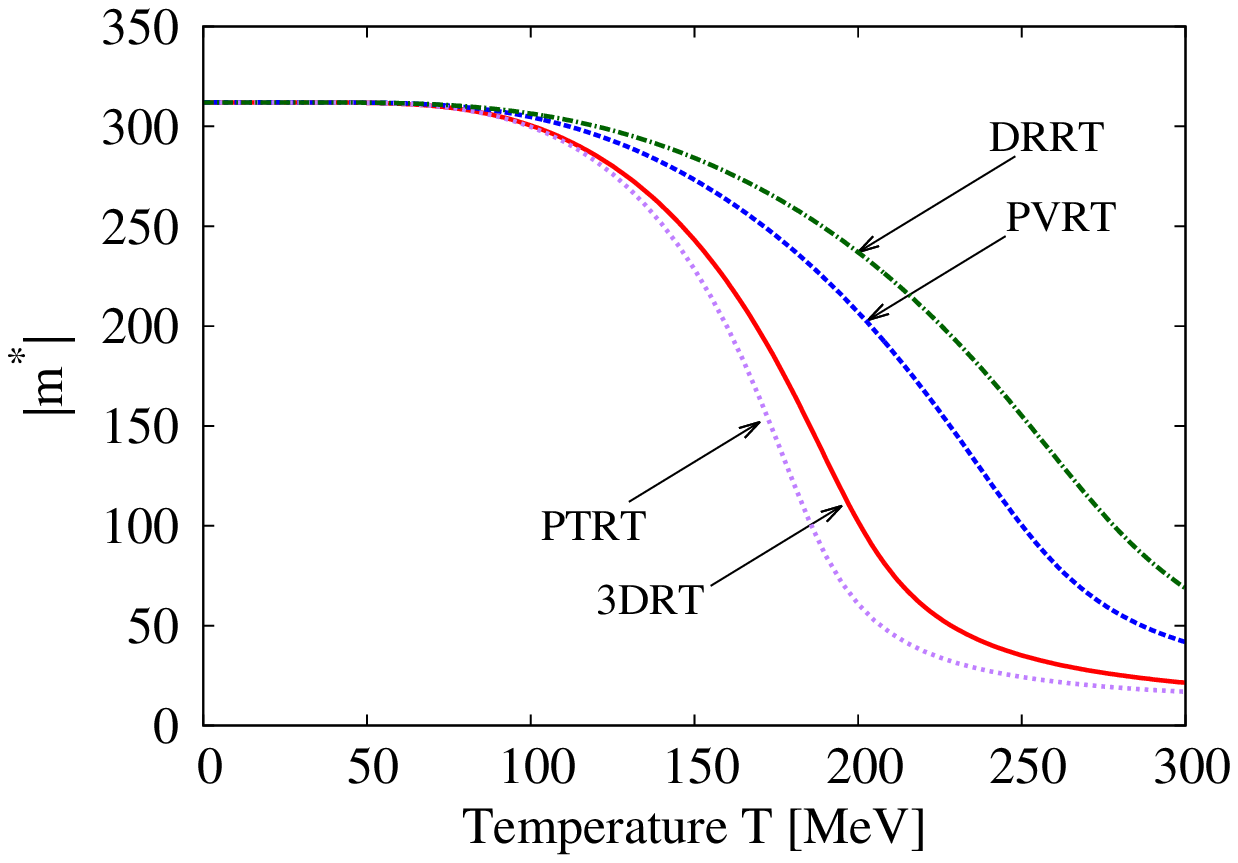} 
  }
   %\hspace{1.0cm} 
  \subfigure{
    \includegraphics[height=4.6cm,keepaspectratio]{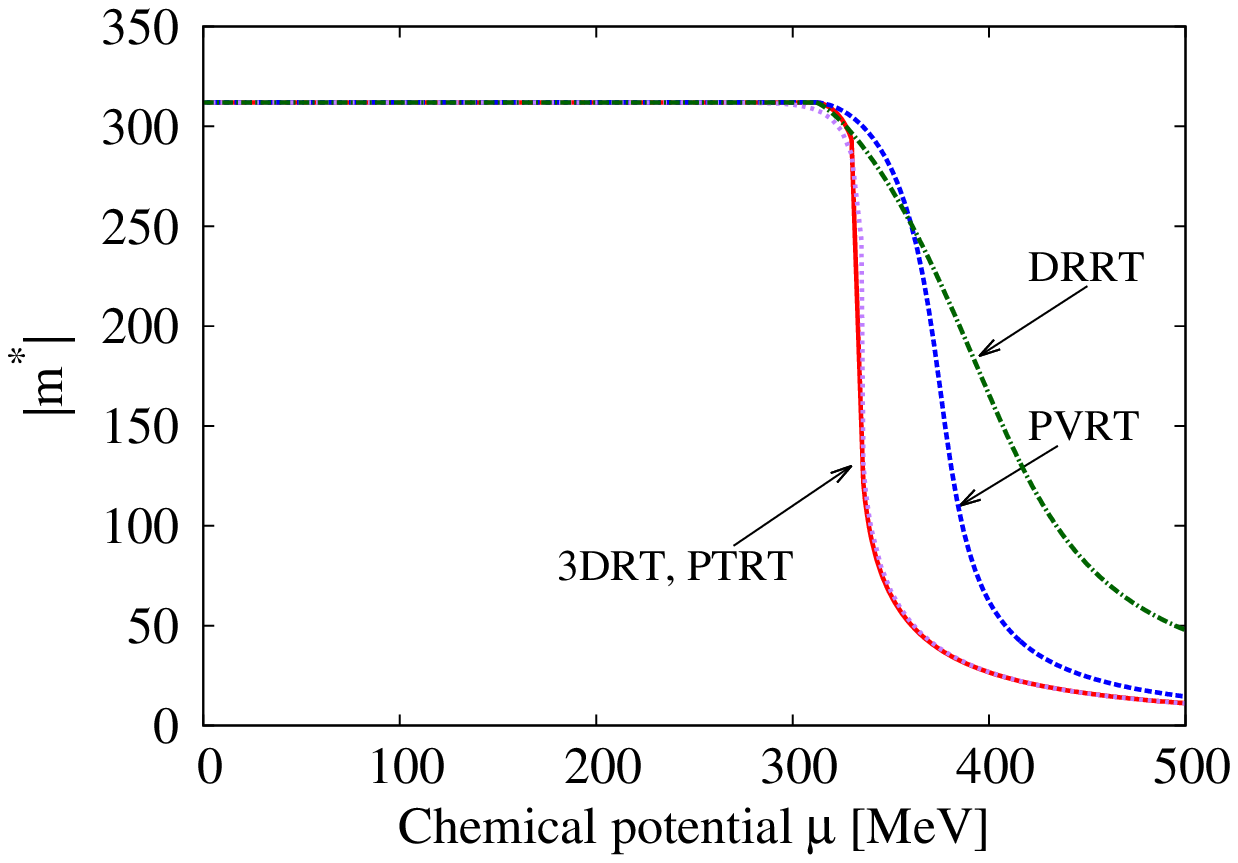} 
  }
  %\vspace{0.7cm}
  \caption{Comparison of the constituent quark mass with fixed parameters
  under $m^*=331$MeV. Left: $\mu=0$. Right: $T=0$ for 3DRT, PVRT, DRRT
  and $T=10$MeV for PTRT.}
  \end{center}
%\vspace{-0.5cm}
\label{fig_MCrt}
\end{figure}
%%%%%%%%%%%%%%%%%%%%%%%%%%%%
Next, we discuss the results with regularizing
both the temperature independent and dependent parts,
3DRT, PVRT, PTRT and DRRT. The behavior of $m^*$ for each 
regularization is shown in 
%Fig. \ref{fig_MCrt}.
%%%%%%
Fig. 11.
%%%%%%
We find that the finite temperature effect becomes smaller or softer than 
3D, PV, PT and DR, respectively.
The behavior of $m^*$ in DRRT is the most closest behavior to the
case of its regularizing only temperature independent part.

%in order of PTRT, 3DRT, PVRT and DRRT. 
%On the other hand,
%the effect of the chemical potential in 
%3DRT, PTRT, and PVRT 
%is very similar to compare with 3D, PT and PV. 
The chemical potential has the similar contribution for
3DRT, PTRT, and PVRT.
The behaviors of $m^*$ in DRRT and DR have %a
large difference.
This difference is caused by the reduction of the momentum
integral dimension, $D$, from $4$ to $3.32$.

%%%%%%%%%%%%%%%%%%%%%%%%%%%%
%\input fig_pd_Crt.tex
\begin{figure}[h!]
 \begin{center}
 %\vspace{-0.5cm}
 \hspace{1.0cm}
    \includegraphics[height=4.6cm,keepaspectratio]{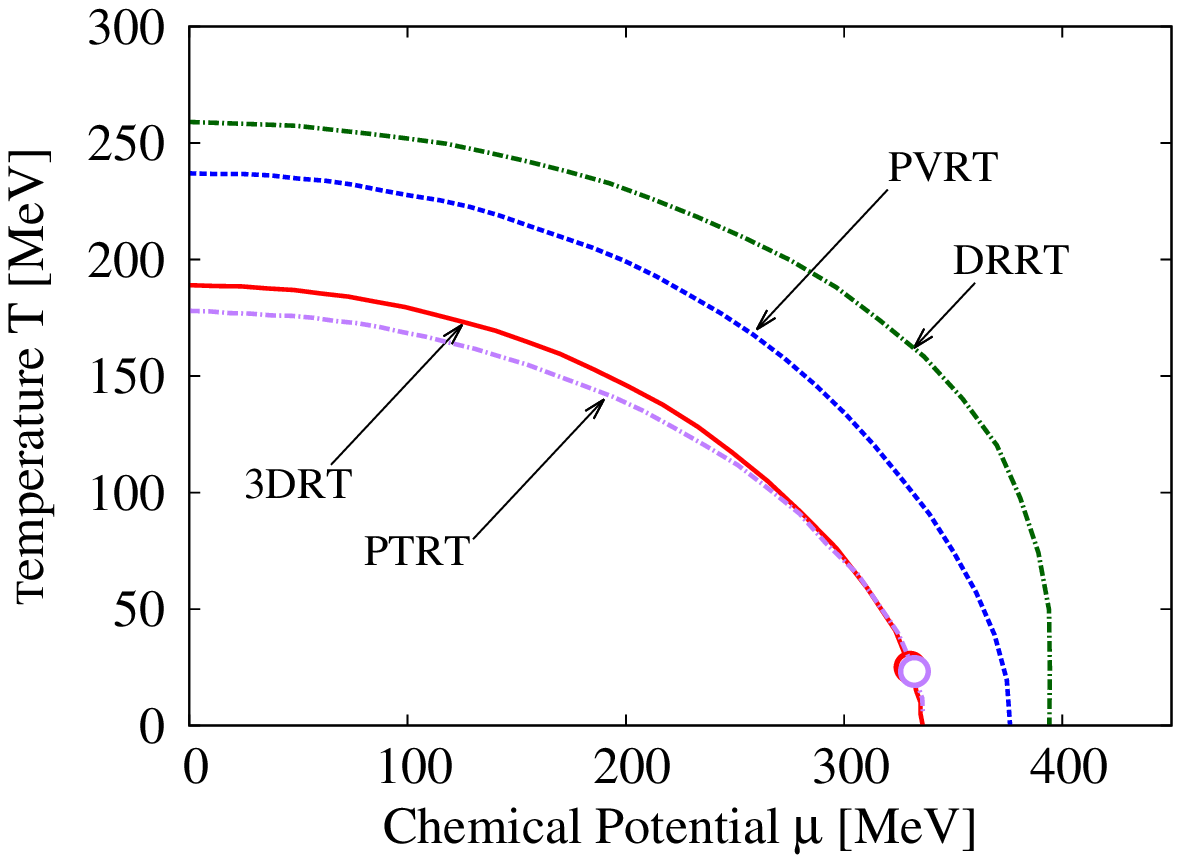} 
  %\vspace{0.7cm}
  \caption{Comparison of the phase diagrams with fixed parameters under
  $m^*=331$MeV.}
  \end{center}
%\vspace{-0.5cm}
\label{fig_pdCrt}
\end{figure}
%%%%%%%%%%%%%%%%%%%%%%%%%%%%
The phase diagram in each regularization is shown in
%Fig. \ref{fig_pdCrt}.
%%%%%%
Fig. 12.
%%%%%%
To compare with
%Fig.\ref{fig_pdC},
%%%%%%
Fig. 10,
%%%%%%
the region of the broken phase enlarges
in PTRT, 3DRT and PVRT for a low chemical potential. However, the critical 
chemical potentials in PTRT, 3DRT and PVRT for a low temperature are
almost equivalent to that in PT, 3D and PV, respectively. 
From the behavior of $m^*$ in
%Fig. \ref{fig_MCrt},
%%%%%%
Fig. 11,
%%%%%%
we observe a larger
critical temperature and chemical potential for DRRT. We display the
location of the critical end points $(\mu_{\rm CP}, T_{\rm CP})$ for each
regularization in Table.\ref{CEP}.
%%%%%%
\begin{table}[h!]
\caption{Critical end point for each regularization}
\label{CEP}
%\begin{ruledtabular}
\begin{center}
\begin{tabular}{ccc}
\hline
Regularization & $\mu_{\rm CP}$ (MeV)  & $T_{\rm CP}$ (MeV) \\
\hline
3D & 330 & 25.0 \\
4D & 333 & 46.8 \\
PT & 330 & 26.0 \\
DR & 289 & 74.7 \\
3DRT & 330 & 25.0 \\
PTRT & 332 & 23.2 \\
\hline
\end{tabular}
\end{center}
%\end{ruledtabular}
\end{table}
%%%

%%%%%%%%%%%%%%%%%%%%%%%%%%%%
\section{Discussions}
\label{sec_discussion}
%%%%%%%%%%%%%%%%%%%%%%%%%%%%
We have introduced the regularization methods then studied the meson
properties and phase diagram in previous sections. In this section, we 
are going to present more detailed discussions on the obtained results.

In comparing the left panels v.s. right panels in
%Fig. \ref{fig_pd},
%%%%%%
Fig. 8,
%%%%%%
one can observe the contribution to apply the regularization procedure
to the thermal correction.
The phase diagram does not show considerable difference for $3$D
and PT cases, while in PV the area of the phase boundary becomes
larger in the right panel when $m_u$ is
large, and the areas in DRRT case is larger than DR case for all the
parameter sets. These difference can be understood through the 
following discussion.

The loop integral,
${\mathcal I}$, essentially
has the following subtracted forms for 
$3$DRT, PVRT, and PTRT,
%%%%%%%
\begin{align}
  &{\mathcal I}_{\rm 3DRT}
  = \int_0^{\infty} \md^3 k \,\,  F(k)
    -\int_{\Lambda}^{\infty} \md^3 k \,\,  F(k) ,\\
  &{\mathcal I}_{\rm PVRT}
  = \int_0^{\infty} \md^3 k \,\,  
     \left[F(k,m^*) - F (k,\Lambda)  \right] ,\\ 
  &{\mathcal I}_{\rm PTRT}
  = \int_0^{\infty} \md^3 k \,\,  
     \left[F(k) - F(k)(1-e^{-(k^2 + m^{*2})/\Lambda^2})
     \right].
\end{align}
%%%%%%%%
Where the typical form of $F(k,m^*)$ is given by $F(k,m^*) = C f(E)/E$
with some constant value, $C$.
Thus the subtracted terms basically relate to the suppression on the high
energy contributions which are expected to be small.
We numerically confirmed that the subtracted parts are small for almost
all the cases, then the phase diagrams does not change drastically.
However, in the PV case with large $m_u$, 
the difference between $F(k,m^*)$ and $F(k,\Lambda)$ is small
since the constituent quark mass becomes comparable to the cutoff
scale, e.g., $m^* = 417$MeV and $\Lambda_{\rm PV}=729$MeV for
$m_u=15$MeV in PV. Consequently,
the thermal contribution strongly suppressed in the PV case
with large $m_u$.

We saw that the area of the phase boundary does
not alter so much in
$3$D, PV and PT regularizations.
The essential reason is that
the infinities appearing from loop integrals are subtracted at high energy.
However in DR, the situation is different since the integral is replaced as
%%%%%%%
\begin{align}
  \int_0^{\infty} \md^4 k \,\,  F(k)
  \to
  \int_0^{\infty} \md^D k \,\,  F(k),
\end{align}
%%%%%%%%
so this modifies the integral kernel
rather than the subtraction of high energy modes. This is the reason
why DR shows considerable difference, if we apply
the regularization to both  temperature independent and 
dependent contributions.

We also saw the location (or existence) of the critical end point is
non-trivial. 
In
%Fig. \ref{fig_pd2},
%%%%%%
Fig. 10,
%%%%%%
the diagram has the critical end point
for $3$D, $4$D, PT and DR. No critical end point appears in the PV.
Thus we find that the PV has weaker tendency of the first order phase
transition than that of the other regularization
methods. Particularly, the temperature of the critical end point in the
DR case is higher then others, which may enable us to conclude that
the DR has the stronger tendency of the first order phase transition.

%%%%%%%%%%%%%%%%%%%%%%%%%%%%
\section{Concluding remarks}
\label{sec_conclusion}
%%%%%%%%%%%%%%%%%%%%%%%%%%%%
We have studied the regularization dependence on the
phase diagram of quark matter on $T-\mu$ plane by using the
NJL model. We have first presented the regularization procedure at
finite temperature and chemical potential, then fitted parameters
within various regularization methods. Thereafter, we have studied the
meson properties and the phase structure.

We find that the model produces the reliable predictions on the
meson properties whose behavior for finite temperature and 
chemical potential does not alter drastically, which indicates that
all the regularizations employed in this paper nicely capture
physics on the meson properties. 
We can conclude that the regularizations are safely adopted. 
In this context the regularization parameter independent 
approach is also interesting \cite{inagaki:2013,inagaki:2014}.

It is expected that observation of the critical end point 
can distinguish a suitable regularization for an effective model of
QCD by comparing with the resulting phase diagrams.
The important difference is the existence of the critical end point.
The model predicts that the critical end
point appears at intermediate chemical potential around
$\mu \simeq 300-400$MeV. This density coincides the one in
which different quark state such as color superconductivity may
occur, there the order of the phase transition might affect crucially
on such the dense states. Moreover the color superconductivity may
be realized in the dense stellar objects, like quark stars and neutron
stars \cite{fujihara:2009}. Therefore the study of the order of the phase
transition has important meaning as well in cosmological observations.
So we believe that the further and more extensive investigations are 
necessary on this subject.

%%%%%%%%%%%%%%%%%%%%%%%%%%%%
\acknowledgments
TI is supported by JSPS KAKENHI Grant Number 26400250.
HK is supported by MOST 103-2811-M-002-087.
%%%%%%%%%%%%%%%%%%%%%%%%%%%%

%%%%%%%%%%%%%%%%%%%%%%%%%%%%
\appendix
%%%%%%%%%%%%%%%%%%%%%%%%%%%%
\section{Analytic expressions for $I^0(p^2)$}
\label{app_I0}
%%%%%%%%%%%%%%%%%%%%%%%%%%%%
Since the $I^0(p^2)$ integral can be evaluated analytically, we will present
the explicit expression for various regularizations.

One needs special care in performing $I(p^2)$ integral since it contains
divergent contribution as seen in the 3D cutoff scheme.
$I_{\rm 4D}^0$ becomes for $m^{*2} > p^2/4$,
%%%%%%%
\begin{align}
  I^0_{\rm 4D}(p^2)
  = \frac{N_c}{4\pi^2}
   \biggl[ 
     & \ln \frac{\Lfd^2 + m^{*2} }{m^{*2}}
        +4a \arctan \left( \frac{1}{2a} \right) \nonumber \\
     & -4b \arctan \left( \frac{1}{2b} \right)
        -\frac{2\Lfd^2}{a p^2} \arctan \left( \frac{1}{2a} \right)
   \biggr],
\end{align}
%%%%%%%%
and for $m^{*2} < p^2/4$,
%%%%%%%
\begin{align}
  I^0_{\rm 4D}(p^2)
  = \frac{N_c}{4\pi^2}
   \biggl[ 
     & \ln (\Lfd^2 + m^{*2} )
        +4a \arctan \left( \frac{1}{2a} \right) \nonumber \\
     & +2 -2 \sum_{\pm} 
        \left(\frac{1}{2} \pm a \right) \ln  \left( \frac{1}{2} \pm a \right) 
   \biggr],
\end{align}
%%%%%%%%
where
%%%%%%%
\begin{align}
 a = \sqrt{ \frac{\Lfd^2 +m^{*2}}{p^2} -\frac{1}{4} }, \quad
 b = \sqrt{ \frac{\Lfd^2}{p^2} -\frac{1}{4}}.
\end{align}
%%%%%%%%

Concerning on $I_{\rm PV}^0(p^2)$, it is convenient that we divide the integral
as
%%%%%%
\begin{equation}
  I_{\rm PV}^0 = I_{\rm PV}^{0(m)} - I_{\rm PV}^{0(\Lambda)},
\end{equation}
%%%%%%
where $I_{\rm PV}^{0(\Lpv)}$ is subtracted part in the original
integral and it becomes
%%%%%%
\begin{align}
  I_{\rm PV}^{0(\Lambda)}
  &=-\frac{N_c}{2\pi^2} \sum_{\pm}
    \left[
      \left( \frac{1}{2} \pm c \right) 
      \ln \left( \frac{1}{2} \pm c \right)
    \right],
\end{align}
%%%%%%
with
%%%%%%%
\begin{align}
 c = \sqrt{\frac{(\Lpv^2 - m^{*2}+k^2 )^2}{4 p^4} 
        - \frac{\Lpv^2}{p^2} }.
\end{align}
%%%%%%%%
As seen above, we need to separately evaluate the integral
$I_{\rm PV}^{0(m)}$ depending on the values of $m^{*2}$ and $p^2$.
It becomes for $m^{*2} > p^2/4$,
%%%%%%
\begin{align}
  I_{\rm PV}^{0(m)}
  = -\frac{N_c}{4\pi^2}
  \left[ 
    \ln \left( m^{*2} \right)
    +4d \arctan\left(\frac{1}{2d}\right) 
  \right],
\end{align}
%%%%%%
and for $m^{*2} < p^2/4$,
%%%%%%
\begin{align}
  I_{\rm PV}^{0(m)}
  &=-\frac{N_c}{2\pi^2} \sum_{\pm}
    \left[
      \left( \frac{1}{2} \pm h \right) 
      \ln \left( \frac{1}{2} \pm h \right)
    \right],
\end{align}
%%%%%%
where
%%%%%%%
\begin{align}
 d = \sqrt{ \frac{m^{*2}}{p^2} -\frac{1}{4}}, \quad
 h = \sqrt{ \frac{1}{4}  - \frac{m^{*2}}{p^2}}.
\end{align}
%%%%%%%%

%%%%%%%%%%%%%%%%%%%%%%%%%%%%
\section{${\rm tr}S^0_{\rm DR}$ and $f_\pi$ for $D \simeq 2, 3, 4$}
\label{app_DR}
%%%%%%%%%%%%%%%%%%%%%%%%%%%%
We arrange
the concrete expressions for ${\rm tr}S^0_{\rm DR}$ and $f_\pi$.

For $D \simeq 2, (D = 2 + 2\epsilon)$ we have
\begin{eqnarray}
{\rm tr}S^0_{\rm DR} &\simeq& \frac{N_c}{2\pi^2} m^* \cdot M_0^2
\left[ \frac1{\epsilon} + \gamma_E - \ln(2\pi) +
\ln\frac{m^{*2}}{M_0^2} \right], \\
f_{\pi {\rm DR}}^2 &\simeq& \frac{N_c}{2\pi} M_0^2
\left[ 1 + \epsilon\left\{\gamma_E - \ln(2\pi) +
\ln\frac{m^{*2}}{M_0^2} \right\}\right].
\end{eqnarray}

For $D \simeq 3, (D = 3 + 2\epsilon)$ we have
\begin{eqnarray}
{\rm tr}S^0_{\rm DR} &\simeq& \frac{N_c}{\sqrt{2}\pi} m^* 
\cdot \sqrt{m^{*2}} M_0
\left[ 1 + \epsilon \left\{\gamma_E + \ln \frac{2}{\pi} -2 
+ \ln\frac{m^{*2}}{M_0^2} \right\}\right], \\
f_{\pi {\rm DR}}^2 &\simeq& \frac{N_c}{2\sqrt{2}\pi}  
\sqrt{m^{*2}} M_0
\left[ 1 + \epsilon \left\{\gamma_E + \ln \frac{2}{\pi}  
+ \ln\frac{m^{*2}}{M_0^2} \right\}\right].
\end{eqnarray}

For $D \simeq 4, (D = 4 - 2\epsilon)$ we have
\begin{eqnarray}
{\rm tr}S^0_{\rm DR} &\simeq& \frac{N_c}{4\pi^2} m^* \cdot m^{*2}
\left[ \frac1{\epsilon} - \gamma_E + \ln(2\pi) +1 +
\ln\frac{M_0^2}{m^{*2}} \right], \\
f_{\pi {\rm DR}}^2 &\simeq& \frac{N_c}{4\pi^2} m^{*2}
\left[ \frac1{\epsilon} - \gamma_E + \ln(2\pi) +
\ln\frac{M_0^2}{m^{*2}} \right].
\end{eqnarray}

%%%%%%%%%%%%%%%%%%%%%%%%%%%%%
%%%%%%%%%%%%%%%%%%%%%%%%%%%%%

%%%
\end{document}